\documentclass[english,12pt]{article}
\usepackage{lmodern}
\usepackage[utf8]{inputenc}
\usepackage{enumerate}
\usepackage[top=1.25in, bottom=1.25in, left=1.25in, right=1.25in]{geometry}
\usepackage{amsopn,amsfonts,amsmath,amssymb,mathrsfs,amsthm,dsfont}
\usepackage[pdftex]{graphicx}
\usepackage{verbatim}
\usepackage[width=\textwidth]{caption}
\usepackage{booktabs}
\usepackage[english]{babel}
\usepackage{subfigure,float,placeins,flafter,longtable,multirow,array}
\usepackage[shortcuts]{extdash}
\usepackage{natbib}
\usepackage{color}
\usepackage[flushleft]{threeparttable}
\usepackage[usenames,dvipsnames,svgnames,table]{xcolor}
\usepackage[hidelinks]{hyperref}
\usepackage{layout}

\usepackage{algorithm}
\usepackage{algorithmicx}
\usepackage{algpseudocode}
\usepackage[usenames,dvipsnames,svgnames,table]{xcolor}
\allowdisplaybreaks

\newtheorem{thm}{Theorem}
\newtheorem{cor}{Corollary}
\newtheorem{prop}{Proposition}
\newtheorem{lem}{Lemma}
\newtheorem{hyp}{Assumption}

\newtheorem{remark}{Remark}[section]

\def\cvx{\succ_{\!\text{cv}}}	
\def\eqd{\stackrel{d}{=}}	
	
\def\B{\mathcal{B}}	
\def\Bcon{\B^{\text{\con}}}
\def\con{\text{con}}
\def\S{\mathcal{S}}

\def\N{{\mathbb N}}

\def\argmin{{\rm argmin}}
\def\CI{\text{CI}_{1-\alpha}}
\def\CR{\text{CR}_{1-\alpha}}

\newcommand{\R}{\mathbb{R}}

\newcommand{\norm}[1]{\left\|#1 \right\|}

\newcommand{\Supp}{\text{Supp}}
\newcommand{\cov}{\text{cov}}
\newcommand{\eps}{\varepsilon}
\newcommand{\deriv}[2]{\partial #1/\partial #2}
\newcommand{\Deriv}[2]{\frac{\partial #1}{\partial #2}}
\newcommand{\indic}[1]{\mathds{1}\left\{#1\right\}}
\newcommand{\indep}{\perp \!\!\! \perp}
\newcommand{\convP}{\stackrel{\mathbb{P}}{\longrightarrow}}

\newcommand{\convD}{\stackrel{d}{\longrightarrow}}

\newcommand{\Cov}{\text{Cov}}

\date{}

\begin{document}

\title{Partially Linear Models under Data Combination\thanks{We thank the Editor, Francesca Molinari, three anonymous referees, Federico Bugni, Nathael Gozlan, Jinyong Hahn, Jim Heckman, Matt Masten, David Pacini, Adam Rosen, Andres Santos, J{\"o}rg Stoye, Martin Weidner, Daniel Wilhelm, Joachim Winter and conference and seminar participants at Aarhus, Duke, Munich, Oxford, S\'eminaire Palaisien, Tilburg, UCLA, the 2021 European Winter Meeting of the Econometric Society, the 2021 Bristol Econometric Study Group, the 2023 IAAE Annual Conference, Econometrics and Optimal Transport Workshop, and the Monash/Princeton/SJTU/SMU Econometrics Conference for useful comments and suggestions. We also thank Hongchang Guo, Zhangchi Ma and Frank Yan for capable research assistance.}}

\author{Xavier D'Haultfoeuille\thanks{CREST-ENSAE, xavier.dhaultfoeuille@ensae.fr. Xavier D'Haultfoeuille thanks the hospitality of PSE where part of this research was conducted.} \and Christophe Gaillac\thanks{Nuffield College and the University of Oxford, christophe.gaillac@economics.ox.ac.uk.} \and Arnaud Maurel\thanks{Duke University, NBER and IZA, arnaud.maurel@duke.edu. Arnaud Maurel thanks the hospitality of the University of Pennsylvania where part of this research was conducted.}}

\maketitle
~\vspace{-1.6cm}
\begin{abstract}
We study partially linear models when the outcome of interest and some of the covariates are observed in two different datasets that cannot be linked. This type of data combination problem arises very frequently in empirical microeconomics. Using recent tools from optimal transport theory, we derive a constructive characterization of the sharp identified set. We then build on this result and develop a novel inference method that exploits the specific geometric properties of the identified set. Our method exhibits good performances in finite samples, while remaining very tractable. We apply our approach to study intergenerational income mobility over the period 1850-1930 in the United States. Our method allows us to relax the exclusion restrictions used in earlier work, while delivering confidence regions that are informative.
\smallskip


\end{abstract}

\newpage

\section{Introduction}

In this paper, we derive partial identification and inference results for a partially linear model, in a context where the outcome of interest and some of the covariates are observed in two different datasets that cannot be merged. Relevant situations include cases where the researcher is interested in the effect of a particular variable that is not observed jointly with the outcome variable, as well as cases where the outcome and covariates of interest are jointly observed but some of the potential confounders are observed in a different dataset.

\medskip
Our analysis focuses on a partially linear model of the following form:
\begin{equation}
	\label{eq:PLM}
E(Y|X)= f(X_c) +  X_{nc}'\beta_0, \quad X=(X_{nc},X_c),
\end{equation}
in a data combination environment where $F_{Y,X_c}$ and $F_{X_{nc},X_c}$ are supposed to be identified, but the joint distribution $F_{Y,X}$ is not. The variable $X_c$ is thus common to both datasets, whereas the variable $X_{nc}\in\R^p$ is only observed in one of the two datasets. In this setup, $\beta_0=(\beta_{01},...,\beta_{0p})'$ is generally not point-identified, and as a result we focus on the identified set of either $\beta_0$ or $\beta_{0k}$ for some $k\in\{1,...,p\}$; the identified set of $f$ can then be deduced from that of $\beta_0$.

\medskip
We first derive a tractable characterization of the identified set of $\beta_0$. Unlike many other models considered in the partial identification literature, our setup does not deliver a tractable characterization of the identified set through the support function \citep[see][for detailed discussions of support functions]{BontempsMagnacARE17,molinari2019econometrics}. However, using Strassen's theorem \citep{strassen1965existence}, a recent result in optimal transport by \cite{backhoff2019existence}, and a convenient characterization of second-order stochastic dominance, we show that this set is convex, compact, includes the origin and can be simply constructed from its radial function.\footnote{The radial function $S$ of a closed, compact convex set $\mathcal{C}$ including the origin is defined, for any $q$ on the unit sphere, by $S(q)=\max_{\lambda q \in\mathcal{C}} \lambda$.} The identified set of $\beta_{0k}$, then, can also be computed at low computational cost by solving an unconstrained convex minimization problem.

\medskip
The characterization of the identified set also implies that point identification may be achieved if $\beta_0=0$, or under a restriction on the unobserved term $Y- f(X_c) -  X_{nc}'\beta_0$. While the latter condition is not directly testable, we show how to assess its plausibility when one has access to a validation sample in which the outcome and covariates are jointly observed.

\medskip
In the partially identified case, the identification region may be reduced by adding restrictions on $f(\cdot)$. The two-sample two-stage least squares estimator (TSTSLS) relies on the assumption $f(X_c)= X_{c,i}'\gamma_0$ for some $\gamma_0$ and $X_c=(X_{c,e}', X_{c,i}')'$. In this context, $X_{c,e}$ (resp. $X_{c,i}$) corresponds to the excluded (resp. included) instruments. This is a leading example that results in point identification. But the exclusion restriction that $E(Y|X)$ does not depend on $X_{c,e}$ may not be credible. We show that alternative restrictions, such as imposing a lower bound on the $R^2$ of the ``long regression'' of $Y$ on $X_{nc}$ and $X_c$ (in a similar spirit as \citeauthor{Oster19}, \citeyear{Oster19}) or shape restrictions such as monotonicity or convexity of $f$, may in practice dramatically reduce the identified set, and allow to, e.g., identify the sign of $\beta_{0k}$.

\medskip
Our identification result is constructive, and readily leads to a simple, plug-in estimator of the identified sets for $\beta_0$ or $\beta_{0k}$. A difficulty arises, however, as the estimator of the radial function is generally not asymptotically normal. To construct asymptotically valid confidence regions on $\beta_0$ or confidence intervals on $\beta_{0k}$, we propose to use subsampling \citep{politis1999subsampling}.

\medskip
Our method is based on a specific characterization of the identified set, and one may wonder whether alternative characterizations would be more convenient. In particular, the identified set can also be expressed through an infinite collection of moment inequalities. Therefore, general approaches for such problems such as that developed by \cite{andrews2017inference} could in principle be used instead. We show through simulations the key computational advantage of relying on the method we propose. With a univariate $X_{nc}$, confidence regions are typically computed in seconds, whereas they take up to 30 seconds with a bivariate $X_{nc}$. Compared to the method of \cite{andrews2017inference}, this corresponds to a dramatic reduction by a factor of more than 1,000 in computational time.

\medskip
We apply our method to study intergenerational income mobility over the period 1850 to 1930 in the United States, revisiting the analysis of \cite{olivetti2015name}. In this context where the main variable and outcome of interest are observed in two different datasets that cannot be linked, we show that the confidence sets obtained using our method are quite informative in practice, while allowing us to relax the exclusion restrictions underlying the TSTSLS approach used in \cite{olivetti2015name}. In the appendix, we consider another application where a key control variable is observed in a separate database. When incorporating sign constraints, our bounds are again very informative.\medskip

\subsubsection*{Related literatures} 
\label{sub:related_literature}

The method we develop in this paper can be used in a broad set of data combination environments. Two such contexts have attracted much attention in the empirical literature.

\medskip
One can use our method to conduct inference on the relationship between a particular covariate and an outcome variable, in situations where both variables are not jointly observed. A large literature on intergenerational income mobility often faces the unavailability of linked income data across generations and relies on exclusion restrictions, as in the application we revisit \citep[see][for a recent survey]{SantavirtaStuhler22}. Data combination issues are also common in consumption research, where income (or wealth) and consumption are often measured in two different datasets \citep{CLP22}. More generally, this type of data combination environment frequently arises in various subfields of empirical microeconomics, including in education and returns to skill estimation \citep{RothsteinWozny13,PP16,garcia2016life,hanushek2020culture}, health \citep{Manski18,RBB22} and labor \citep{ACI20}. A leading example that has attracted much interest in the literature is one where the researcher seeks to combine experimental data with another observational dataset, in particular situations where data on long-term outcomes is not available in the experimental data.

\medskip
Our approach can also be used to conduct inference on the causal effect of a variable of interest, in a setup where some of the confounders are observed in an auxiliary dataset. As such, our paper expands the range of data environments in which unconfoundedness is a credible assumption, complementing a literature that focuses on evaluating its reasonableness in the absence of data combination (see, e.g., \citeauthor{AET05}, \citeyear{AET05}; \citeauthor{Oster19}, \citeyear{Oster19}; \citeauthor{DMP22}, \citeyear{DMP22}).

\medskip
From a methodological standpoint, our paper is connected to the seminal article of \cite{CM02} and subsequent work by \cite{MP06}. They consider the issue of identifying the ``long regression'', in our context $E(Y|X_c,X_{nc})$, in the same data combination set-up as here. Importantly though, these two papers focus on deriving the identification region for $E(Y|X_c,X_{nc})$, but do not address the issue of inference. They also consider a setup where the covariates $X_{nc}$ have a discrete distribution with finite support, while we allow $X_{nc}$ to be continuously distributed. On the other hand their setup is entirely nonparametric, whereas we focus on a model that is linear in the covariates $X_{nc}$ and without interaction terms with $X_c$. The linearity assumption plays an important role in our ability to derive a tractable inference method. The absence of interaction further implies that in our set-up, and in contrast with these two papers, the identified set shrinks as one considers different values of $X_c$.

\medskip
Our paper is also related to \cite{pacini2019two} and \cite{hwang2022bounding}. Both papers construct bounds on the best linear predictor of $Y$ on $X$ in a similar data combination framework as here. We show that if one is ready to impose the usual assumption that the model is partially linear, large identification gains may be achieved, possibly up to point identification. \cite{hwang2022bounding} also considers a set-up where some of the $X$'s are only observed with $Y$ but not with $X_{nc}$, a case we do not study in this paper.

\medskip
More generally speaking, our paper relates to the broader literature on data combination problems in econometrics and statistics. We refer the reader to  \cite{RM2007} for a survey of this literature and to \cite{fan2014identifying}, \cite{FSS16}, \cite{BLL16}, and \cite{ACI20} for recent contributions. Contrary to ours, most of these papers impose restrictions that entail point identification.

\medskip
Within the data combination literature, our paper is technically closest to \cite{DGM21}. Though that paper considered the entirely different context of rational expectation testing, we also relied therein on Strassen's theorem to obtain a characterization of the null hypothesis of rational expectations. Importantly, we extend here our previous main result in a highly non-trivial way, by relying in particular on \cite{backhoff2019existence} to handle multivariate  $X_{nc}$. Also, we previously based our inference on \cite{andrews2017inference}. In contrast, a key contribution of our paper lies in the novel and tractable inference method that we derive.

\medskip
Finally, by developing in this data combination context a feasible inference method that can be implemented at a very limited computational cost, our paper also adds to the growing set of papers that propose tractable computational methods for partially identified models (see \citeauthor{BontempsMagnacARE17}, \citeyear{BontempsMagnacARE17} and \citeauthor{molinari2019econometrics}, \citeyear{molinari2019econometrics} for recent surveys). In particular, our paper fits into the strand of the literature that uses tools from optimal transport to devise computationally tractable identification and inference methods for partially identified models (\citeauthor{GalichonHenry11}, \citeyear{GalichonHenry11}; \citeauthor{galichon2016optimal}, \citeyear{galichon2016optimal}). By characterizing the sharp identified set based on the radial function, a novel approach in the partial identification literature, we show that it is possible to achieve very substantial tractability gains in this context, relative to a more standard characterization in terms of many moment inequalities.


\subsubsection*{Organization of the paper} 
\label{sub:organization}

The remainder of the paper is organized as follows. In Section \ref{sec2} we present our main identification results for the two-sample partially linear model described above. Section \ref{sec3} studies estimation and inference for this model. In Section \ref{sec:appli}, we apply our method to intergenerational income mobility in the United States. Section \ref{sec:concl} concludes. The Appendix of the paper gathers additional results on robustness to measurement errors, identification in models with heterogeneous effects of $X_{nc}$ on $Y$, and a test for point-identification. It also presents our second application to the black-white wage gap in the United States. Monte Carlo simulation results, additional material on the application, and the proofs are collected in the online Appendix. Some complements of the proofs appear in supplementary material available in our working paper version \citep[see][]{DGM22}. Finally, our inference method can be implemented using our companion R package, \texttt{RegCombin}, available at  \href{https://CRAN.R-project.org/package=RegCombin}{CRAN.R-project.org/package=RegCombin}.


\section{Identification}
\label{sec2}

Before presenting our main identification results, we introduce some notation that will be used throughout the paper. We let  $\|\cdot\|$, $0_p$ and $\S_p$ denote respectively the  usual Euclidean norm in $\R^p$, the vector $0$ and the unit sphere in $\R^p$; we may omit the index $p$ in the absence of ambiguity. For any cumulative distribution function (cdf) $F$ defined on $\R$, we let $F^{-1}(t)=\inf\{x: F(x)\geq t\}$ denote its generalized inverse and $\overline{F}=1-F$ be the corresponding survival function. For any random variable $A$, we let $\Supp(A)$ be its support, $F_A$ denote its cdf. and $V(A)$ its variance, if defined. We also let $\cvx$ denote the convex ordering, namely, for two random variables $A$ and $B$ with $E[|A|]<\infty$ and $E[|B|]<\infty$, $A\cvx B$ if $E[\phi(A)]\geq E[\phi(B)]$ for all convex functions $\phi$.\footnote{\label{foot:exist_Esp}Even though we may have $E[|\phi(A)|]=\infty$, $E[\phi(A)]$ is always well-defined because $E[\max(0,-\phi(A))] <\infty$, since there exists $a,b$ such that for all $x$, $\phi(x)\ge a+bx$.} We write $A\not\cvx B$ when $A\cvx B$ does not hold. Finally, for any sets $C$ and $C'$, we denote by $\partial C$ the boundary of $C$ and by $d_H(C,C')$ the Hausdorff distance between $C$ and $C'$, defined by
$$d_H(C,C')=\max\left(\sup_{c'\in C'} \inf_{c\in C} ||c- c'||, \; \sup_{c\in C} \inf_{c'\in C'}  ||c- c'||\right).$$ 	

\subsection{Identification without common regressors}\label{sec:without_common}

\subsubsection{A tractable characterization of the identified set} 
\label{ssub:characterization_result}

We first consider a linear model and derive the sharp identified set of $\beta_0$ in the absence of common regressors observed in both datasets. We suppose that we observe from two samples that can not be merged the distributions of the outcome, $F_Y$, and covariates, $F_X$. We maintain the following assumption:
\begin{hyp}
	\label{hyp:mom}
	We have $E(Y^2)<\infty$, $E(\|X\|^2)<\infty$, $V(Y)>0$ and $V(X)$ is non-singular. Moreover, $E(Y|X)=\alpha_0+X'\beta_0$ for some $(\alpha_0,\beta_0)\in\R\times \R^p$.
\end{hyp}

\medskip
We focus hereafter on the identified set $\B$ of $\beta_0$. Since $\B$ is the set of all vectors in $\R^p$ that are compatible with the model and the marginal distributions of $Y$ and $X$, we have
\begin{equation}\label{eq:def_B}
  \B = \left\{ \beta \in  \R^p: \exists\; \text{r.v. } (\widetilde{X}, \widetilde{Y}): \ E(\widetilde{Y}_0|\widetilde{X}_0)=\widetilde{X}_0'\beta, \ \widetilde{X}\eqd  X, \ \widetilde{Y} \eqd Y\right\},
\end{equation}
where, for any random variable $A$ with $E[|A|]<\infty$, we let $A_0=A-E(A)$ and we have used that $E(Y|X)=\alpha_0+X'\beta_0$ for some $\alpha_0$ is equivalent to $E(Y_0|X_0)=X_0'\beta_0$. Now, our goal is to express $\B$ to make it amenable to (simple) estimation. To this end, we define, for any  $\alpha\in (0,1)$, $F$ and $G$ cdfs with expectation 0, the following functions:
\begin{align}
R(\alpha, F,G)  & =\frac{\int_{\alpha}^1 F^{-1}(t)dt}{\int_{\alpha}^1  G^{-1}(t)dt}, \label{eq:def_R} \\
S(F, G)  & =\inf_{\alpha\in (0,1)}R(\alpha, F, G). \notag
\end{align}
These two functions play an important role in our analysis. Remark that, since $F$ and $G$ are cdfs of mean zero distributions, $\int_{\alpha}^1 F^{-1}(t)dt$ and $\int_{\alpha}^1 G^{-1}(t)dt$ are both positive, so that the ratio of superquantiles $R(\alpha, F,G) $ is well-defined, with $R(\alpha, F,G)>0$ and $S(F,G)\geq 0$. Theorem \ref{thm:main} is our main identification result.

\begin{thm}\label{thm:main} Suppose that Assumption \ref{hyp:mom} holds. Then
\begin{equation}\label{eq:caract_B}
\B =  \left\{\lambda q: \; q\in \S, \; 0\leq \lambda \leq S(F_{Y_0}, F_{X'_0q})\right\}.
\end{equation}
$\B $ includes $0_p$ and is a convex, compact subset of $ \B^V = \left\{\beta\in \R^p:\ \beta'V(X)\beta \leq V(Y)  \right\}$.
\end{thm}

We now give a sketch of the proof of \eqref{eq:caract_B}. Let $\B'$ denote the set on the right-hand side of \eqref{eq:caract_B}. First, one can show that by definition of $S(F_{Y_0}, F_{X_0'q})$,
$$\B' =  \left\{\beta\in\R^p: \forall \alpha \in (0,1), \, \int_\alpha^1 F^{-1}_{X_0'\beta}(t)dt \leq \int_\alpha^1 F^{-1}_{Y_0}(t)dt \right\}.$$
This, in turn, is equivalent to $F_{X_0'\beta}$ dominating $F_{Y_0}$ at the second order \citep[see, e.g.][]{de2006stochastic}, implying that
$$\B'=\left\{\beta\in\R^p: Y_0\cvx X_0'\beta\right\}.$$
The inclusion $\B \subset \B'$ then follows essentially from Jensen's inequality. As a side remark, note that we can also express $\B'$ through infinitely many moment inequality restrictions:
\begin{equation}
\B'=\left\{\beta\in\R^p: \, E\left[\max(0,Y_0-t)\right]\geq E\left[\max(0,X_0'\beta-t)\right] \; \forall t\in \R\right\}.
\label{eq:caract_B_mom}
\end{equation}
This equality directly follows from Fubini-Tonelli, applied to the standard characterization of the second-order stochastic dominance condition, namely $\int_{-\infty}^y F_{Y_0}(t)dt \ge \int_{-\infty}^y F_{X_0'\beta}(t)dt$ $\forall y\in \R$. We return to this alternative characterization of the identified set in Subsections~\ref{sec:simus1} and \ref{subsubsec:CPU_time} of the online appendix, where we document the computational advantages of using our characterization instead.

\medskip
	The inclusion $\B'\subset \B$ is more intricate to prove. Assume $\beta \in \B'$. By what precedes, $Y_0\cvx X_0'\beta$. Then, by Strassen's theorem \citep[Theorem 8 in][]{strassen1965existence},
\begin{equation}
	\inf_{(\widetilde{Y},\widetilde{X}^\beta): \widetilde{Y}\eqd Y, \, \widetilde{X}^\beta \eqd X'\beta} E\left[\left|\widetilde{X}^\beta_0 - E[\widetilde{Y}_0 | \widetilde{X}^\beta_0]\right|\right]=0. \label{eq:Strassen}
\end{equation}
This result was already used in \cite{DGM21} to characterize the restrictions on $F_Y$ and $F_\psi$ entailed by the rational expectation hypothesis $E(Y|\psi)=\psi$, where $\psi$ denotes the subjective expectations on an outcome $Y$. Importantly though, when $X$ is multivariate, \eqref{eq:Strassen} is not sufficient to conclude that $\B'\subset \B$, as the $\sigma$-algebras generated by $X$ and $X'\beta$ are not equal in general. Nonetheless, we prove, using in particular Theorem 1.3 in \cite{backhoff2019existence}, that for $\beta \in \B'$,\footnote{We thank Nathael Gozlan for his help in obtaining \eqref{eq:inf_univ_multiv}.} 
\begin{align}
	\label{eq:inf_univ_multiv}
 \inf_{(\widetilde{Y},\widetilde{X}): \widetilde{Y}\eqd Y, \, \widetilde{X} \eqd X} E\left[\left|\widetilde{X}_0'\beta - E[\widetilde{Y}_0 | \widetilde{X}_0]\right|\right]  \leq & \inf_{(\widetilde{Y},\widetilde{X}^\beta): \widetilde{Y}\eqd Y, \, \widetilde{X}^\beta \eqd X'\beta} E\left[\left|\widetilde{X}_0^\beta - E[\widetilde{Y}_0 | \widetilde{X}_0^\beta]\right|\right] .	
\end{align}
Together, \eqref{eq:Strassen}, \eqref{eq:inf_univ_multiv}, and the existence of a minimizer on the left-hand side of \eqref{eq:inf_univ_multiv} \citep[Theorem 1.2 in][]{backhoff2019existence}, imply that we can find random variables $\widetilde{Y}$ and $\widetilde{X}$ such that $E[\widetilde{Y}_0 | \widetilde{X}_0]=\widetilde{X}_0'\beta$, $\widetilde{Y}\eqd Y$ and $\widetilde{X}\eqd X$. Thus, $\beta\in \B$.

\medskip
Turning to the second part of the theorem, $0_p\in\B$ follows by noting that one can always rationalize, from the sole knowledge of their marginal distributions, that $X$ and $Y$ are independent. That $\B \subset \B^V$ comes from the inclusion $\B \subset \B'$, combined with the fact that $Y_0\cvx X_0'\beta$ implies $V(Y)\ge V(X'\beta)$. Hence, $\B$ is included in a bounded ellipsoid. The equality $\B =\B^V$ occurs for instance when $Y$ and $X$ are normally distributed. Otherwise, $\B$ may be substantially smaller than $\B^V$, as we illustrate below. In such cases, $\B^V$ remains a natural benchmark as it is very simple to characterize using $V(Y)$ and $V(X)$ only, and straightforward to estimate.

\medskip
\begin{remark}
Using the exact same reasoning as above, one can prove that without any linear restriction on the conditional expectation, the identified set for $E[Y_0|X_0]$ is $\{g: Y_0\cvx g(X_0)\}$. Similarly, if we only impose that $m(x):=E[Y_0|X_0=x]$ belongs to a linear space $\mathcal{Z}$ of functions, then the identified set for $m$ is $\{\lambda q:q\in \mathcal{Z}: \, E[|q(X_0)|]=1, E[q(X_0)]=0,\, 0\le \lambda\le S(F_{Y_0},F_{q(X_0)})\}$.\footnote{We thank a referee for pointing out this extension.}
\end{remark}

\begin{figure}[H]
	\begin{center}
	\includegraphics[scale=1]{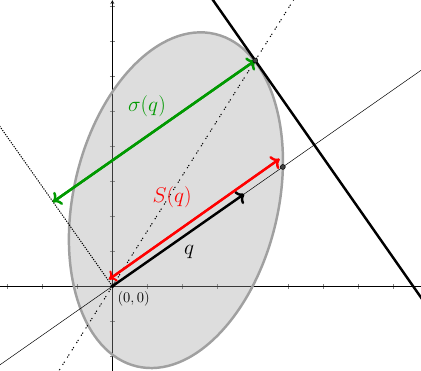}		
	\end{center}
	\caption{Two characterizations of a closed convex set including the origin, either through its support function $\sigma$ (green), or through the radial function $S$ (red).}
	\label{fig:convex}
\end{figure}

\paragraph{Radial vs. support function characterization of the identified set.}
A key takeaway from Equation~\eqref{eq:caract_B} is that the identified set admits a very simple expression as a function of $S$, which is the inverse of the Minkowski gauge function of $\B$ \citep[see, e.g., Definition 1.2.4 p.137 and Proposition 3.2.4  p.157][]{hiriart2012fundamentals}, also known as the radial function of $\B$. This function differs from the support function $\sigma$ of $\B$, defined by $\sigma(q,F_{Y_0}, F_{X_0})=\sup_{ b\in \B} \ q'b$. The difference between these two functions is illustrated in Figure \ref{fig:convex}.

\medskip
The partial identification literature has largely relied on support functions, as these are powerful tools that uniquely characterize their convex sets. But the radial function also uniquely characterizes convex sets if, as is the case here, these sets include the origin.\footnote{More generally, star-shaped sets are fully characterized by the radial function (and a given point, $0_p$ in our setup). See \cite{Molchanov17}, p.156, for more details on this point.} Importantly, this approach allows us to characterize the sharp identified set by minimizing a simple function over the interval $(0,1)$. In contrast, the support function approach will generally be significantly less tractable in our context as it would require solving a high-dimensional constrained optimization problem. Namely, using the characterization of the identified set given in Equation~(\ref{eq:caract_B_mom}) above, the support function can be obtained by solving the following program:
\begin{align} \label{eq:direct_supp}
 \sigma(q,F_{Y_0}, F_{X_0})&=\sup_{b \in\R^p} q'b
\quad \textrm{s.t.} \inf_{t\in\R} E\left[\max(0,Y_0-t)\right] - E\left[\max(0,X_0'b-t)\right] \geq 0,
\end{align}
where the constraint itself involves an optimization problem. Simulation results indicate that using the radial function rather than the support function approach does result in very large computational gains, see Online Appendix \ref{subsubsec:CPU_time} for details on this.

\paragraph{Partial identification of subcomponents of $\beta_0$.}
The support function still plays a key role in our context when one is interested in a component of $\beta_0=(\beta_{0,1},...,\beta_{0,p})'$, say $\beta_{0,k}$. The following result shows that we can actually recover this function at a low computational cost once $S$ is known. Hereafter, we let $e_k$ denotes the $k$-th element of the canonical basis in $\R^p$ and use the convention $1/0=\infty$ and $1/\infty=0$.

\begin{cor}\label{cor:sub}
	Suppose that Assumption \ref{hyp:mom} holds. Then, the identified set $\B_k$ of $\beta_{0,k}$ satisfies $\B_k=[-\sigma(-e_k, F_{Y_0}, F_{X_0}),\sigma(e_k, F_{Y_0}, F_{X_0})]$. Moreover,
	\begin{equation}
	\sigma(e_k, F_{Y_0}, F_{X_0}) =	\frac{1}{\inf_{q\in \R^p: q_k=1} 1/S(F_{Y_0},F_{X_0'q}) }.\label{eq:link_S_sig}
	\end{equation}
	The same holds with $\sigma(-e_k, F_{Y_0}, F_{X_0})$, after replacing $q_k=1$ by $q_k=-1$.
\end{cor}

\medskip
We use the expression \eqref{eq:link_S_sig} of the support function, rather than the simpler expression $\sigma(e_k, F_{Y_0}, F_{X_0}) =	\sup_{q\in \R^p: q_k=1} S(F_{Y_0},F_{X_0'q})$, because $q\mapsto 1/S(F_{Y_0},F_{X_0'q})$ is convex
(see the proof of Proposition \ref{prop:epsilon}, which also applies when $\eps=0$), whereas $q\mapsto S(F_{Y_0},F_{X_0'q})$ may not be concave. It follows that one can recover the support function $\sigma$, and in turn the sharp bounds on $\beta_{0,k}$, by simply minimizing a convex function over $\R^{p-1}$.

\subsubsection{Point identification} 
\label{subs:comparison}

In some cases, our approach yields point identification of the parameters of interest, or subcomponents of it. Proposition~\ref{prop:point_ident} below presents two such cases under which the identified sets $\B$ and $\B_1$, respectively, boil down to a singleton.

\begin{prop}
	\label{prop:point_ident}
	Suppose that Assumption \ref{hyp:mom} holds and let $\phi$ be a convex function such that $E[\phi(Y)]<\infty$. Then:
	\begin{enumerate}
		\item If for all $\beta\ne 0_p$, $E[\phi(X'\beta)]=\infty$, then $\B =\{\beta_0\}=\{0_p\}$.
		\item If $E[\phi(X_1\beta_1)]  = \infty$ for all $\beta_1 \ne 0$ and $E[\phi(X'_{-1}\beta_{-1})]$ $< \infty$ for all $\beta_{-1}\in\R^{p-1}$,  then $\B_1=\{\beta_{0,1}\}=\{0\}$.
	\end{enumerate}
\end{prop}

Recall from our main identification result above that the identified set $\B$ always includes the origin. The first point of Proposition \ref{prop:point_ident} further establishes point identification of $\beta_0=0_p$ when, basically, $Y$ has lighter tails than any linear index of $X$. The second point is similar but focuses on a subcomponent instead: if $Y$ and $X_{-1}'\beta_{-1}$ have lighter tails than  $X_1$, then $\beta_{0,1}=0$ is point identified. As an example of function $\phi$ for which Proposition \ref{prop:point_ident} holds, one might consider for instance $\phi(x)=|x|^a$ for some $a>2$ (in which case $X'\beta$ or $X_1$ have heavy tails), or $\phi(x)=\exp(a |x|^b)$ for some $a,b>0$ (in which case $X'\beta$ or $X_1$ have exponential tails).

\medskip
To illustrate Point 1 of Proposition \ref{prop:point_ident}, suppose that $p=1$, $X$ follows a Laplace distribution (with density $\exp(-|x|)/2$ on $\R$) and $Y\sim\mathcal{N}(0,1)$. Then, by using $\phi(x)=\exp(|x|^{3/2})$, it follows from Point 1 of Proposition \ref{prop:point_ident} that $\beta_0=0$ is point identified in this case. On the other hand, the variance restrictions only set identify $\beta_0$, with an identified set given by $\B^V=[-1/\sqrt{2},1/\sqrt{2}]\simeq [-0.707,\,0.707]$. This example illustrates the (in this case point-) identifying power of higher-order moments of the distributions of $X$ and $Y$.


\subsection{Identification with common regressors}\label{sec:with_commonD}

We now turn to the frequent situation where some regressors are observed in both datasets. Namely, suppose we observe regressors $X_c$ that are common to both datasets, and assume that the partially linear model \eqref{eq:PLM} holds:
\begin{equation*}
E(Y|X)= f(X_c) +  X_{nc}'\beta_0, \quad X=(X_{nc},X_c),
\end{equation*}

The key here is to note, following \cite{Robinson88}, that this case is equivalent to the previous setup without common regressors once we compute the following residuals, for all $x$ in the support of $X_c$:
\begin{align*}
	X^x & = X_{nc} - E(X_{nc} | X_c=x), \\
	Y^x & = Y - E(Y|X_c=x).
\end{align*}
It directly follows that $\beta_0$ satisfies $E(Y^x|X^x)=X^x{}'\beta_0$, which allows us to use the characterization of the identified set without common regressors obtained in Section~\ref{sec:without_common}. \\

Let $\B^c$ and $\mathcal{F}$ denote the identified sets of $\beta_0$ and $f$, respectively. We have the following characterization of $\B^c$ and $\mathcal{F}$:

\begin{prop}\label{prop:common}
	Suppose that $E(Y^2)<\infty$, for all $x\in \Supp(X_c)$, $E(X^xX^x{}' | X_c=x)$ is nonsingular and \eqref{eq:PLM} holds. Then:
	\begin{align*}
\B^c & = \left\{\lambda q: q\in \S, \; 0\leq \lambda\leq \overline{S}(F_{Y,X_c}, F_{X_{nc}'q,X_c})\right\}, \\
\mathcal{F} & =  \left\{x\mapsto E(Y|X_c = x) - E(X_{nc}|X_c=x)'\beta: \; \beta \in  \B^c\right\},	
	\end{align*}
	where $\overline{S}(F_{Y,X_c}, F_{X_{nc}'q,X_c}) = \inf_{x\in\Supp(X_c)} S(F_{Y^x|X_c=x}, F_{X^x{}'q|X_c=x})$. $\B^c$ includes $0_p$, is compact and convex.
\end{prop}

It is possible to extend \eqref{eq:PLM} by including interaction terms. Notably, such specification allows for heterogeneous effects of $X_{nc}$ on $Y$, which can be important in practice \citep[see, e.g.,][pp.1110-1111]{hausman2016fiscal}. We consider this extension in Appendix \ref{app:interact}. Another interesting extension corresponds to cases where $E(Y|X)=f(X_c)+X_{nc}'\beta_0+X_a'\delta_0$ and we observe in a first dataset $(Y, X_a, X_c)$ and in a second dataset, $(X_c, X_{nc})$. This setup leads to qualitatively different results. For instance, if there is no common regressors and $(Y, X_a)$ and $X_{nc}$ are Gaussian, one can show that the sharp identified set of $(\beta_0,\delta_0)$ is not convex and does not include $0_{p+r}$ (with $r$ the dimension of $X_a$). We refer the reader to \cite{hwang2022bounding} for outer bounds on the best linear predictor in this setup and leave its study for future research.

\subsection{Identifying power of additional restrictions}
\label{subs:id_power_additional_restrictions}

We now consider additional restrictions that may reduce the identified set, in some cases resulting in point identification of the parameters of interest.

\subsubsection{Lower bound on the $R^2$ of the long regression} 
\label{ssub:lower_bound_on_the_r_2}

A first way to reduce the identified set is to use a lower bound on the predictive power of $X_{nc}$ and $X_c$ with respect to $Y$. To formalize this idea, we assume that $R^2_\ell$, the coefficient of determination of the ``long'' regression of $Y$ on $X_{nc}$ and $X_{c}$ is higher than a certain threshold. This threshold may be absolute (e.g., 0.1) or relative to $R^2_s:=V(E(Y|X_c))/V(Y)$, the $R^2$ of the ``short'' regression of $Y$ on $X_c$, which is directly identified from the data. This is in the same spirit as \cite{Oster19}, who suggests fixing $R^2_\ell/R^2_s$ to 1.3. Note that
$$f(X_c)+X_{nc}'\beta= E(Y|X_c) + (X_{nc}- E(X_{nc}|X_c))'\beta,$$
and the two components on the right-hand side are uncorrelated. Thus,
$$R^2_\ell = \frac{V(E(Y|X_c)) + \beta'E(V(X_{nc}|X_c))\beta}{V(Y)} = R^2_s + \frac{\beta'E(V(X_{nc}|X_c))\beta}{V(Y)}.$$
Then, if one imposes a lower bound $\underline{R}^2$ on $R^2_\ell$ such that $\underline{R}^2 \geq R^2_s$, the identified set on $\beta$ becomes
$$\left\{\lambda q: q\in \S, \, \left(\frac{(\underline{R}^2 - R^2_s) V(Y)}{q'E(V(X_{nc}|X_c))q}\right)^{1/2} \le \lambda \le \overline{S}(F_{Y,X_c}, F_{X_{nc}'q,X_c}) \right\},$$
provided that $E(V(X_{nc}|X_c))$ is nonsingular. This restriction has three key attractive features. First, one can in practice motivate this restriction based on a ``validation sample'', namely a subset of the population or another population (e.g., a different country than that under investigation), for which we identify the joint distribution of the outcome and covariates, and thus the $R^2$ of the ``long'' regression. Second, imposing a lower bound such that $\underline{R}^2>R^2_s$ allows one to exclude $0_p$ from the identified set. Third, the identified set still admits a very simple expression.


\subsubsection{Linear shape restrictions}
\label{subsubs:id_power_shape_restrictions}

Another way to narrow the identified set $\B^c$ with common regressors is to impose some constraints on $f(\cdot)$. Shape restrictions such as monotonicity or convexity often follow from economic theory; see \cite{Matzkin94} and \cite{chetverikov2018econometrics} for econometric reviews, and \cite{Tripathi00} and \cite{AbrevayaJiang05} for their use and testability with partially linear models. We characterize here the identified set when we impose such restrictions on $f$.

\medskip
We model these restrictions by $[Rf](r)\ge \underline{c}(r)$ for all $r\in\mathcal{R}$, with $R$ a known linear operator, $\underline{c}$ a known, real function and $\mathcal{R}$ the domain of $[Rf]$ and $\underline{c}$. For instance, if $X_c$ is discrete such that $\Supp(X_c)=\{x_{c,1},...,x_{c,K}\}\subset \R$, with $K>1$ and $x_{c,1}<...<x_{c,K}$, considering $[Rf](r)=f(x_{c,r+1})-f(x_{c,r})$ for $r\in\mathcal{R}=\{1,...,K-1\}$ (resp. $[Rf](r)=(f(x_{c,r+2})-f(x_{c,r+1}))/(x_{c,r+2} - x_{c,r+1}) - (f(x_{c,r+1})-f(x_{c,r}))/(x_{c,r+1} - x_{c,r})$ for $r\in\mathcal{R}=\{1,...,K-2\}$ with $K>2$) and $\underline{c}(r)=0$ corresponds to imposing that $f$ is non-decreasing (resp. convex). When $X_c$ is continuous, the same two constraints can be imposed by considering $[Rf](r)=f'(r)$ and $[Rf](r)=f''(r)$, with $\mathcal{R}=\Supp(X_c)$.

\medskip
This framework also accommodates restrictions on the magnitude of the effect of $X_c$ on $Y$. Namely, suppose for simplicity that $X_c$ is binary and consider $[Rf](1)=-[Rf](2)=f(x_{c,2})-f(x_{c,1})$  with $\mathcal{R}=\{1,2\}$ and $\underline{c}(1)=\underline{c}(2)=\underline{c}\ge 0$. The extreme case $\underline{c}=0$ corresponds to $X_c$ having no effect on $Y$, as in the two-sample two-stage least squares strategy (see the next subsection for a related, more general point identification result in this context). More generally, this corresponds to the constraint that the magnitude of the effect of $X_c$ is bounded by the cutoff $\underline{c}$, $|f(x_{c,2})-f(x_{c,1})| \leq \underline{c}$.\footnote{If $X_c$ has $K>2$ points of support, the same idea can be generalized by imposing restrictions on $|f(x_{c,k})-f(x_{c,j})|$ for specific pairs $(j,k)\in\{1,...,K\}^2$, $j\ne k$.} By increasing $\underline{c}$, one can therefore study how the identified set varies when relaxing the exclusion restriction, in a similar spirit to, e.g., \cite{masten2018identification}.

\medskip
Hereafter, we denote by $m_Y(\cdot)=E[Y|X_c=\cdot]$, $m_{X_{nc}}(\cdot)=E[X_{nc}|X_c=\cdot]$ and
\begin{align*}
	\underline{S}^c(m_Y,m_{X_{nc}},q) & = \sup_{\substack{r\in \mathcal{R}:\\ [Rm'_{X_{nc}}q](r)\le 0}} \lim_{u\downarrow 0} \frac{[Rm_Y - \underline{c}](r)+u}{[Rm'_{X_{nc}}q](r)-u^2}, \\
		\overline{S}^c(m_Y,m_{X_{nc}},q) & = \inf_{\substack{r\in \mathcal{R}:\\ [Rm'_{X_{nc}}q](r)\ge 0}} \lim_{u\downarrow 0} \frac{[Rm_Y - \underline{c}](r)+u}{[Rm'_{X_{nc}}q](r)+u^2},
\end{align*}
where we let $\sup\emptyset=-\inf\emptyset=-\infty$ and we note that the two functions above may be infinite. We introduce limits to deal with the cases where $[Rm'_{X_{nc}}q](r)=0$. Proposition~\ref{prop:shape_f} characterizes the identified sets of $\beta_0$ and $f$ under such shape restrictions.

\begin{prop}
	Suppose that the conditions of Proposition \ref{prop:common} hold and $[Rf](r)\ge \underline{c}(r)$ for all $r\in\mathcal{R}$. Then, the identified sets $\Bcon$ and $\mathcal{F}^{\con}$ of $\beta_0$ and $f$ satisfy
\begin{align*}
\Bcon & =\left\{\lambda q: q\in \S^+, \; \underline{S}^{\con}(q, F_{Y,X_c},F_{X_{nc},X_c}) \le \lambda \le \overline{S}^{\con}(q, F_{Y,X_c},F_{X_{nc},X_c}) \right\}, \\
\mathcal{F}^{\con} & =  \left\{x\mapsto E(Y|X_c = x) - E(X_{nc}|X_c=x)'\beta: \; \beta \in  \Bcon\right\},	
	\end{align*}
where $\S^+=\S\cap\{(x_1,...,x_p)\in\R^p:x_1\ge 0\}$ and
\begin{align*}
	\underline{S}^{\con}(q,F_{Y,X_c},F_{X_{nc},X_c}) & = \max\bigg(-\overline{S}(F_{Y,X_c},F_{-X'_{nc}q,X_c}),  \underline{S}^c(m_Y,m_{X_nc},q) \bigg), \\
	 \overline{S}^{\con}(q, F_{Y,X_c},F_{X_{nc},X_c}) & = \min\bigg(\overline{S}(F_{Y,X_c},F_{X'_{nc}q,X_c}), \overline{S}^c(m_Y,m_{X_nc},q) \bigg).
\end{align*}
$\Bcon$ is compact, convex but does not include $0_p$ if for some $r\in\mathcal{R}$, $[Rm_Y - \underline{c}](r)<0$.
\label{prop:shape_f}
\end{prop}

In contrast to our baseline identification results in the absence of additional restrictions, the resulting identified set may exclude the origin. This illustrates the practical importance of imposing these types of shape restrictions in contexts where these are likely to hold. Suppose for instance that $p=1$, $X_c$ is binary ($\Supp(X_c)=\{0,1\}$), $\mathcal{R}=\{1\}$ and $[Rf](1)=f(1)-f(0)$, namely we impose that $f$ is non-decreasing. If $f(1)-f(0)< (m_{X_{nc}}(0)-m_{X_{nc}}(1))\beta_0$, then $m_Y(1)<m_Y(0)$. As a result, $0\not\in\Bcon$. The condition $f(1)-f(0)< (m_{X_{nc}}(0)-m_{X_{nc}}(1))\beta_0$ holds for instance if $m_{X_{nc}}$ is decreasing and $\beta_0$ is positive and large enough.

\begin{remark}
While we focus here on the identifying power of each type of restrictions considered separately, researchers may in some contexts want to jointly impose several of these restrictions and consider the intersection of the associated identified sets. In the particular cases of the shape restrictions and the restrictions on the $R^2$ considered above, the identified sets share the same structure. Thus, the identified set resulting from both types of constraints can be simply computed by replacing the lower bound on $\lambda$ by the maximum of the lower bounds of the initial sets, and proceeding symmetrically for the upper bound.
\end{remark}

\subsubsection{Functional form restrictions involving common regressors} 
\label{ssub:linear_restrictions_involving_common_regressors}

One may alternatively be willing to impose functional form restrictions on $f$. The following proposition shows that this may yield point identification.

\begin{prop}
	Suppose that $E(Y^2)<\infty$, $E[\|X\|^2]<\infty$ and $f$ belongs to a vector space $\mathcal{G}$. Then, if for all $\gamma\ne 0$,  $m_{X_{nc}}'\gamma\not\in\mathcal{G}$, $\beta_0$ and $f$ are point identified.
\label{prop:point_f}
\end{prop}

This proposition encompasses several popular restrictions. We consider in particular three such restrictions, for which the key point-identifying condition $m_{X_{nc}}'\gamma\not\in\mathcal{G}$ has a simple interpretation:

\begin{enumerate}

\item $f(X_c)=f_1(X_{i,c})$, with $X_c=(X_{i,c}',X_{e,c}')'$. This restriction, which is implicit in, and central to the two-sample two-stage least squares strategy, states that conditional on $(X_{nc}, X_{i,c})$, $Y$ is mean-independent of $X_{e,c}$. In such a case, $m_{X_{nc}}'\gamma\not\in\mathcal{G}$ for all $\gamma\ne 0$ basically means that $m_{X_{nc}}(X_c)$ varies with $X_{e,c}$. To see this, consider the simple case where $m_{X_{nc}}(X_c)=m_i(X_{i,c})+ \Pi X_{e,c}$, for some function $m_i$ and a $p\times q$ matrix $\Pi$. Then, $m_{X_{nc}}'\gamma\not\in\mathcal{G}$ is equivalent to $\Pi$ having rank $p$, which is the usual rank condition in linear instrumental variable models.

\item $f(X_c)=X_c'\gamma_0$. Under this linearity restriction on $f(\cdot)$, $m_{X_{nc}}'\gamma\not\in\mathcal{G}$ for all $\gamma\ne 0$ basically means that $m_{X_{nc}}(X_c)$ is nonlinear in $X_c$ (the two notions are actually equivalent if $X_{nc}\in\R$). Note that  this point identification result fully relies on the linearity of $f(\cdot)$ combined with the nonlinearity of $E(X_{nc}|X_c)$, and is thus akin to, e.g., the identification of sample selection models without instruments exploiting the nonlinearity of the inverse Mill's ratio. Also, this result does not apply when $X_c$ is binary, since in this case $m_{X_{nc}}(X_c)$ is necessarily linear in $X_c$.

\item $f(X_c)=\sum_{j=1}^J f_j(X_{j,c})$, with $X_c=(X_{1,c},...,X_{J,c})'$. Under this additivity restriction on $f(\cdot)$,
$m_{X_{nc}}'\gamma\not\in\mathcal{G}$ for all $\gamma\ne 0$ means that $m_{X_{nc}}(X_c)$ is not additive in $X_c$. If for instance  $X_c=(X_{1,c},X_{2,c})$ with $X_{1,c}, X_{2,c}$ both binary, $m_{X_{nc}}'\gamma\not\in\mathcal{G}$ for all $\gamma\ne 0$  holds if in the regression of $X_{nc}$ on $X_{1,c}, X_{2,c}$ and $X_{1,c}\times X_{2,c}$, the coefficient of $X_{1,c}\times X_{2,c}$ is not zero.
\end{enumerate}


\subsubsection{Tail conditions} 
\label{ssub:tail_conditions}

Finally, if one is ready to impose a relative tail condition between the error term $U:=Y_0-X_0'\beta_0$ and $X_0'\beta_0$, the identified set is considerably reduced. For simplicity, we assume here that there are no common regressors but Proposition \ref{prop:point_ident2} readily extends to accomodate such regressors.

\begin{prop}
Suppose that Assumption \ref{hyp:mom} holds. Then:
\begin{enumerate}
	\item If there exists a convex function $\phi$ such that $E[\phi(U\lambda)]< E[\phi(X_0'\beta_0 \lambda)]=\infty$ for all $\lambda>1$, the identified set of $\beta_0$ is included in $\partial\B$;
	\item $X\in\R$, $E[\phi(U\lambda)]<E[\phi(X\lambda)]=\infty$ for all $\lambda >0$ and it is known that  $\beta_0>0$, $\beta_0$ is point identified.
\end{enumerate}
	\label{prop:point_ident2}
\end{prop}

With $X\in \R$, the condition $E[\phi(U\lambda)]<E[\phi(X\lambda)]=\infty$ for all $\lambda>0$ holds for instance if $E[|U|^a]<E[|X|^a]=\infty$ for some $a>2$. More generally, the condition $E[\phi(U\lambda)]<E[\phi(X_0'\beta_0 \lambda)]=\infty$ basically imposes that $X_0'\beta_0$ has fatter tails than $U$. In this sense, this condition is similar to those in Proposition \ref{prop:point_ident} above.

\medskip
\paragraph{Testability.}
Note that we cannot test the condition $E[\phi(U\lambda)]<E[\phi(X_0'\beta_0 \lambda)]=\infty$  for some convex function $\phi$ and all $\lambda>1$, simply because $U$ is not identified. On the other hand, we can assess the plausibility of $\beta_0 \in \partial \B$ using a validation sample, as defined above. Denoting by $(Y_v, X_v)$ the variables corresponding to this validation sample, it becomes possible to test whether the corresponding parameter $\beta_v=V(X_v)^{-1}\cov(X_v,Y_v)$ is at the boundary of the identified set one would get from the sole knowledge of $F_{Y_v}$ and $F_{X_v}$. Provided that $\beta_v\ne 0$, this condition is indeed equivalent to $\norm{\beta_v}= S(F_{Y_{v 0}}, F_{X'_{v0} \beta_v/\norm{\beta_v}})$ or, in simpler terms,
$$S(F_{Y_{v0}}, F_{X'_{v0} \beta_v})=1.$$
We consider a statistical test of this condition in Appendix \ref{app:test_point_id}, and apply it in Section \ref{sec:appli} below.


\subsection{Numerical illustration} 
\label{sub:illustration}

We illustrate the previous results by considering the following model:
$$Y = \gamma_{0,0} + X_c^{1.3} \gamma_{0,1}  + X_{nc,1}\beta_{nc,1} + X_{nc,2}\beta_{nc,2} + U, \; U|X\sim\mathcal{N}(0,9).$$
We set the coefficients as follows:  $\gamma_{0,0}=-0.1$, $\gamma_{0,1} = 0.3$, $\beta_{nc,1} = 1$ and  $\beta_{nc,2}=1$. The variables $X$ are transformations of $(N_1,N_2,N_3)'$, which is supposed to follow a multivariate normal distribution with mean 0 and covariance matrix $$ \Sigma = \left(\begin{array}{ccc}
1 & -0.3 & -0.8  \\
-0.3&  1& -0.1 \\
-0.8 &  -0.1 & 1
\end{array} \right).$$
Specifically, the common regressor is given by $ X_c = \sum_{k=1}^K (k-1) 1\{ c_{k-1} \leq N_1 \leq c_k \}$, $K=4$, $c_0 = -\infty $, $c_1,\dots, c_{K-1}$, are respectively the quantiles of order 0.1, 0.37, 0.67 and 0.9 of the standard normal, and $c_K=\infty$. We consider two cases for the regressors that are observed in one of the datasets only, $X_{nc}$. In the first case,  $(X_{nc,1}, X_{nc,2})=(N_2, \exp(N_3))$ and in the second, $(X_{nc,1}, X_{nc,2})=(\exp(N_2), \exp(N_3))$.

\begin{figure}[H]
\centering
	\subfigure[$X_{nc,1}$ and $\ln(X_{nc,2})$ normal]{\includegraphics[width=0.4\linewidth, height=0.25\textheight]{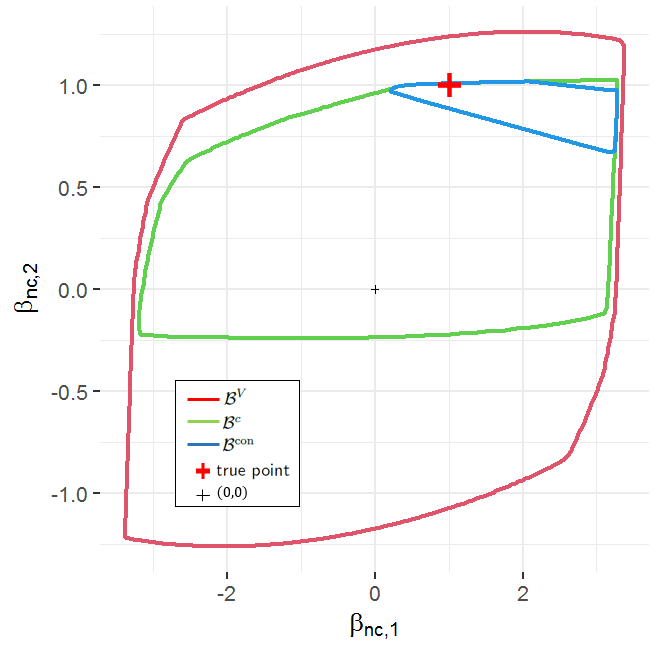}
		\label{fig:norm1}}
	\quad
	\subfigure[$X_{nc,1}$ and $X_{nc,2}$ lognormal]{\includegraphics[width=0.4\linewidth, height=0.25\textheight]{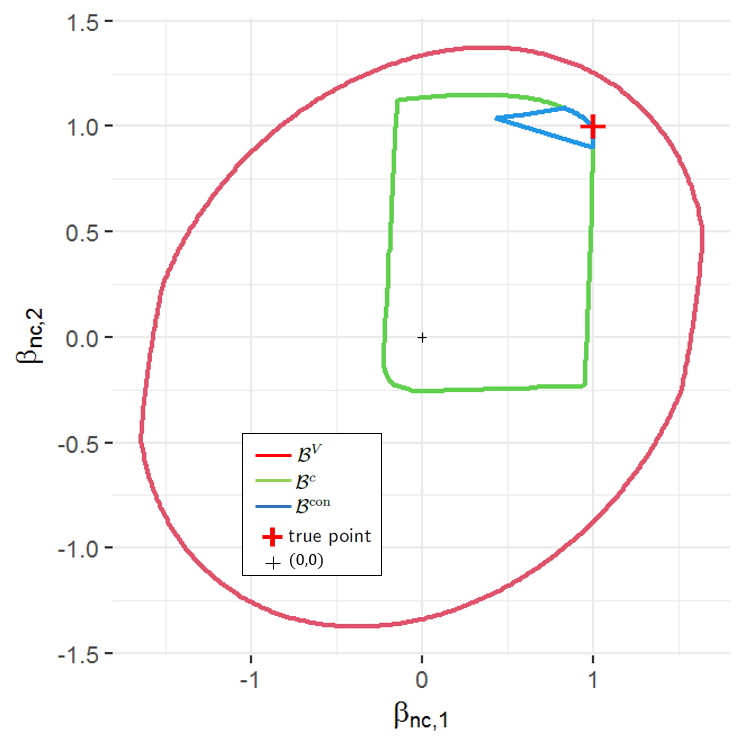}
		\label{fig:lognorm}}
\caption*{\footnotesize Note: the sets are obtained using a sample of size 100,000 and taking the convex hull of the set obtained from a uniform grid of 1,000 directions on the 2 dimensional sphere. $\Bcon$ uses both convexity and monotonicity constraints on $X_c$.}
	\caption{Identification regions for different distributions of $(X_{nc,1},X_{nc,2})$}
	\label{fig:Sp30}
\end{figure}

\medskip
Figure \ref{fig:Sp30} displays several identified sets for each of the two data-generating processes (DGPs) described above, each of them being associated with particular restrictions. Namely, the set in red, denoted by $\B^V$, is obtained from the variance restrictions only:
$$\B^V=\left\{\beta: \beta'V(X^0)\beta\leq V(Y^0)\right\}\cap \left\{\beta: \beta'V(X^1)\beta\leq V(Y^1)\right\},$$
where $X^x$ and $Y^x$ are defined as in Section \ref{sec:with_commonD}. Hence, $\B^V$ is the intersection of two ellipses. The set in green, $\B^c$, is obtained as in Proposition \ref{prop:common} and relies on the restrictions $E(Y^x|X_{nc},X_c=x)=X^{x}{}'\beta_0$ for $x\in\{0,1\}$. Finally, the set in blue, $\Bcon$, is a subset of $\B^c$ that imposes both convexity and monotonicity constraints on $X_c$.

A couple of comments are in order. In case (a) the restrictions implied by the model are much more informative than the variance restrictions, because of the non-normality of $X_{nc,2}$, and in particular the fact that it has fatter tails than the residuals $U$. The true point is at the boundary of $\B^c$, illustrating Proposition \ref{prop:point_ident} applied conditional on $X_c=0$ and $X_c=1$. In this case, the shape restrictions are sufficient to imply that $0_2\not\in \Bcon$ but also to rule out that $\beta_{nc,1}=0$ as well as $\beta_{nc,2}=0$. The identified set $\B^c$ is reduced further in case (b), as a result of the fatter tails of both $X_{nc,1}$ and $X_{nc,2}$. Like in case (a), the shape constraints on $f(X_c)$ allow to reduce dramatically the identified set.

\begin{figure}[H]
	\centering
	\subfigure[Convexity constraint on $f(\cdot)$]{\includegraphics[width=0.4\linewidth, height=0.25\textheight]{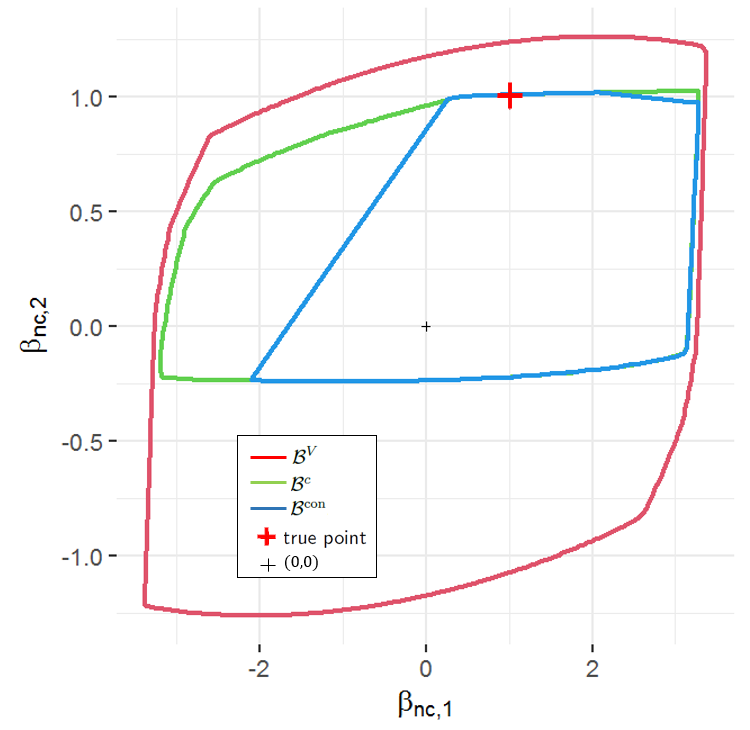}
		\label{fig:norm1_monot_conv}}
	\quad
	\subfigure[$R_l^2 \geq r R_s^2$ constraints]{\includegraphics[width=0.4\linewidth, height=0.25\textheight]{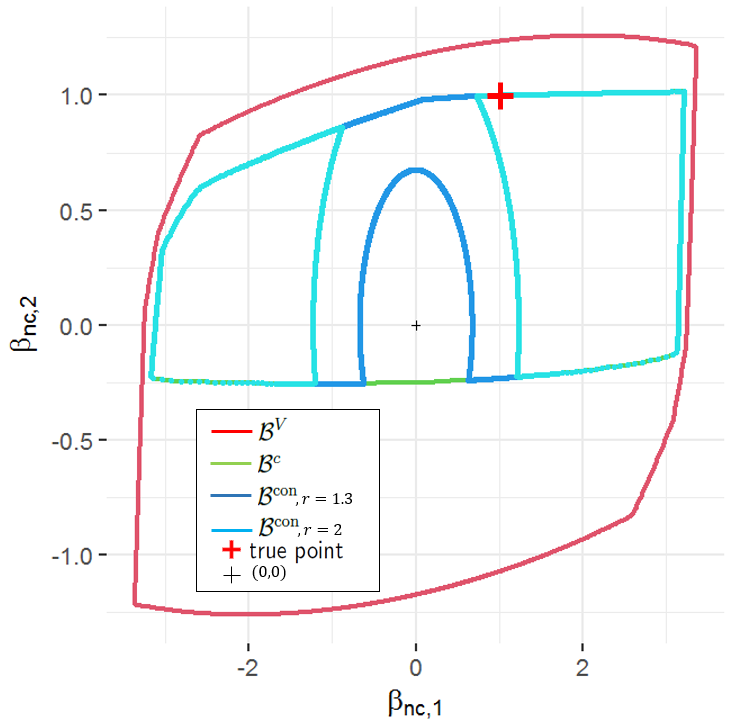}
		\label{fig:norm1_monot_convr2}}
	\caption*{\footnotesize Note: For Panel \ref{fig:norm1_monot_conv} and $\mathcal{B}^V$ and $\mathcal{B}^c$ in Panel \ref{fig:norm1_monot_convr2}, the sets are obtained as in Figure \ref{fig:Sp30}. For the constraints $R_l^2 \geq r R_s^2$ in Panel \ref{fig:norm1_monot_convr2}, we use 1,500 directions and no convexification. For this DGP, the true values of the $R^2$ of the long and short regressions are $R_l^2 = 0.307$ and $R_s^2 =0.107 $.  }
	\caption{Identification regions for different shape restrictions}
	\label{fig:Sp31}
\end{figure}

\medskip
Figure \ref{fig:Sp31} presents convexity constraints and different  constraints on the $R^2$ on the first DGP. While unlike case (a) of Figure \ref{fig:Sp30}, convexity constraints alone fail to reject $0_2\not\in \Bcon$, imposing a constraint of the form  $\underline{R}^2 \geq r R_s^2$ with $r>1$ rejects it by definition. In the latter case, the identified set is no longer convex, allowing to exclude some directions from the identified set and providing an informative lower bound on $|\beta_{nc,1}|$. Overall, that the sharp identified sets $\B^c$ are much more informative than the identified set $\B^V$ based on the variance restrictions highlights the importance of using all of the restrictions implied by the model. Another takeaway from these numerical illustrations is that sign constraints can be very informative in practice, resulting in significant shrinkage of the identified set.

\subsection{Regularization}
\label{sub:regul}

An issue for estimation and inference on $\B$ is that when $\alpha\to 0$ or $\alpha\to 1$, $R(\alpha,F,G)$ is a ratio of two terms tending to 0. It follows that its plug-in estimator may become very unstable. To regularize the problem, we consider an outer set of $\B$ based on the removal of extreme values of $\alpha$. We will focus on this outer set when we turn to estimation and inference in Section~\ref{sec3}. Specifically, we define, for any $\eps\in (0,1/2)$,
\begin{align}
S_\eps(F,G) &=\min_{\alpha \in [\eps,1-\eps]} R(\alpha,F,G), \label{eq:def_S_eps}\\
\B_\eps & = \left\{\lambda q: q\in \S, 0\leq \lambda \leq S_\eps(F_{Y_0},F_{X_0'q})\right\}.\notag
\end{align}
Note that for all $F,G$, $\alpha \mapsto R(\alpha, F,G)$ is continuous on $[\eps,1-\eps]$. Thus, the minimum in \eqref{eq:def_S_eps} is well-defined. Proposition~\ref{prop:epsilon} below describes some properties of $\B_\eps$ and relates it to the sharp identified set $\B$.

\begin{prop}\label{prop:epsilon}
	Suppose that Assumption \ref{hyp:mom} holds. Then:
	\begin{enumerate}
		\item For all $\eps \in (0,1/2)$, $\B_\eps$ includes $0_p$, is compact and convex;
	 	\item For all $0<\eps < \eps'<1/2$, $\B \subset \B_\eps \subset \B_{\eps'}$ and $\cap_{\eps\in (0,1/2)} \B_\eps =\B$;
		\item Suppose that $F_Y$ is continuous and $U:=Y_0-X_0'\beta_0$ satisfies
		\begin{equation}
\forall \lambda>0,\quad \lim_{t\to\infty} \sup_s \frac{\overline{F}_{\|X_0\|}(\lambda t)}{\overline{F}_{U|X_0'\beta_0=s}(t)}=0, \; \lim_{t\to\infty} \sup_s \frac{\overline{F}_{\|X_0\|}(\lambda t)}{\overline{F}_{-U|X_0'\beta_0=s}(t)}=0.			
			\label{eq:lighter_tails_X}
		\end{equation}
		Then, there exists $\eps_0\in (0, 1/2)$ such that for all $\eps\in(0,\eps_0]$, $\B=\B_{\eps}$.
	\end{enumerate}
\end{prop}
	
The first part of Proposition~\ref{prop:epsilon} states that the regularized set $\B_\eps$, for all $\eps \in (0,1/2)$, preserves the compactness and convexity of the sharp identified set $\B$. The second part states that $\B_\eps$ is always a superset of $\B$, which is arbitrarily close to $\B$ as $\eps\downarrow 0$. The third part states that if, basically, the tails of $\|X_0\|$ are thinner than those of $U$ (Condition (11)), the set $\B_\eps$ coincides with the sharp set $\B$ for $\eps$ small enough. Condition \eqref{eq:lighter_tails_X} holds in particular if $X$ has a bounded support and $\Supp(U)=\R$, or if $U$ is symmetric and has a tail index larger than that of $\|X_0\|$. Note that $\B=\B_{\eps}$ may hold even without \eqref{eq:lighter_tails_X}. For instance, if both $U$ and $X$ are normally distributed, it is easy to check that $\B_\eps=\B$ for all $\eps\in (0,1/2)$.

\medskip
On the other hand, when $U$ has thinner tails than $X'\beta_0$, $\B_\eps$ will be a strict superset of $\B$ for $\eps$ large enough. In such cases, and under additional restrictions, we provide upper bounds on the Hausdorff distance between $\B$ and $\B_\eps$ in Proposition \ref{prop:Hausdorff}. Intuitively, these bounds inform us about the maximal possible loss, in terms of 
identification, that is due to regularization.

\begin{prop}
	\label{prop:Hausdorff}
	Suppose that Assumption \ref{hyp:mom} holds and let $U:=Y_0-X_0'\beta_0$. Then:
	\begin{enumerate}
		\item\label{prop:H1} Assume that $X$ has an elliptical distribution with nonsingular variance matrix $\Sigma$, a density with respect to the Lebesgue measure and $\lim\inf_{|x|\to\infty} |x|^{1+c} f_{X_0'\beta_0}(x)>0$ for some $c>1$. Suppose also that $\lim\sup_{x\to\infty}$ $x^d \overline{F}_{|U|}(x)<\infty$ for some $d>c$. Then, there exists $K_1>0$ such that for $\eps$ small enough,
		$$d_H(\B,\B_\eps)\le K_1 \eps^{\frac{1/c-1/d}{1+1/d}}.$$
		\item Assume that $\beta_0=0_p$, $\lim\inf_{x\to\infty} \inf_{q\in\S} x^c \overline{F}_{X_0'q}(x)>0$ and $\lim\sup_{x\to\infty} x^d \overline{F}_{|U|}(x)$ $<\infty$ for some $d>c$. Then, there exists $K_2>0$ such that for $\eps$ small enough,
$$d_H(\B,\B_\eps)\le K_2 \eps^{1/c-1/d}.$$
	\end{enumerate}
\end{prop}

The tail conditions imposed in Proposition \ref{prop:Hausdorff} are basically the opposite as in Point 3 of Proposition \ref{prop:epsilon}, as they imply that $\|X\|$ has fatter tails than $U$. The assumption	 that $X$ has an elliptical distribution in Point 1 allows us to relate $S_\eps(F_{Y_0}, F_{X_0'q})-S(F_{Y_0}, F_{X_0'q})$, for any $q\in\S$, with  $S_\eps(F_{Y_0},F_{X_0'\beta_0})-S(F_{Y_0},F_{X_0'\beta_0})$, but it is not necessary to obtain an upper bound on $S_\eps(F_{Y_0},F_{X_0'\beta_0})-S(F_{Y_0},F_{X_0'\beta_0})$.

\medskip
In the two cases of Proposition \ref{prop:Hausdorff}, we produce upper bounds on the Hausdorff distance between $\B$ and $\B_\eps$ that are, up to some constants, power of the regularization parameter $\eps$. The upper bounds are close to 0 when $\eps$ is small, in line with Point 2 of Proposition \ref{prop:epsilon}. They are also closer to 0 the smaller $c$ is, i.e. the fatter the tails of $X'\beta_0$ (or $X'q$) are, or the larger $d$ is, i.e. the thinner the tails of $U$ are.


\section{Inference}
\label{sec3}

We now consider the estimation of the identified set, and how to conduct inference on the parameters of interest $\beta_0$. As in the previous section, we first consider the case without common regressors before showing how to incorporate such regressors and combine them with additional constraints. We conclude this section by discussing some computational aspects of our procedure. We illustrate the finite sample performances of our inference method in Online Appendix~\ref{secMC}.

\subsection{No common regressors}
\label{Inference_baseline}

\subsubsection{Estimation of the identification region and confidence region} 
\label{ssub:estimation_of_the_identification_region_and_confidence_region}

We rely on random samples from the distributions of $Y$ and $X$.

\begin{hyp}
	We observe $(Y_1,...,Y_{n_Y})$ and $(X_1,...,X_{n_X})$, two independent samples of i.i.d. variables with the same distribution as $Y$ and $X$, respectively.
	\label{hyp:samples}
\end{hyp}

For any $q\in\S$, let $\widehat{F}_Y$ and $\widehat{F}_{X'q}$ denote the empirical cdf of $Y$ and $X'q$ and let $\widehat{F}_{Y_0}(t)=\widehat{F}_Y(t+\overline{Y})$ and $\widehat{F}_{X'_0q}(t)=\widehat{F}_{X'q}(t+\overline{X}'q)$. We simply estimate $R(\alpha, F_{Y_0},F_{X_0'q})$ and $S_\eps(F_{Y_0},F_{X_0'q})$ by their empirical counterpart $R(\alpha, \widehat{F}_{Y_0}, \widehat{F}_{X'_0q})$ and $S_\eps(\widehat{F}_{Y_0}, \widehat{F}_{X'_0q})$. It turns out that these functions can be computed quickly, as detailed in Section \ref{subsec:computation} below. We then also simply estimate the identified set $\B_\eps$ by plug-in:
$$ \widehat{\B}_\eps := \left\{\lambda q:   q\in \S, \ 0 \leq \lambda \leq S_{\eps}\left(\widehat{F}_{Y_0}, \widehat{F}_{X_0'q}\right)\right\}.$$

Next, we build confidence regions on $\beta_0$. The asymptotic distribution of $S_{\eps}\left(\widehat{F}_{Y_0}, \widehat{F}_{X_0'q}\right)$ is not Gaussian in general, so we rely on subsampling \citep{politis1999subsampling}. One could alternatively use the numerical bootstrap, see the discussion pp. 18-19 in the first version of \cite{DGM22}.

\medskip
Let $n=(n_Xn_Y)/(n_X+n_Y)$ and let $b_n$ denote the size of the subsample. For any estimator $\widehat{\theta}$, let $\widehat{\theta}^*$ denotes its subsampling counterpart. For a nominal coverage of $1-\alpha$, the confidence region on $\beta_0$ we consider is given by
$$\CR(\beta_0)= \left\{\lambda q: \ q\in\S, \ 0\leq \lambda \leq S_{\eps}\left(\widehat{F}_{Y_0}, \widehat{F}_{X_0'q}\right) - \widehat{c}_{\alpha,\eps}(q) n^{-1/2}\right\},$$
where $\widehat{c}_{\alpha,\eps}(q) $ is the quantile of order $\alpha$ of the distribution of $b_n^{1/2}[ S_\eps(\widehat{F}^*_{Y_0},\widehat{F}^*_{X_0'q}) -  S_{\eps}(\widehat{F}_{Y_0}, $ $\widehat{F}_{X_0'q})]$, conditional on the data.

\medskip
\paragraph{Inference on subcomponents of $\beta_0$.}
In practice, one is often interested in conducting inference on subcomponents of $\beta_0$. In view of \eqref{eq:link_S_sig}, the identified (outer) set $\B_{k,\eps}$ of $\beta_{0,k}$ corresponding to $\B_\eps$ satisfies
\begin{equation}\label{eq:Bk}
\B_{k,\eps}=[-\sigma_\eps(-e_k, F_{Y_0}, F_{X_0}),\, \sigma_\eps(e_k, F_{Y_0}, F_{X_0})],
\end{equation}
where $\sigma_\eps(\cdot, F_{Y_0}, F_{X_0})$ denotes the support function associated to $q\mapsto S_\eps(F_{Y_0},F_{X_0'q})$ and $e_k$ is the $k$-th element of the canonical basis of $\R^p$. To construct confidence intervals on $\beta_{0k}$, we first estimate $\sigma_\eps(\cdot, F_{Y_0}, F_{X_0})$ by
\begin{equation}
\sigma_\eps(e, \widehat{F}_{Y_0}, \widehat{F}_{X_0}) = \frac{1}{\inf_{q\in \R^p: q'e=1} 1/S_{\eps}\left(\widehat{F}_{Y_0}, \widehat{F}_{X_0'q}\right)},	
	\label{eq:sig_hat}
\end{equation}
see Corollary \ref{cor:sub}. Then, denoting by $\widetilde{c}_{\beta,\eps}(e)$ the quantile of order $\beta\in (0,1)$ of the distribution of  $b_n^{1/2}( \sigma_{\eps}(e, \widehat{F}^*_{Y_0}, \widehat{F}^*_{X_0}) -  \sigma_\eps(e, \widehat{F}_{Y_0}, \widehat{F}_{X_0}))$, conditional on the data, the confidence interval we consider for $\beta_{0,k}$ is
$$\CI(\beta_{0,k})=  \left[\left(-\sigma_\eps(-e_k, \widehat{F}_{Y_0}, \widehat{F}_{X_0}) + \frac{\widetilde{c}_{\alpha,\eps}(-e_k)}{n^{1/2}}\right)^-,
\left(\sigma_\eps(e_k, \widehat{F}_{Y_0}, \widehat{F}_{X_0}) -  \frac{\widetilde{c}_{\alpha, \eps}(e_k)}{n^{1/2}}\right)^+ \right],$$
where $x^-=\min(0,x)$ and $x^+=\max(0,x)$. The rationale for using $(\cdot)^-$ and $(\cdot)^+$ is to ensure that $0\in \CI(\beta_{0,k})$: recall that without constraints, $0\in \B_{k,\eps}$. The advantage, then, is that we can still use the quantiles of order $\alpha$ while maintaining coverage even under point identification, as formally shown in Theorem \ref{thm:inference} below.

\medskip
\paragraph{Choice of the regularization parameter $\eps$.}
Because $S_\eps(F_{Y_0},F_{X_0'q}) \geq S(F_{Y_0},F_{X_0'q})$, the confidence regions and intervals above are conservative in general. To gain in efficiency, we suggest using several $\eps$, and, basically, keep the one leading to the smallest confidence regions or intervals. We distinguish the cases $p=1$, where we can adapt the choice to the direction $q\in\mathcal{S}$ while preserving the convexity of $\widehat{\mathcal{B}}_\eps$, from the case $p>1$. When $p=1$, let us define, for $q\in\mathcal{S}=\{-1,1\}$,
\begin{equation}\label{eq:tradeoff}
 \eps(q) = \underset{\eps \in \mathcal{E}}{\argmin}  \  S_\eps(\widehat{F}_{Y_0}, \widehat{F}_{X_0'q}) -  \widehat{c}_{\alpha,\eps}(q)n^{-1/2},
\end{equation}
where $\mathcal{E}$ is a finite grid in $(0,1/2]$. Hence, $\eps(q)$ simply minimizes the boundary value of the confidence region in the direction $q\in\mathcal{S}$. This idea is similar to that of \cite{chernozhukov2013intersection} in the context of intersection bounds.

\medskip
Now consider the case $p>1$. If one focuses on confidence intervals on $\beta_{0k}$, we need to choose the parameter $\eps$ that appears in $\sigma_\eps(\pm e_k, F_{Y_0}, F_{X_0})$. To this end, we simply use $\eps(q)$ as  given above, with $q=\pm e_k$. If we are interested instead in the set $\B$ itself, we recommend using $\underline{\eps} = \min_{q \in \mathcal{Q}} \eps(q)$, where $\mathcal{Q}$ is a finite subset of $\mathcal{S}$.


\subsubsection{Consistency and validity of the confidence region} 
\label{ssub:consistency_and_validity_of_the_confidence_region}

The following theorem shows that $\widehat{\B}_\eps$ is consistent for $\B_\eps$, in the sense of the Hausdorff distance, under mild regularity conditions.

\begin{thm}\label{thm:consistency}
Suppose that Assumptions \ref{hyp:mom}-\ref{hyp:samples} hold. Then, as $n\to\infty$,
$$d_H\left(\widehat{\B}_\eps, \B_\eps\right)\convP 0.$$	
\end{thm}

\medskip
Next, we establish the asymptotic validity of $\CR(\beta_0)$ and $\CI(\beta_{0,k})$, under Assumptions \ref{hyp:for_CR} and \ref{hyp:for_CI} respectively. Assumption \ref{hyp:eps_q} (resp. \ref{hyp:eps_e}) is used to establish the asymptotic validity of $\CR(\beta_0)$ (resp. $\CI(\beta_{0,k})$) using $\eps(q)$ or $\underline{\eps}$ (resp. $\eps(\pm e_k)$), as defined above, instead of a fixed $\eps$.

\begin{hyp} (Regularity conditions for $\CR(\beta_0)$)
$E[\|X\|^2]<\infty$, $E[Y^2]<\infty$. Also, for all $q\in\S$, there exists $\eps'\in (0,\eps)$ such that $F_{X'q}$ and $F_Y$ are continuous and strictly increasing on $[F_{X'q}^{-1}(\eps'), F_{X'q}^{-1}(1-\eps')]$ and $[F_Y^{-1}(\eps'), F_Y^{-1}(1-\eps')]$ respectively.
	\label{hyp:for_CR}
\end{hyp}

\begin{hyp} (Regularity conditions for $\CI(\beta_{0,k})$)
$E[\|X\|^2]<\infty$, $E[Y^2]<\infty$. Also, there exists $\eps'\in (0,\eps)$ such that  for all  $(\alpha,\alpha')\in[\eps',1-\eps']^2$, there exists a strictly increasing and continuous function $m$ such that $m(0)=0$ and
\begin{align}
\sup_{q\in\S} \left|F_{X'q}^{-1}(\alpha')-F_{X'q}^{-1}(\alpha)\right|& <m(|\alpha'-\alpha|),\label{hyp:reg_X_CI} \\	
 \left|F_Y^{-1}(\alpha')-F_Y^{-1}(\alpha)\right|& < m(|\alpha'-\alpha|).	\notag
\end{align}
Finally, for all $e=\pm e_k$ $(k=1,...,p)$, either (i) $\sigma_\eps(e, F_{Y_0}, F_{X_0})>\sigma(e, F_{Y_0}, F_{X_0})$, (ii) $q \mapsto [qS_\eps(F_{Y_0}, F_{X_0'q})]'e$ admits a unique maximizer on $\S$, or (iii) for all $q_m\in \arg\max_{q\in\S} [qS_\eps(F_{Y_0}, F_{X_0'q})]'e$, $a\mapsto R(a, F_Y, F_{X'q_m})$ admits a unique minimizer on $[\eps,1-\eps]$.
\label{hyp:for_CI}
\end{hyp}

\begin{hyp} (Regularity conditions for the validity of $\CR(\beta_0)$ based on data-dependent $\eps$)
For all $q\in\mathcal{S}$, we either have (i) $S_\eps(F_{Y_0},F_{X_0'q})>S(F_{Y_0},F_{X_0'q})$ for all $\eps\in \mathcal{E}$, or (ii) $a\mapsto R(a,F_{Y_0},F_{X_0'q})$ admits a unique minimizer on $(0,1)$.
\label{hyp:eps_q}
\end{hyp}

\begin{hyp} (Regularity conditions for the validity of $\CI(\beta_{0,k})$ based on data-dependent $\eps$)
For all $e=\pm e_k$ $(k=1,...,p)$, we either have (i) $\sigma_\eps(e, F_{Y_0},F_{X_0})>\sigma(F_{Y_0},F_{X_0})$ for all $\eps\in \mathcal{E}$ or (ii) for all $q_m\in \arg\max_{q\in\S} [qS_{\eps_{j_0}}(F_{Y_0}, F_{X_0'q})]'e$, $a\mapsto R(a,F_{Y_0},F_{X_0'q_m})$ admits a unique minimizer $a(q_m)$ on $(0,1)$, with $a(q_m)\in [\eps_{j_0},1-\eps_{j_0}]$ and $\eps_{j_0}:=\max\{\eps\in\mathcal{E}: \sigma_\eps(F_{Y_0},F_{X_0'q})=\sigma(F_{Y_0},F_{X_0'q})\}$.
\label{hyp:eps_e}
\end{hyp}

The second part of Assumption \ref{hyp:for_CR} holds if for all $q\in\S$, the distributions of $X'q$ and $Y$ are continuous with respect to the Lebesgue distribution and their support is a (possibly unbounded) interval. The first part of Assumption \ref{hyp:for_CI} is basically a reinforcement of Assumption \ref{hyp:for_CR} to ensure that some of our results hold uniformly over $q$. This is needed when we consider the support function, as this function implies an optimization over $q$. A sufficient condition for \eqref{hyp:reg_X_CI} is that, for all $q\in\S$, $X'q$ admits a density $f_{X'q}$ with respect to the Lebesgue measure and $\inf_{(q,\alpha)\in\S\times[\eps,1-\eps]} f_{X'q}(F_{X'q}^{-1}(\alpha))>0$. The conditions (ii) and (iii) in Assumption \ref{hyp:for_CI} are sufficient conditions for the continuity of the asymptotic distribution of $n^{1/2}\left(\sigma_\eps(e, \widehat{F}_{Y_0}, \widehat{F}_{X_0})- \sigma_\eps(e,F_{Y_0},F_{X_0})\right)$, which is necessary for the validity of subsampling.

\medskip
Assumption \ref{hyp:eps_q} can accomodate DGPs where the tails of $\|X_0\|$ are thinner than those of $U$ (which may correspond to $a\mapsto R(a,F_{Y_0},F_{X_0'q})$ admitting a unique minimum) but also DGPs for which the opposite holds (since in this case we can have $S_{\eps}(F_{Y_0},F_{X_0'q})>S(F_{Y_0},F_{X_0'q})$ for all $\eps\in (0,1/2)$ and $q$). For instance, one can check that it holds if $Y=c+X+U$ with $c\in\R$, $X\indep U$, and either $X$ follows a Laplace distribution while $U$ is uniform, or the other way around. But it fails to hold when both $X$ and $Y$ are Gaussian, since then $a\mapsto R(a,F_{Y_0},F_{X_0'q})$ is actually constant. Assumption \ref{hyp:eps_e} is basically similar to Assumption \ref{hyp:eps_q} but somewhat more complicated, as we consider therein the support function instead of the radial function.

\medskip
\begin{thm}\label{thm:inference}
Fix $(\eps,\alpha)\in (0,1/2)^2$ and suppose that $n_X/(n_X+n_Y)\to \mu\in (0,1)$, $b_n\to\infty$, $b_n/n\to 0$ and Assumptions \ref{hyp:mom}-\ref{hyp:samples} hold. Then:
\begin{enumerate}
	\item If Assumption \ref{hyp:for_CR} also holds,
	\begin{equation}
\underset{\beta  \in  \B}{\inf} \  \underset{n\to \infty}{\liminf} P\left(\beta \in \CR(\beta_0)\right) \geq
 1-\alpha,
		\label{eq:valid_CR}
	\end{equation}
	with equality if $\B=\B_\eps$. Moreover, if Assumption \ref{hyp:eps_q} also holds, \eqref{eq:valid_CR} is still true if we use $\eps(q)$ (when $p=1)$ or $\underline{\eps}$ (when $p>1$) instead of $\eps$.
	\item If Assumption \ref{hyp:for_CI} also holds,
	\begin{equation}
	\underset{n\to \infty}{\liminf} \underset{\beta_k  \in  \B_k}{\inf} \  P\left(\beta_k \in \CI(\beta_{0,k})\right) \geq
	1-\alpha,
		\label{eq:valid_CI}
	\end{equation}
	with equality if $\B_k=\B_{k,\eps}$.  Moreover, if Assumption \ref{hyp:eps_e} also holds, \eqref{eq:valid_CI} is still true if we use $\eps(e_k)$ and $\eps(-e_k)$ instead of $\eps$.
\end{enumerate}
\end{thm}

To prove \eqref{eq:valid_CR}-\eqref{eq:valid_CI}, we first show the weak convergence of
$$\sqrt{n}\left(R(\alpha, \widehat{F}_{Y_0},\widehat{F}_{X_0'q}) - R(\alpha, F_{Y_0},F_{X_0'q})\right),$$
seen as a process indexed by either $\alpha$ or $(\alpha,q)$. The convergence in distribution of $S_\eps(\widehat{F}_{Y_0}, \widehat{F}_{X_0'q})$ and $\sigma_\eps(e, \widehat{F}_{Y_0}, \widehat{F}_{X_0})$, and in turn \eqref{eq:valid_CR}-\eqref{eq:valid_CI}, then essentially follows by the Hadamard directional differentiability of the minimum and maximin maps, shown respectively by \cite{carcamo2019directional} and \cite{firpo2021uniform}.

\medskip
Our results for a fixed $\eps>0$ extend to the data-dependent $\eps(q)$ and $\underline{\eps}$, under the additional conditions provided above. Note that one could avoid these conditions by using sample splitting, with one subsample used to choose $\eps(q)$ or $\underline{\eps}$ and the other to construct the confidence regions/intervals. One drawback of this alternative solution, though, is that it increases the size of confidence regions/intervals, to a point that we may lose the benefits of using a data-dependent rather than a fixed $\eps$.


\subsection{Common regressors and possible constraints}
\label{subsec:common_reg}

We now turn to inference on $\beta_0$ with common regressors $X_c$. Recall from Proposition \ref{prop:common} that the identified set on $\beta_0$ is
$$\B^c  = \left\{\lambda q: q\in \S, \; 0\leq \lambda\leq \overline{S}(F_{Y,X_c}, F_{X_{nc}'q,X_c})\right\},$$
with $\overline{S}(F_{Y,X_c}, F_{X_{nc}'q,X_c}) = \inf_{x\in \Supp(X_c)} S(F_{Y^x|X_c=x}, F_{X^x{}'q|X_c=x})$.

\medskip
Let us first assume that $X_c$ has a finite support. Let $\widehat{F}_{Y^x|X_c=x}$ and $\widehat{F}_{X^x{}'q|X_c=x}$ denote the empirical estimators of $F_{Y^x|X_c=x}$ and $F_{X^x{}'q|X_c=x}$, respectively. Following the same logic as above, we estimate $\overline{S}(F_{Y,X_c}, F_{X_{nc}'q,X_c})$ by
$$\widehat{\overline{S}}(q, F_{Y,X_c},F_{X_{nc}'q,X_c}) = \min_{x \in \Supp(X_c)} S_\eps(\widehat{F}_{Y^x|X_c=x},\widehat{F}_{X^x{}'q |X_c=x}).$$
Let $\widehat{c}^c_{\alpha,\eps}(q)$ be the quantile of order $\alpha\in(0,1)$ of the distribution of $b_n^{1/2}(\widehat{\overline{S}}{}^*(q, F_{Y,X_c},F_{X_{nc},X_c})$ $- \widehat{\overline{S}}(q, F_{Y,X_c}, F_{X_{nc},X_c}))$, conditional on the data. For a nominal coverage of $1-\alpha$, the confidence region on $\beta_0$ we consider is
$$\CR^c(\beta_0)= \left\{\lambda q: q\in\S, \ 0 \le \lambda \le  \widehat{\overline{S}}(q, F_{Y,X_c},F_{X_{nc},X_c}) - \widehat{c}^c_{\alpha,\eps}(q)n^{-1/2} \right\}.$$

\medskip
With continuous common regressors, one can adapt the earlier arguments using sieve estimation. Specifically, suppose that Model \eqref{eq:PLM} holds and consider a linear sieve approximation of $f(\cdot)$ by a step function $x_c\mapsto \sum_{k=1}^{K_n} \indic{x_c\in I_{n,k}}\gamma_k$ for some partition $(I_{n,k})_{k=1...K_n}$ of the support of $X_c$ and with $K_n$ tending to infinity at an appropriate rate. Then, one can construct a confidence region on $\beta_0$ by following a similar logic as above.\footnote{Establishing the asymptotic validity of such a confidence region would require to handle both the bias stemming from the approximation of $f(\cdot)$ and the increasing complexity of the approximation. We leave this analysis for future research.}

\medskip
We now discuss how to conduct inference under constraints on the $R^2$ or shape restrictions, as considered in Subsections~\ref{ssub:lower_bound_on_the_r_2} and \ref{subsubs:id_power_shape_restrictions} respectively. The main difference with above is that for a given direction $q\in\S$, both the lower and upper bounds on the identified set need to be estimated. As before, we can estimate them with plug-in estimators. The only substantive difference is that in the confidence regions, we need to account for the variability of both bounds. For instance, with shape restrictions, we can consider the following confidence region:
\begin{align*}
\CR^{con}(\beta_0)= \bigg\{& \lambda q: q\in\S, \ \widehat{\underline{S}}{}^{con}(q, F_{Y,X_c},F_{X_{nc},X_c}) +
 \widehat{\underline{c}}{}^{con}_{1-\alpha/2,\eps}(q)n^{-1/2} \le \lambda \\
 &  \le  \widehat{\overline{S}}{}^{con}(q, F_{Y,X_c},F_{X_{nc},X_c}) - \widehat{\overline{c}}{}^{con}_{\alpha/2,\eps}(q)n^{-1/2} \bigg\},	
\end{align*}
where $\widehat{\underline{c}}^{con}_{\delta,\eps}(q)$ is the quantile of order $\delta$ of $b_n^{1/2}(\widehat{\underline{S}}{}^{con*}(q, F_{Y,X_c},F_{X_{nc},X_c})$ $- \widehat{\underline{S}}{}^{con}(q, F_{Y,X_c},$ $F_{X_{nc},X_c}))$, conditional on the data and similarly for $\widehat{\overline{c}}{}^{con}_{\delta,\eps}$. We conjecture  that with a finite number of constraints, $X_c$ finitely supported and if $[Rm_{X_{nc}}'q](r)\ne 0$ for all $r\in\mathcal{R}$,
$\CR^{con}(\beta_0)$ is pointwise asymptotically conservative. Alternatively, one could use the formulation of our problem with shape constraints as a set of infinitely many moments inequalities. While generally far less tractable that our baseline approach, confidence intervals based on the inversion of the test of these many moment inequalities have uniformly correct asymptotic size \citep{andrews2017inference}.

\subsection{Computational aspects}
\label{subsec:computation}

We first discuss how to efficiently compute $S_\eps(\widehat{F}_{Y_0},\widehat{F}_{X_0'q})$. Let $Y_{(1)}<...<Y_{(m_y)}$ represent the $m_y\le n_y$ distinct, ordered values of the $(Y_i)_{i=1,...,n_y}$ and let $W_{(j)}^Y=\#\{i:Y_{i}=Y_{(j)}\}/n_y$. Let us also define $I^Y=\{\sum_{j=1}^i W_{(j)}^Y: i=1,...,m_y-1\}$. We define similarly $W_{(j)}^{X'q}$ and $I^{X'q}$. By construction, the numerator $\widehat{f}^Y(\alpha):=\int_\alpha^1 \widehat{F}^{-1}_{Y_0}(t)dt$ of $R(\alpha, \widehat{F}_{Y_0}, \widehat{F}_{X'_0q})$ is  linear on
all intervals $[\sum_{j=1}^i W_{(j)}^Y, \sum_{j=1}^{i+1} W_{(j)}^Y]$ ($i=0,...,m_y-1$). Moreover, for any $\alpha=\sum_{j=1}^i W_{(j)}^Y\in I^Y$,
\begin{equation}
\widehat{f}^Y(\alpha) = \sum_{j=i+1}^{m_y} W^Y_{(j)}\left(Y_{(j)} - \overline{Y}\right).	
	\label{eq:numer_Rhat}
\end{equation}
The same holds for the denominator $\widehat{f}^{X'q}(\alpha) $ of $R(\alpha,\widehat{F}_{Y_0}, \widehat{F}_{X'_0q})$. As a result, $R(\alpha, \widehat{F}_{Y_0},$ $\widehat{F}_{X'_0q})$ is of the form $(a \alpha + b)/(c\alpha + d)$ on intervals between two consecutive values of $I^Y \cup I^{X'q}$. Now, observe that the minimum of such a function is reached at one of the endpoints of the interval. As a result, we can compute $S_\eps(\widehat{F}_{Y_0},\widehat{F}_{X_0'q})$ using the following algorithm:
\begin{enumerate}
	\item Compute $\widehat{f}^Y(\cdot)$ on $I^Y$ using \eqref{eq:numer_Rhat} and let $\widehat{f}^Y(0) = \widehat{f}^Y(1) =0$. Proceed similarly with $\widehat{f}^{X'q}(\cdot)$;
	\item Interpolate linearly $\widehat{f}^Y(\cdot)$ (resp. $\widehat{f}^{X'q}(\cdot)$) on $\{\eps, 1-\eps\}\cup I^{X'q}$ (resp. $\{\eps, 1-\eps\}\cup I^Y$).
	\item Compute $S_\eps(\widehat{F}_{Y_0},\widehat{F}_{X_0'q})= \min_{\alpha\in\{\eps,1-\eps\}\cup I^Y\cup I^{X'q}} \widehat{f}^Y(\alpha)/\widehat{f}^{X'q}(\alpha)$.
\end{enumerate}

To compute $\sigma_\eps(\pm e_k, F_{Y_0}, F_{X_0})$, we solve \eqref{eq:sig_hat}, in which $q \mapsto 1/S_\eps(\widehat{F}_{Y_0},\widehat{F}_{X_0'q})$ is also convex. In practice, we use  the BFGS quasi-Newton method implemented in the R package \texttt{optim}, using as a starting point the considered direction $e$.

\medskip
Finally, the exact computation of $\widehat{\B}_\eps$ and $\CR(\beta_0)$  requires the computation of $S_\eps(\widehat{F}_{Y_0}, \widehat{F}_{X_0'q})$ and $\widehat{c}_{\alpha,\eps}(q)$ for all $q\in\S$, which is in practice infeasible if $p>1$ as $\mathcal{S}$ is infinite. Instead, we suggest to (i) fix a grid $\widetilde{\S}\subset \S$; (ii) compute $S_\eps(\widehat{F}_{Y_0}, \widehat{F}_{X_0'q})$ and $\widehat{c}_{\alpha,\eps}(q)$ for each $q\in\widetilde{\S}$; (iii) construct an approximation of $\widehat{\B}_\eps$ and $\CI$ by computing the convex hulls of $\{S_\eps(\widehat{F}_{Y_0}, \widehat{F}_{X_0'q})q: \ q\in\widetilde{\mathcal{S}}\}$ and  $\left\{\left(S_\eps(\widehat{F}_{Y_0}, \widehat{F}_{X_0'q}) - \widehat{c}_{\alpha,\eps}(q) n^{-1/2}\right)q:\right.$ $ \left. q\in\widetilde{\mathcal{S}}\right\}$, respectively.\footnote{The convex hull of $n$ points in $\R^p$ can be computed efficiently by the quickhull algorithm \citep{barber1996quickhull}, which  requires around $n^{p/2}$ operations.} The resulting sets, $\widetilde{\B}_\eps$ and $\widetilde{\text{CR}}_{1-\alpha}(\beta_0)$ say, are convex, inner approximations of $\widehat{\B}_\eps$ and $\CR(\beta_0)$, and satisfy,  as $d_H(\S,\widetilde{\mathcal{S}})\to 0$, $d_H(\widetilde{\B}_\eps, \widehat{\B}_\eps)  \to 0$ and $d_H(\widetilde{\text{CR}}_{1-\alpha}(\beta_0), \CR(\beta_0))  \to 0$.

\medskip
The computation of the estimated set, the confidence regions  on $\beta_0$ and $\gamma_0$ in the specification $f(X_c) =X_c'\gamma_0$ (where $X_c$ is the vector of all dummy variables associated with a finitely supported variable) and the confidence intervals on the corresponding subcomponents are implemented in our companion R package \texttt{RegCombin}. The package also handles shape restrictions and lower bound on the $R^2$ of the long regression, as well as combinations of these. The \texttt{RegCombin} vignette, available through the description of the package on CRAN, provides additional details about the implementation, including the choice of the tuning parameters $\mathcal{E}$ and $b_n$.

\section{Application to intergenerational mobility in the United States}
\label{sec:appli}

We now apply our method to conduct inference on the intergenerational income mobility over the period 1850 to 1930 in the United States, revisiting the influential analysis of \cite{olivetti2015name} on this question. We follow their paper and focus on the father-son and father-son-in-law intergenerational income elasticities. We conduct our analysis using 1 percent extracts from the decennial censuses of the United States, over the period 1850 to 1930 (1850-1930 IPUMS).\footnote{We refer the reader to Section 2 of \cite{olivetti2015name} for a detailed discussion of the data used in the analysis. Note that they estimate the evolution of the intergenerational income mobility over a longer time window (1850 to 1940) than we do. We confine our analysis to the period 1850-1930 as the 1940 portion of the data (1\% extract of the IPUMS Restricted Complete Count Data) is not publicly available.}

\medskip
An important feature of the historical Census data used in this analysis is that father's and son's (as well as son-in-law's) incomes are not jointly observed. \cite{olivetti2015name} address this measurement issue by predicting, for any given child (John, say)  observed in one of the Census datasets, their father's log earnings using the mean log earnings of fathers whose children have the same first name (namely, John). \citeauthor{olivetti2015name} then estimate in a second step the intergenerational elasticity by regressing son's log earnings on the predicted father's log earnings computed from the previous step. This procedure boils down to a two-sample two-stage least squares estimator (TSTSLS).\footnote{Another limitation of the data used in \cite{olivetti2015name} and in this application is that it does not allow us to directly calculate the intergenerational elasticity in income. Instead, we follow the baseline specification of \cite{olivetti2015name} and proxy income using an index of occupational standing available from IPUMS (OCCSCORE), which is constructed as the median total income of the persons in each occupation in 1950.} The corresponding exclusion restriction that the son's first name does not predict his log earnings, once we control for his father's log earnings, may nonetheless be problematic; see \cite{SantavirtaStuhler22} for a critical review of the empirical literature using TSTSLS in this context of intergenerational mobility. For the periods 1860-1880 and 1880-1900 only, the IPUMS Linked Representative Samples link fathers and sons using information on first and last names, which allows us to estimate more directly the father-son elasticity using OLS.\medskip

Using our notation and consistent with \cite{olivetti2015name}, the population parameter of interest here is given by
$$\theta_0:=\frac{\Cov(Y,X_{nc})}{V(X_{nc})} = \beta_0 + \left(\frac{\Cov(X_c,X_{nc})}{V(X_{nc})}\right)'\gamma_0,$$
where $Y$ denotes the son's (or son-in-law's) log-income, $X_{nc}$ the father's log-income and $X_c$ the vector of indicators corresponding to the son's (or son-in-law's) first names observed in both datasets. The second equality follows from \eqref{eq:PLM}, since $X_c$ is discrete and thus $f(X_c)=X_c'\gamma_0$ for some $\gamma_0$. In what follows, we report the upper bound of the estimated identified set and confidence interval on $\theta_0$.

\medskip
Even though  the sample sizes as well as the number of common regressors $X_c$ are quite large, our method can still be implemented at a very reasonable computational cost. For instance, for the sample of sons over the first period (1850-1870), the computation of the confidence intervals only takes less than 4 minutes with our R package. As expected, computational time is highest for the period 1910-1930 associated with the largest number of observations, with $n > 100,000$ for both samples of $Y$ and $X_{nc}$. Nonetheless, our inference procedure remains tractable in this case too, with a computational time of about 11 minutes.\footnote{These CPU times are obtained using our companion R package, parallelized on 20 CPUs on an Intel Xeon Gold 6130 CPU 2.10GHz with 382Gb of RAM.} Overall, this illustrates the applicability of our method, which can be easily implemented even in this type of rich and high-dimensional data environment.

\medskip
Figures \ref{fig:appli_res10}-\ref{fig:appli_res_w10} and Table \ref{tab:OP_DGM} below display the results, for the father-son as well as father-son-in-law elasticities, obtained using our approach, the TSTSLS and, for the sample of sons over the years 1860-1880 and 1880-1900, the OLS.\footnote{In practice we need to restrict the set of first names included in $X_c$ to avoid very uncommon occurrences that are perfect predictors of the outcome variable $Y$. In our baseline specification, we implement this by restricting $X_c$ to the set of first names that account for at least 0.01\% of the observations in the pooled sample, and appear at least 10 times in either of the samples. We discuss in the following the robustness of our results to alternative cutoffs.} Specifically, we report in Figures \ref{fig:appli_res10}-\ref{fig:appli_res_w10} the estimated upper bounds of the identified sets (in solid red) and the confidence intervals (dashed red) obtained with our method, the TSTSLS estimates and confidence intervals (solid and dashed blue, resp.) as well as, for 1860-1880 and 1880-1900 and the sample of sons only, the OLS estimates and confidence intervals (solid and dashed green, resp.).

\medskip
A first conclusion from these results is that the upper bounds of the confidence intervals associated with our method range, depending on the periods, between 0.48 and 0.61 (0.51 and 0.6) for the sample of sons (sons-in-law). These values of the intergenerational 
coefficient are all well below the natural upper bound of 1. Also, even though the estimates vary depending on the data and econometric specification being used, most of the existing point estimates of the father-son income elasticity range between 0.40 and 0.50 \citep{olivetti2015name}. Overall, this clearly indicates that our method leads to informative inference on the parameter of interest.

\medskip
Second, consider the two cases where the linked data is available (1860-1880 and 1880-1900 for the sample of sons). Results in Table~\ref{tab:OP_DGM} indicate that the corresponding OLS estimates of the intergenerational income elasticities are quantitatively very close to the estimated upper bound of our identified set. Recall that, from Proposition~\ref{prop:point_ident2} in Section~\ref{ssub:tail_conditions}, the upper bound of our identified set ($\overline{\theta}_0$, say) plays a special role: under an additional restriction on the distributions of $X_{nc}$ and the error term, $\theta_0$ is actually point identified and equal to $\overline{\theta}_0$.\footnote{Proposition~\ref{prop:point_ident2} is obtained without  $X_c$. Yet, it can be combined with Proposition \ref{prop:common} to show that $\beta_0$, and in turn $\gamma_0$ (and thus $\theta_0$ here) are point identified with such $X_c$.} In other words, the results from these two periods support the hypothesis that the restriction on the distributions of $X_{nc}$ and the error term guaranteeing point identification of $\theta_0$ by $\overline{\theta}_0$ hold.

\begin{figure}[H]
	\begin{centering}
	\subfigure[For sons]{\includegraphics[scale=0.19]{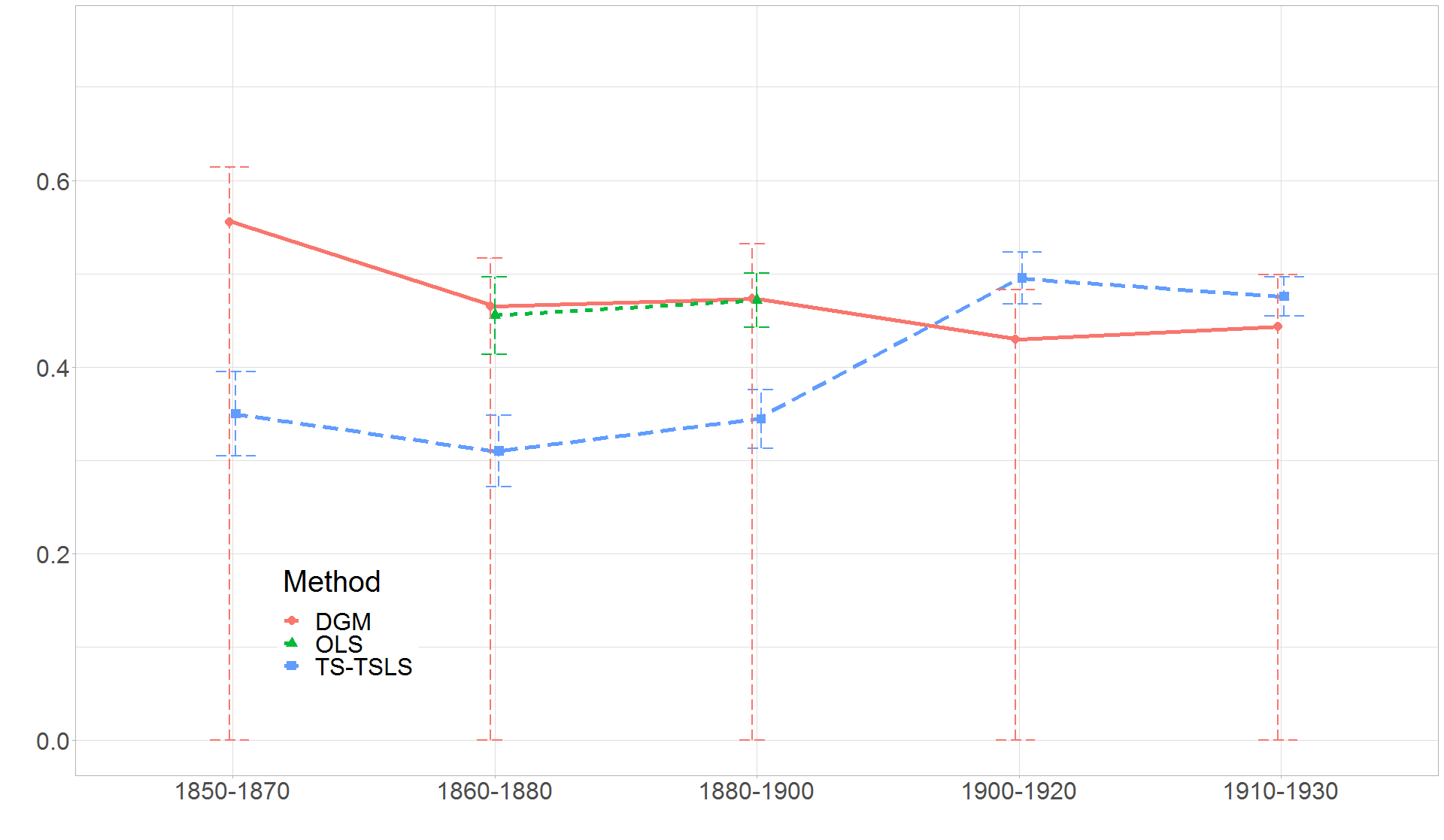}
		\label{fig:appli_res10}}
	\quad
	\subfigure[For sons-in-law]{\includegraphics[scale=0.4]{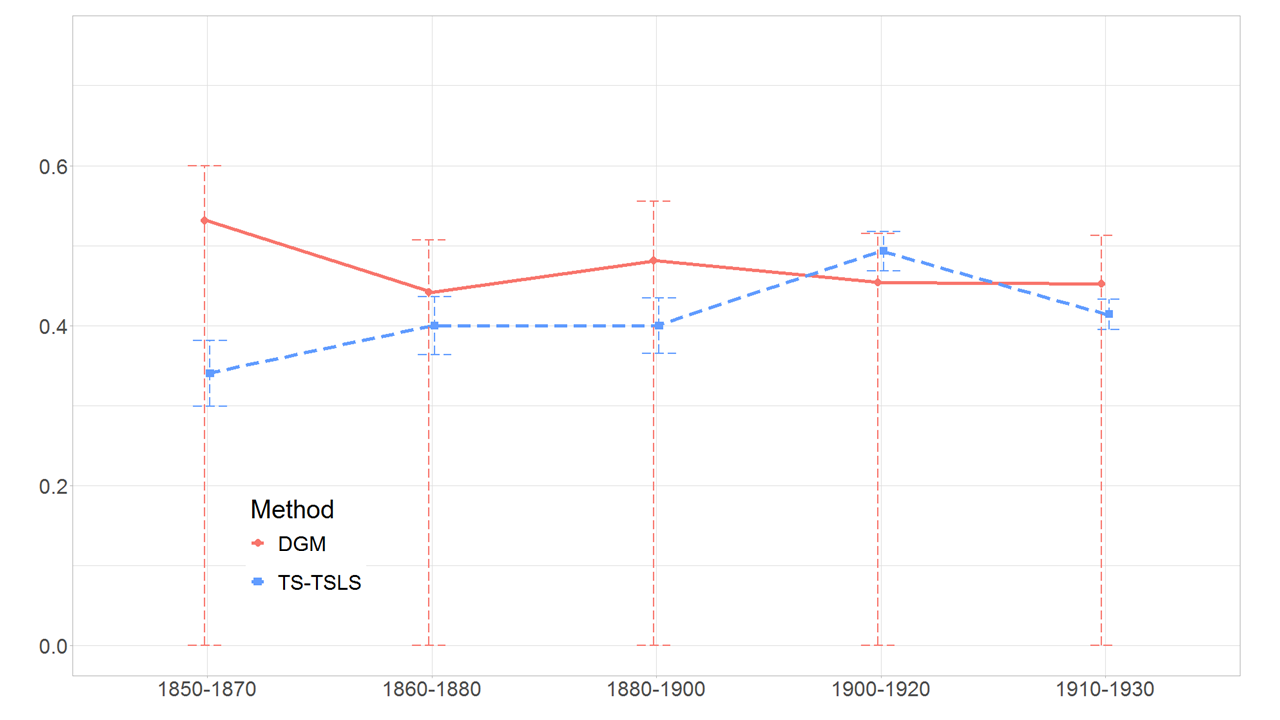}\label{fig:appli_res_w10}}
	\end{centering}
	{\footnotesize \\[2mm] Note: for readability and because 0 is a natural lower bound, the y-axis starts at 0, even though the lower bounds of our confidence intervals without restrictions are negative (see Table \ref{tab:OP_DGM}).}
	\caption{Intergenerational income correlation using different methods.}
	\label{fig:appli_10}
\end{figure}

\begin{table}[H]
	\centering
	\scalebox{0.58}{
		\begin{tabular}{rlllllllll}
			\toprule
			Sample:		& \textbf{1850-1870} &  & \textbf{1860-1880} &  & \textbf{1880-1900} &  & \textbf{1900-1920} &  & \textbf{1910-1930} \\ 	
			\textbf{Sons}				 &  &  &  &  &  &  &  &  &  \\
			\midrule
		DGM, set &[-0.555,0.555] &  & [-0.465,0.465] &  & [-0.473,0.473] &  & [-0.430,0.430] &  & [-0.443,0.443] \\
		DGM, CI. &  [-0.614,0.614] &  & [-0.517,0.517] &  & [-0.532,0.532] &  & [-0.483,0.483] &  & [-0.499,0.499] \\
			&  &  &  &  &  &  &  &  &  \\
		DGM, $\underline{R}^2 \geq 1.3 R_s^2$, set & [0.081,0.555] &  & [0.075,0.465] &  & [0.075,0.473] &  & [0.076,0.430] &  & [0.071,0.443] \\
		DGM, $\underline{R}^2 \geq 1.3 R_s^2$,  CI. & [0.033,0.617] &  & [0.034,0.519] &  & [0.039,0.527] &  & [0.044,0.477] &  & [0.047,0.491] \\
		&  &  &  &  &  &  &  &  &  \\
		DGM, $\underline{R}^2 \geq 2 R_s^2$,  set & [0.163,0.555] &  & [0.151,0.465] &  & [0.153,0.473] &  & [0.164,0.430] &  & [0.159,0.443] \\
		DGM, $\underline{R}^2 \geq 2 R_s^2$,  CI. & [0.095,0.601] &  & [0.093,0.506] &  & [0.102,0.513] &  & [0.127,0.459] &  & [0.132,0.477] \\
	&  &  &  &  &  &  &  &  &  \\
	TSTSLS, pt. & 0.350 &  & 0.310 &  & 0.344 &  & 0.495 &  & 0.476 \\
	TSTSLS, CI. &  [0.305,0.395] &  & [0.272,0.348] &  & [0.313,0.376] &  & [0.468,0.523] &  & [0.454,0.497] \\
    	&  &  &  &  &  &  &  &  &  \\
Test of equality, p-value	& $<$0.001 &  & 0.001   &  & 0.001   &  &0.999&  & 0.014  \\
 (Stat.; critical val. 95\%)        & (28.52; 15.16)  &  & (25.21; 12.37) &  & (25.92; 28.30) &  &   (15.09; 17.69)   &  &  (8.18; 6.79)\\
		&  &  &  &  &  &  &  &  &  \\
		OLS, pt. & &  & 0.455 &  & 0.472 &  &  &  & \\
		OLS, CI. &  &  & [0.414,0.497] &  & [0.443,0.501] &  & &  & \\
		&  &  &  &  &  &  &  &  &  \\
Test pt identification, p-value& &  & 0.147 &  &  0.003 &  &  &  & \\
(Stat.; critical val. 95\%)       &  &  & (9.21,17.03) &  & (23.06,6.33)  &  & &  & \\
			&  &  &  &  &  &  &  &  &  \\
		Number of names $X_c$& 225 &  & 261 &  & 382 &  & 514 &  & 598 \\
			Sample sizes $Y$ and $X_{nc}$ & (39,734; 34,603) & & (55,728; 47,014) & & (85,340; 73,999) &  &(116,986; 102,053)  & &  (131,089; 116,328)\\ 		
			\midrule
			Sample:	& \textbf{1850-1870} &  & \textbf{1860-1880} &  & \textbf{1880-1900} &  & \textbf{1900-1920} &  & \textbf{1910-1930} \\ 	
			\textbf{Sons-in-law}				 &  &  &  &  &  &  &  &  &  \\
			\midrule
	DGM, set & [-0.531,0.531] &  & [-0.442,0.442] &  & [-0.481,0.481] &  & [-0.454,0.454] &  & [-0.452,0.452] \\
	DGM, CI. &[-0.601,0.600] &  & [-0.507,0.507] &  & [-0.554,0.555] &  & [-0.515,0.515] &  & [-0.513,0.513] \\
			&  &  &  &  &  &  &  &  &  \\
	DGM, $\underline{R}^2 \geq 1.3 R_s^2$, set &  [0.089,0.531] &  & [0.085,0.442] &  & [0.075,0.481] &  & [0.073,0.454] &  & [0.062,0.452] \\
	DGM, $\underline{R}^2 \geq 1.3 R_s^2$, CI. &  [0.030,0.605] &  & [0.036,0.505] &  & [0.029,0.552] &  & [0.037,0.513] &  & [0.033,0.509] \\
	&  &  &  &  &  &  &  &  &  \\
	DGM, $\underline{R}^2 \geq 2 R_s^2$, set & [0.186,0.531] &  & [0.178,0.442] &  & [0.159,0.481] &  & [0.164,0.454] &  & [0.146,0.452] \\
	DGM, $\underline{R}^2 \geq 2 R_s^2$, CI.  & [0.115,0.596] &  & [0.114,0.490] &  & [0.105,0.534] &  & [0.122,0.499] &  & [0.113,0.496] \\
				&  &  &  &  &  &  &  &  &  \\
			TSTSLS, pt. &0.340 &  & 0.400 &  & 0.400 &  & 0.493 &  & 0.414 \\
			TSTSLS, CI. & [0.299,0.381] &  & [0.364,0.436] &  & [0.365,0.434] &  & [0.469,0.518] &  & [0.395,0.433] \\
			&  &  &  &  &  &  &  &  &  \\
			Test of equality, p-value	& $<$0.001 & & 0.998  & &  0.012 & & 1 & &   1 \\
			(Stat.; critical val. 95\%)        &(23.12; 9.67) & & (5.87; 18.53) & & (13.08; 12.94)&& (8.03; 13.28) & & (8.33; 13.07)\\
					&  &  &  &  &  &  &  &  &  \\
			Number of names $X_c$& 155 &  & 212 &  & 323 &  & 468 &  & 545 \\
			Sample sizes $Y$ and $X_{nc}$ & (25,760; 33,256) & & (32,970; 45,800) & & (49,068; 71,141) &  &(73,425; 99,871)  & &  (85,122; 112,763)\\ 	
			\bottomrule
			\multicolumn{10}{p{680pt}}{{\footnotesize Notes: Dependent variable $Y$ is son's (or son-in-law's) log income. Common regressors $X_c$ are dummies for the first names appearing more than 0.01\% in the pooled dataset and 10 times in both datasets. ``DGM, set'' and ``DGM, CI.'' refer to the estimated identified set and 95\% confidence interval, respectively, obtained with our method.  ``TSTSLS, pt.'' and ``TSTSLS, CI.'' refer to the TSTSLS point estimate and 95\% confidence interval, respectively. The test of equality between the TSTSLS ($\beta_{TSTSLS}$) estimates and DGM ($\beta_{DGM}$) upper bound estimates is performed using subsampling with 1,000 replications. The statistic  (``Stat.") is $n^{1/2}\widehat{\theta}$, where $\widehat{\theta}=\widehat{\beta}_{TSTSLS} - \widehat{\beta}_{DGM}$ and $n=n_yn_x/(n_y+n_x)$, $n_y$ and $n_x$ being the respective sample sizes of $Y$ and $X_{nc}$. The critical value corresponds to the  $1-\alpha$ quantile of the distribution of  $b_n^{1/2}|\widehat{\theta}^* - \widehat{\theta}|$, where $\widehat{\theta}^*$ is a subsampled version of  $\widehat{\theta}$ and $b_n$ is the subsample size. The sample sizes where the joint distribution is observed for both periods 1860-1880 and 1880-1900 are respectively 3,947 and 9,076. The $R^2$ on the short and long regressions are respectively 0.04 and  0.18  for 1860-1880, and 0.02 and 0.17 for 1880-1900. The test for point identification is performed with the $\eps$ selected in \eqref{eq:tradeoff}, however this choice appears conservative on simulations. Taking $\eps/2$ yields p-values of 0.69 for the period 1860-1880 and 0.04 for 1880-1900.}}
		\end{tabular}
	}
	\caption{Intergenerational income correlation for sons using different methods.}
	\label{tab:OP_DGM}
\end{table}

\medskip
Besides, the fact that we do not reject at standard levels the null hypothesis of point identification with our formal test described in Section \ref{app:test_point_id} for the period 1860-1880 (p-value of 0.147) provides suggestive evidence in this direction.\footnote{Simulation results available from the authors upon request indicate that our choice of $\eps$ tends to be conservative for the test of point identification. One would not reject either at the 1\% level the null hypothesis for the period 1880-1900 with a less conservative choice of $\eps$ (e.g. we obtain a p-value of 0.04 using $\eps/2$).} Under this assumption, our results are informative not only on the maximal father-son elasticity coefficient for a given period of time, but also on its evolution. It follows in particular that our estimates point to a mild decrease in this elasticity coefficient for sons between 1850 and 1930.

\medskip
Third, the results from the equality test reported in Table~\ref{tab:OP_DGM} indicate that the TSTSLS estimates are in several cases statistically distinguishable from the estimated upper bounds of our identified sets. This includes, for the sample of sons, all periods except 1900-1920, and the periods 1850-1870 and 1880-1900 for the sample of sons-in-law. Besides, for the sample of sons in particular, the TSTSLS estimates exhibit a sharp increase, while our estimated upper bound decreases between the periods 1880-1900 and 1900-1920. In that sense, our results offer suggestive evidence that the intergenerational income correlation might have been more stable at the beginning of the 20th century than what one would infer from the TSTSLS estimates.

\medskip
Fourth, we also report in Table~\ref{tab:OP_DGM} the estimated identified set and confidence intervals associated with our method when we impose a lower bound on the $R^2$ of the long regression, namely $\underline{R}^2 \geq 1.3 R_s^2$ or $\underline{R}^2 \geq 2 R_s^2$ . Imposing any of these restrictions, which are satisfied for the periods 1860-1880 and 1880-1900 for which the linked data is available, results in substantially tighter confidence intervals. In particular, for the sample of sons, the confidence intervals obtained under the restriction $\underline{R}^2 \geq 2 R_s^2$ allow us to reject values of the intergenerational income elasticity coefficient smaller than 0.13 and larger than 0.48 for the years 1910-1930.

\medskip
We consider in Tables~\ref{tab:robust_m} and \ref{tab:robust_w}, and Figure \ref{fig:Seps1} in online Appendix~\ref{app:appli} several robustness checks. They relate to the set of first names that we include as controls in our estimation procedure (Panel A), the choice of $\eps$ (Panel B and Figure \ref{fig:Seps1}), and restrictions of the sample to the set of individuals whose first name is included in the set of controls $X_c$ (Panel C). Throughout the tables, we focus on the upper bound of the estimated identified set (``DGM, set'') and of the confidence interval (``DGM, CI.'').

\medskip
The main takeaway from Table~\ref{tab:robust_m} and Figure \ref{fig:Seps1} is that, for the sample of sons, the results from our inference procedure are qualitatively, and in most cases quantitatively, robust to these different sensitivity analyses. The one case that exhibits more sensitivity is the specification where we control for the first names that account for at least 0.02\% of the sample, instead of 0.01\% in our baseline specification. The upper bound of our confidence interval for the period 1900-1920 increases in this case from 0.48 to 0.58, the results remaining, however, stable for the other periods. The results for the sample of sons-in-law (Table~\ref{tab:robust_w}) are also, for most periods at the exception of the same limit for 1900-1920, qualitatively, and in some cases quantitatively similar across specifications. The main difference with the sample of sons is that the choice of $\eps$ does appear to matter more for the sons-in-law, a limitation that one should keep in mind when interpreting the findings for this subgroup. Nonetheless, to the extent that our baseline choice of $\eps$ (see Section~\ref{subsec:computation}) is motivated by the theory and is found to perform well in our Monte Carlo simulation exercises, we do not view this as particularly worrisome.

\FloatBarrier

\section{Conclusion}\label{sec:concl}

We study the identification of and inference on partially linear models, in an environment where the outcome of interest and some of the covariates are observed in two different datasets that can not be matched. This setup arises in particular when one is interested in the effect of a variable that is not observed jointly with the outcome variable, or in cases where potential confounders are observed in a different dataset from the one including the outcome and regressor of interest. In such situations, researchers often rely on strong assumptions to point identify their parameters of interest. Our approach offers a useful alternative when such assumptions are debatable. The application shows that in addition to its tractability, our method is able to deliver informative bounds. Finally, beyond the model considered in this paper, our analysis suggests that the radial function is an appealing tool in partial identification problems where the support function proves difficult
to compute.

\newpage
\linespread{1}\selectfont
\bibliographystyle{chicago}
\bibliography{bib_1}
\newpage
\linespread{1.3}\selectfont
\appendix

\section{Additional theoretical results} 
\label{sec:additional_theoretical_results}

\subsection{Measurement errors} 
\label{app:ME}

We have assumed so far that the outcome and covariates are perfectly observed. However, measurement errors are pervasive in survey data. We now explore the robustness of the identified set proposed earlier to measurement errors on the outcome and covariates, which we denote by $Y^*$ and $X^*$. Specifically, consider a situation where both the covariates and the outcome are measured with error, such that:
\begin{equation}\left\{
	\begin{array}{rl}
		X=X^*+\xi_X, & \xi_X\indep X^*,\\[2mm]
		Y=Y^*+\xi_Y, & \xi_Y\indep (X^*, Y^*).		
	\end{array}\right.
	\label{eq:ME2}
\end{equation}\\
We introduce a new set, $\B^*$, which is defined as the original identified set $\B$ after replacing the observed measurement error-ridden covariates and outcome $(X,Y)$ by their latent counterparts $(X^*,Y^*)$.

\begin{prop}
	If Assumption \ref{hyp:mom} is satisfied with $(X,Y)$ replaced by $(X^*,Y^*)$, \eqref{eq:ME2} holds and for all $\beta\in \B^*$, $\xi_{Y_0}\cvx \xi_{X_0}'\beta$, then $\B^* \subset \B$.
	\label{prop:ME}
\end{prop}

The proof is in our supplementary material. This proposition establishes that the identified set is robust to measurement errors in the following sense: if (centered) measurement errors on the outcome $Y^*$ second-order stochastically dominate those on the linear index $X_0^*{}'\beta$ for all $\beta\in \B^*$, the identified set $\B$ based on the observed covariates $X$ and outcome $Y$ always contains the true value of the parameter of interest.\footnote{\label{foot:meas_QE} This result and underlying assumptions are closely related to the robustness to measurement errors on the beliefs of the test of rational expectations proposed in \cite{DGM21} (Subsection 2.2.4).} To better understand the above domination condition, suppose that $p=1$, $\xi_Y\sim\mathcal{N}(0,\sigma^2_Y)$ and $\xi_X\sim\mathcal{N}(0,\sigma^2_X)$. Then, recalling that any $\beta\in \B^*$ satisfies the variance restriction $\beta^2 V(X^*) \leq  V(Y^*)$, a sufficient condition for the dominance condition $\xi_{Y_0}\cvx \xi_{X_0} \beta$ is $\sigma^2_Y \ge [V(Y^*)/V(X^*)] \sigma^2_X$. In our application  for instance, $Y^*$ and $X^*$ are the log earnings of fathers and sons (or sons-in-law), respectively, so $V(Y^*)\simeq V(X^*)$ and $\sigma^2_Y\simeq \sigma^2_X$ seem credible. This suggests that the key domination condition from Proposition~\ref{prop:ME} is likely to hold in this context.


\subsection{Identification of a model with interaction terms} 
\label{app:interact}

Let $X_c = (X_{1,c},X_{-1,c})$ and $X_{nc}=(X_{1,nc},X_{-1,nc})$. We consider here the following model
$$E(Y|X)=f(X_c) + X_{nc}'\beta_0+X_{1,nc}X_{1,c}\delta_0.$$
First define, for $x\in\Supp(X_{1c})$ and $q\in\R^p$ ($q\ne 0_p$),
\begin{align*}
  \overline{S}_x(q, F_{Y,X_c},F_{X}) & = \inf_{x_{-1,c}\in\Supp(X_{-1,c}|X_{1,c}=x)} S(F_{Y|X_{-1,c}=x_{-1,c},X_{1,c}=x}, F_{X_{nc}'q|X_{-1,c}=x_{-1,c},X_{1,c}=x})\\
  \B_x & =\bigg\{\lambda q: \, q\in\mathcal{S}, 0\le \lambda \le \overline{S}_x(q,F_{Y,X_c},F_X)\bigg\}.
\end{align*}
Proposition \ref{prop:common} applied to the subpopulation $X_{1,c}=x$ implies that
for all $x\in \Supp(X_{1,c})$, $\beta_0  + x\delta_0 e_1 \in \B_x$, where $e_1=(1,0,\dots,0)'\in\R^p$. Because the converse also holds, the identified set $\B^{\delta \beta}$ of $(\delta_0, \beta_0)$ is
\begin{equation}
\B^{\delta \beta} = \{(\delta,\beta): \forall\,x\in \Supp(X_{1,c}),\; \beta  + x\delta e_1 \in \B_x\}.
	\label{eq:carac_beta_delta}
\end{equation}
The sets $\B_x$ are convex and include $0_p$. Hence, $\mathcal{B}^{\delta \beta}$ is convex too, and also includes $0_{p+1}$. Moreover, because $\B_x$ are compact, any $(\delta, \beta)\in\B^{\delta \beta}$ satisfies, for any $(x,x')\in\Supp(X_{1,c})^2$, $x\ne x'$,
\begin{equation}
|\delta| |x-x'| \le \|\beta  + x \delta e_1\| + \|\beta + x' \delta e_1\| \le M_x+M_{x'},	
	\label{eq:bound_delta}
\end{equation}
for some $M_x, M_{x'}>0$. Moreover,
$$\|\beta\| \le \|\beta  + x \delta e_1\| + |x||\delta| \le M_x + \frac{|x|}{|x-x'|}(M_x+M_{x'}),$$
which implies that $\B^{\delta \beta}$ is also compact. Thus, $\B^{\delta \beta}$ can also be described by its radial function, which we denote by $S(q,F_{Y,X_c},F_{X})$. Moreover, it follows from \eqref{eq:carac_beta_delta} that
$$ S(q,F_{Y,X_c},F_{X}) = \inf_{x\in\Supp(X_{1,c})} \overline{S}_x(q_{-1}+x q_1 e_1,F_{Y,X_c},F_X).$$


\subsection{Test for point-identification} 
\label{app:test_point_id}

We develop here a statistical test that can be used to check whether $\beta_0\in\partial\B$. Following the discussion in Subsection \ref{ssub:tail_conditions}, this boils down to testing for
\begin{equation}
H_0:\; S(F_{Y_{v0}}, F_{X'_{v0} \beta_v})=1 \quad \text{against} \quad H_1: S(F_{Y_{v0}}, F_{X'_{v0} \beta_v}) >1,	
	\label{eq:test_bord}
\end{equation}
where we recall that the joint distribution of the validation data $(X_v,Y_v)$ is observed and $\beta_v=V(X_v)^{-1}\cov(X_v,Y_v)$. We consider a statistical test based on i.i.d. data $(X_{vi},Y_{vi})_{i=1,...,n}$. The test statistic is
$$T = b_n^{1/2}\left(S_\eps(\widehat{F}_{Y_{v0}}, \widehat{F}_{X'_{v0} \widehat{\beta}_v})-1 \right),$$
where $\widehat{\beta}_v$ is the OLS estimator of $\beta_v$. The critical value is then $q_{1-\alpha}(T^*)$, the quantile of order $1-\alpha$  (defined conditional on the data) of
$$T^* = n^{1/2}\left(S_\eps(\widehat{F}^*_{Y_{v0}}, \widehat{F}^*_{X'_{v0} \widehat{\beta}^*_v})-S_\eps(\widehat{F}_{Y_{v0}}, \widehat{F}_{X'_{v0} \widehat{\beta}_v})\right),$$
where $\widehat{F}^*_{Y_{v0}}$, $\widehat{F}^*_{X_{v0}'q}$ and $\widehat{\beta}^*_v$ are the subsampling counterpart of $\widehat{F}_{Y_{v0}}$, $\widehat{F}_{X_{v0}'q}$ and $\widehat{\beta}_v$, respectively. We establish the asymptotic properties of the test under the following assumption.

\begin{hyp}
$E[\|X_v\|^{2+\delta}]<\infty$ for some $\delta>0$, $E[Y^2_v]<\infty$, $\beta_v\ne 0$ and $S(F_{Y_{v0}}, F_{X'_{v0} \beta_v}) = S_\eps(F_{Y_{v0}},$ $ F_{X'_{v0} \beta_v})$. Also, there exists $\mathcal{V}\subset \S$, compact and including a ball of positive radius centered at $\beta_v/\norm{\beta_v}$, and $\eps'\in (0,\eps)$ such that for all  $(\alpha,\alpha')\in[\eps',1-\eps']^2$, there exists $c>0$ and a strictly increasing and continuous function $m$ such that $m(0)=0$ and
\begin{align*}
	\inf_{q\in\mathcal{V}} \left|F_{X'q}^{-1}(\alpha')-F_{X'q}^{-1}(\alpha)\right|& > c |\alpha'-\alpha|,\\	
\sup_{q\in\mathcal{V}} \left|F_{X'q}^{-1}(\alpha')-F_{X'q}^{-1}(\alpha)\right|& <m(|\alpha'-\alpha|),\\	
 \left|F_Y^{-1}(\alpha')-F_Y^{-1}(\alpha)\right|& < m(|\alpha'-\alpha|).
\end{align*}	
	\label{hyp:test_bord}
\end{hyp}
\vspace{-1cm}

Up to the condition $S(F_{Y_{v0}}, F_{X'_{v0} \beta_v}) = S_\eps(F_{Y_{v0}}, F_{X'_{v0} \beta_v})$ on which we come back below, Assumption \ref{hyp:test_bord} is very close to the first part of Assumption \ref{hyp:for_CI}, but it is weaker as we require that it holds over $\mathcal{V}$ instead of $\mathcal{S}$.

\begin{prop}
	Suppose that $b_n\to\infty$, $b_n/n\to 0$ and Assumptions \ref{hyp:mom}-\ref{hyp:samples} and \ref{hyp:test_bord} hold. Then:
	\begin{enumerate}
		\item If $H_0$ in \eqref{eq:test_bord} holds, $\lim_{n\to\infty} P(T>q_{1-\alpha}(T^*))= \alpha$.
		\item If $H_1$ in \eqref{eq:test_bord} holds, $\lim_{n\to\infty} P(T>q_{1-\alpha}(T^*))=1$.
	\end{enumerate}
\label{prop:test_bord}
\end{prop}

The proof is in our supplementary material. Note that if $H_0$ holds but $S(F_{Y_{v0}}, F_{X'_{v0} \beta_v}) < S_\eps(F_{Y_{v0}}, F_{X'_{v0} \beta_v})$, $T$ will tend to infinity and $H_0$ will be rejected. Because we are testing here the validation of the tail condition described above, failing to reject $H_0$ under the alternative is more of an issue than wrongly rejecting $H_0$. Thus, potential over-rejection is arguably not as problematic as in other more standard contexts, such as testing the null of no effect of a treatment.



\section{Application to the black-white wage gap} 
\label{sec:appli_BWWG}

We apply our method to estimate the black-white wage gap among young males in the United States using the 1979 panel of the National Longitudinal Survey of Youth (NLSY79), revisiting the seminal work of \cite{neal1996role} on this question.
Considering the same restrictions as \cite{neal1996role} leads to a sample of size $n=1,675$.\footnote{We refer the reader to \cite{neal1996role} for a detailed discussion on the data.} We focus on the following model :
$$ Y = \gamma_{c,0} +  X_{c,1} \gamma_{c,1} +  X_{c,2} \gamma_{c,2}  + X_{nc} \beta_{nc}  + \epsilon, \quad E\left[ \epsilon | X_c ,X_{nc}\right] = 0, $$
where $Y$ is the mean log wage in 1990-1991, $X_{nc}$ denotes the AFQT and $X_{c,k}$, $k=1,2$ are dummy variables for being black or Hispanic. While $(Y,X_c,X_{nc})$ is jointly observed in the NLSY79 dataset, we proceed in the following as if AFQT, which is used in \cite{neal1996role} to control for pre-market factors, was not observed jointly with wages. This setup, which mimics the data environments in several other countries, allows us to directly compare the confidence intervals based on our partial identification approach with the ones obtained from the oracle OLS specification.

\medskip
Results in Table \ref{tab:NLSY79} below show the effect on our bounds when we impose different sets of constraints, namely i) a negative sign constraint on the coefficient $\gamma_{c,1}$ associated with the black indicator as well as a positive sign constraint on the coefficient $\beta_{nc}$ associated with the AFQT, ii) the latter constraints combined with the constraint $\underline{R}^2\geq 1.3 R^2_s$, and iii) the sign constraints i) combined with a less conservative bound $\underline{R}^2\geq 2 R^2_s$. Focusing on the main coefficient of interest $\gamma_{c,1}$, these results indicate that imposing these constraints on the $\underline{R}^2$ results in an identified set and confidence interval that are quite informative. Notably, the lower bound of the confidence interval is equal to $-.25$ and $-.2$ respectively in cases ii) and iii), against $-.17$ (i.e. a 17 log points wage penalty) for the OLS estimator. Taken together, these results show that our method is able to deliver confidence intervals that are very informative in practice.

\medskip

\begin{table}[ht]
	\centering
	\scalebox{0.70}{
		\begin{tabular}{rlllll}
			\hline
			& OLS & \multicolumn{4}{c}{DGM} \\
			\cline{3-6}
	\textbf{Constraints}		&  & Without & \multicolumn{3}{c}{With signs constraints}\\
		\cline{4-6}
		&  &   & Only & And  $\underline{R}^2 \geq 1.3 R^2_s$  & And  $\underline{R}^2 \geq 2 R^2_s$\\
			& (1) & (2) & (3)  & (4)  & (5)\\
	\textbf{Omitted variable} $X_{nc}$  & &   &  & &\\
    AFQT    &  0.150 & [-0.437,0.437] & [0,0.154] & [0.045,0.154] &[0.082,0.154]  \\
		CI	& [0.11,0.19] & [-0.522,0.522] & [0,0.215] & [0.004,0.211] & [0.010,0.207] \\
	\textbf{Common variables} $X_{c}$ &  &  &  & &  \\
			Black & -0.076 & [-0.664,0.318] & [-0.173,0] & [-0.123,0]& [-0.081,0] \\
			CI & [-0.171,0.02] & [-0.847,0.507] & [-0.304,0] & [-0.247,0] & [-0.199,0]  \\
			Hispanic & 0.016 & [-0.334,0.266] & [-0.034,0.071] & [-0.003,0.071]&   [0.022,0.071]\\
			CI & [-0.083,0.116] & [-0.506,0.450] & [-0.197,0.226] & [-0.186,0.225]& [-0.174,0.223] \\
			\bottomrule
		\multicolumn{6}{p{550pt}}{{\footnotesize Notes: $Y$ is average log wage in 1990 and 1991,  $X_{nc}$  is the AFQT, $X_c$ are dummies for being Black or Hispanic. The sample size is $n=1,675$, which is randomly split in two to artificially create a dataset where we observe $(Y,X_c)$ and another one with $(X_{nc},X_c)$. The first column presents the OLS estimates on the full dataset, where the 95\% CI have been multiplied by $\sqrt{2}$ to make it comparable with the DGM procedure using only half of it. The second column (2) presents the DGM estimates without constraints. Column (3) is the DGM estimates with a negative sign constraint on the coefficient of Black and a positive one on the coefficient of AFQT. Column (4) and (5) gather the DGM estimates with the latter constraints plus a lower bound constraint $R^2$ of the long regression: $\underline{R}^2 \geq r R^2_s$, with respectively $r=1.3$ and $r=2$. The $R^2$ on the short and long regressions are respectively 0.051 and 0.142.}}
		\end{tabular}
	}
	\caption{Bounds on the wage gap under different constraints for NLSY79}
	\label{tab:NLSY79}
\end{table}


\newpage
\pagenumbering{arabic}
\begin{center}
{\huge Online Appendix}
	
\end{center}

\section{Monte Carlo simulations}
\label{secMC}

In this section we study the finite sample performances of our inference method through Monte Carlo simulations. We first consider the baseline case where no common regressor is available, before evaluating the performance of our method in the presence of a common regressor. Finally, we discuss the computational time of our procedure compared to a many moment inequality-based alternative.

\subsection{Univariate case without common regressors}
\label{sec:simus1}

We first explore the finite sample performances of our inference method with $p=1$ and no $X_c$, considering the following DGP:
$$Y =  X_{nc}\beta_0+ U, \quad \beta_0=1, \; X_{nc}\indep U.$$
Then, we either assume that $X_{nc} \sim \mathcal{N}(0,1.5)$ and $U \sim \mathcal{N}(0,1)$, referred to in the following as the normal case, or $X_{nc} \sim \Gamma(1,2)$ and $U \sim \Gamma(0.4,2)$, which we refer to as the gamma case.

\medskip
We compare the finite sample performances of our inference method with those based on \citeauthor{andrews2017inference}, \citeyear{andrews2017inference}, henceforth AS. Specifically, recall from \eqref{eq:caract_B_mom} above that
$$\B = \left\{\beta\in\R^p: \, E\left[\max(0,Y_0-t)\right]\geq E\left[\max(0,X_{nc0}'\beta-t)\right] \; \forall t\in \R\right\}.$$
Hence, $\B$ is characterized by infinitely many moment inequalities. We then construct confidence regions for $\beta_0$ by inverting tests that these moment inequalities hold.\footnote{These tests involve several tuning parameters. Following the recommendation of AS (and using their notation), we fix $\epsilon=0.05$ and $\eta=10^{-6}$. To fix $b_0$ and $\kappa$, we follow the same procedure as in \cite{DGM21}, which yields $b_0=0.5$ and $\kappa=10^{-4}$. To construct a confidence region on $\beta_0$, we first fix a few directions $(q_1,...,q_n)$ in $\S$. Then, for $q=q_k$, we compute by a bisection method the maximal $\lambda\in\R^+$ such that the test of the moment inequalities at $\beta=\lambda q$ is not rejected.}

\medskip
In Table~\ref{tab:MCp1} below, we report the average bounds, across all 500 simulations, of the estimated identified sets and the 95\% confidence intervals associated with each of the five different sample sizes (Column “Bounds”) obtained with our method (“DGM”) and by applying \cite{andrews2017inference} (“AS”). In order to isolate sampling uncertainty, we report for each sample size and separately for our method and AS what we call the excess length (“Ex. length”), namely the mean difference between the length of the confidence sets and that of the identified set. We also report the coverage rates across simulations (“Coverage”). Finally, we report the average, across all simulations, of the estimates of the identified set $\mathcal{B}_{\eps(q)}$, where $\eps(q)$ is given by \eqref{eq:tradeoff} and thus varies from one simulation to another.

\begin{table}[H]
\scalebox{0.75}{
		\centering
	\begin{tabular}{rcccccccc}
		\toprule
		& \multicolumn{4}{c}{\textbf{DGM}} &  & \multicolumn{3}{c}{\textbf{AS }}\\
			\cmidrule{2-5} 	\cmidrule{7-9}
	Sample size	& Bounds & Ex. length & Coverage & 	$\widehat{\B}_{\eps(q)}$&   & Bounds &   Ex. length  & Coverage \\
	\midrule	
			\textbf{Normal} &  & &  &&  & & & \\
Identified set	 &    [-1.202,1.202]&  & & &  &  [-1.202,1.202]  &  & \\ 	
\cmidrule{2-2} \cmidrule{7-7}
400 & [-1.305,1.307] & 0.208 & 0.938 &  [-1.202,1.202] &   & [-1.374,1.367]& 0.337 & 0.983  \\
800 & [-1.280,1.280] & 0.156 & 0.942 &   [-1.202,1.202]&   &  [-1.329,1.328] &  0.253 & 0.985 \\
1,200 & [-1.266,1.267] & 0.129 & 0.940 &  [-1.202,1.202] &   & [-1.301,1.301] & 0.198 & 0.978  \\
2,400 & [-1.246,1.247] & 0.089 & 0.948 &  [-1.202,1.202]&    & [-1.268,1.270]&  0.134 & 0.975  \\
4,800 & [-1.234,1.235] & 0.065 & 0.936 &  [-1.202,1.202] &     & [-1.251,1.250]&  0.097 & 0.980  \\
\midrule
\textbf{Gamma} &  & &  &  & & & & \\
Identified set	 &[-0.025,1.046] & & &&   & [-0.025,1.046]  &  & \\
\cmidrule{2-2} \cmidrule{7-7}
400 & [-0.758,1.357] & 1.043 & 1 &  [-0.464,1.287] &   & [-0.538,1.343]& 0.809 & 1 \\
800 & [-0.603,1.302] & 0.834 & 0.996  & [-0.340,1.257] &  & [-0.466,1.313]  & 0.707 & 1 \\
1,200 & [-0.546,1.28] & 0.754 & 1 & [-0.293,1.247] &   & [-0.438,1.302] & 0.668 & 1  \\
2,400 & [-0.458,1.243] & 0.629 & 1 & [-0.237,1.220] &   &  [-0.391,1.277]  & 0.596 & 1  \\
4,800 & [-0.391,1.213] & 0.532 & 1  & [-0.199,1.197] &   &  [-0.362,1.258]  & 0.548 & 1  \\
		\bottomrule
	\end{tabular}
}
\caption*{\footnotesize{Notes: results obtained with 500 simulations.  Column ``Bounds" reports either the identified set or the average of the bounds of the 95\% confidence intervals over simulations. ``Ex. length" is the excess length, i.e. the average length of the confidence region minus the length of the identified set.  Column ``Coverage'' displays the minimum, over $\beta\in \B$, of the estimated probability that $\beta\in\CR(\beta_0)$.  Column ``$\widehat{\B}_{\eps(q)}$'' displays the average, across all simulations, of the estimates of the identified set $\mathcal{B}_{\eps(q)}$, where $\eps(q)$ is given by \eqref{eq:tradeoff}. We use 1,000 subsampling (resp. bootstrap) replications to compute the confidence intervals for the DGM (resp. AS) method.}}
\caption{Finite sample performances for $p=1$}
\label{tab:MCp1}
\end{table}

A couple of remarks are in order. First, as expected, the 95\% confidence intervals shrink with the sample sizes $n$. For both DGPs and all sample sizes, comparing the identified set with the confidence intervals indicates that identification uncertainty clearly dominates sampling uncertainty. This is especially striking for the normal case, which yields a substantially wider identified set, but also holds in the gamma case, where the regressor $X_{nc}$ has thicker tails. In particular, considering the excess length in the normal case, the confidence set is only between 8.6\% (for $n=400$) and 2.7\% (for $n=4,800$) wider than the identified set. In the gamma case, the confidence set ranges between 20.7\% and 14.8\% larger than the (regularized) identified set ($\mathcal{B}_{\varepsilon}$).

\medskip
Second, the coverage of our confidence intervals is good: coverage rates are always larger than 93.6\%. Third, our inference method generally performs similarly or better than AS, delivering consistently tighter confidence sets for sufficiently large sample size. For example, in the normal case, the excess length of the confidence set is reduced by around 30\% to 39\% depending on the sample sizes. In the gamma case, the two methods are close, AS doing slightly better only for sample sizes smaller than $n=4,800$. These results are consistent with our inference method exploiting the specific geometric structure of the identified set. This could also be due to the fact that we do not need to bear the cost, in terms of statistical power, of incorporating potentially many non-binding inequality constraints.

\medskip
Finally, the good finite sample performances of our inference method offers supporting evidence that our choice of the regularization parameter $\eps(q)$, given by \eqref{eq:tradeoff} and motivated in Section \ref{subsec:computation} above, is appropriate. In the normal case where $\B_\eps=\B$ for all $\eps$, $\eps(q)$ remains close to 0.5 for all sample sizes. In contrast, in the gamma case where the minimum of $R(\cdot, F_{Y_0}, F_{X_0'q})$ is reached at $\eps=0$ for both $q=1$ and $q=-1$, $\eps(q)$ tends to 0 as $n$ tends to infinity. Overall, the results suggest that the chosen $\eps(q)$ achieves a good balance between identification (a large $\eps$ leading to an increase in $\B_\eps$) and statistical uncertainty (a small $\eps$ leading to more volatility when estimating $S_\eps$ and thus larger quantiles $\widehat{c}_{\alpha,\eps}$).

\subsection{Multivariate case without common regressor}\label{sec:simus2}

We now consider the multivariate case ($p=2$) with the following DGP:
\begin{eqnarray}
Y &=& \gamma_{0} +  X_{nc}'\beta_0 + U, \;  U|X_{nc} \sim \mathcal{N}(0,4).
\end{eqnarray}
We set the coefficients as follows: $\gamma_{0} = -0.1$, $\beta_{0,1}=1$, and $\beta_{0,2}=1$. The variables $X_{nc}$ follow a multivariate normal distribution with mean 0 and covariance matrix
$$ \Sigma = \left(\begin{array}{cc}
  1& -0.2  \\
  -0.2 & 1
\end{array} \right).$$
 We report in Table~\ref{tab:MCp2} below the performances of our inference method, applied to the first component of $\beta_0$, for the same sample sizes as above, along with the identified set of the projection. These results were obtained using 500 simulations. We restrict to the first component of $\beta_0$ as the results are very similar for the second component. The main takeaway of this table is that our inference method exhibits similar finite-sample performances to the ones discussed in the univariate ($p=1$) normal case. In particular, the excess length of the confidence sets relative to the identified set tends to be quite small, and declines as $n$ gets larger.

\begin{table}[H]
	\centering	
	\scalebox{0.9}{
	\begin{tabular}{rccc}
		\toprule
Average		& Bounds & Excess length & Coverage  \\
	\cmidrule{2-4}
	Identified set & [-2.367,\;2.367]  & & \\
		\cmidrule{2-2}
		Sample size & & & \\
400 & [-2.599,\;2.599] & 0.465 & 0.94 \\
800 & [-2.555,\;2.554] & 0.376 & 0.962 \\
1,200 & [-2.523,\;2.522] & 0.312 & 0.96 \\
2,400 & [-2.496,\;2.497] & 0.26 & 0.982 \\
4,800 & [-2.475,\;2.474] & 0.217 & 0.986\\
		\bottomrule
	\end{tabular}
}
\caption*{\footnotesize{Notes: results obtained with 500 simulations.  Column ``Bounds" reports either the identified set or the average of the bounds of the 95\% confidence intervals over simulations. ``Excess length" is the average length of the confidence region minus the length of the identified set.  Column ``Coverage'' displays the minimum, over $\beta_1 \in \B_1$, of the estimated probability that $\beta_1\in\CI(\beta_{0,1})$. We use 200 subsampling replications to compute the confidence intervals.}}
\caption{Finite sample performances for $\beta_{0,1}$ with $p=2$}
\label{tab:MCp2}
\end{table}

\subsection{Case with a common regressor and possible constraints}\label{sec:simusXc}

We now examine the performances of our inference method in the presence of a common regressor. Namely, we consider the DGP:
$$Y = X_c \gamma_0 + X_{nc}\beta_0  + U,   \;  U|X \sim \mathcal{N}(0,4).$$
We set the coefficients as follows: $\gamma_0 = 0.3$ and $\beta_0 = 1$. The covariates are transformations of $(N_1,N_2)'$, which follows a multivariate normal distribution with mean 0 and covariance matrix $$ \Sigma = \left(\begin{array}{cc}
1 & 0.8   \\
0.8&  1.5 \\
\end{array} \right).$$
Specifically, the common regressor is given by $ X_c = \indic{ N_1 \leq 0.3 }$, and the regressors observed in one of the datasets only are such that $X_{nc}=N_2$.\medskip

We report in Table~\ref{tab:MCp1_Xnc} the performances of our inference method applied to the parameters $\beta_0$ and $\gamma_0$ along with the identified sets, with or without imposing the sign constraint $\gamma_0 \geq 0$. For $\beta_0$, coverage ranges between 95.4\% and 97\%. Similar to the baseline case without common regressors, the excess length of the confidence interval relative to the identified set declines as $n$ grows, and becomes quite small for the largest sample sizes. For instance, for $n=4,800$, our confidence interval is only 4\% larger than the identified set, highlighting again the limited role of sampling uncertainty in this context. The sign constraint reduces considerably the confidence interval, allowing to reject that $\beta_0=0$ for all the considered sample sizes. \medskip

Similar comments apply to the results on $\gamma_0$. The coverage rate is always over 96.4\%. The length of our confidence intervals is between 8\% and 30.5\% larger than the one of the unconstrained identified set. Note that the upper bounds of our confidence intervals on $\gamma_0$ are larger with the sign constraint than without it as in the former case we use critical values based on quantiles of order $\alpha/2$ and $1-\alpha/2$ to ensure that the confidence region $\CR^{con}(\beta_0)$ is asymptotically conservative.\medskip

Table \ref{tab:MCp1_Xnc0} illustrates the performances of our inference method on $\beta_0$, using the same DGP as above except for $\gamma_0$ which is set equal to zero, and compare them to the TSTSLS confidence intervals which are valid under this particular DGP. We implement our inference method without imposing the constraint that  $\gamma_0 = 0$. A couple of remarks are in order. First, the coverage with our method ranges between 94.8\% and 99.4\%, with the exception of one case ($n=400$ and the constraint $\gamma_0 \geq 0$, where the coverage is 92.2\%). Second, while the bounds obtained without imposing the sign constraint are substantially larger than the TSTSLS ones, which rely on the constraint $\gamma_0 =0$, the non-negativity constraint $\gamma_0 \geq 0$ does result in significantly tighter confidence intervals. In particular, the lower bounds on $\beta_0$ become close to the TSTSLS ones, and exclude 0.

\begin{table}[H]
	\centering
	\scalebox{0.75}{
		\begin{tabular}{rccccccc}
			\toprule
			& \multicolumn{3}{c}{Without sign constraint} &  & \multicolumn{3}{c}{With the constraint $\gamma_0 \geq 0$}  \\
			\cmidrule{2-4} 		\cmidrule{6-8} 	
			Average & Bounds &  Excess length &  Coverage &  & Bounds &  Excess length &  Coverage   \\
\midrule
		&	\multicolumn{6}{c}{Parameter $\beta_0$} \\
		\cmidrule{2-8} 	
			Identified set	  &  [-2.125,\;2.125]    & &  &  & [0.768,\;2.125]  & &    \\
			\cmidrule{2-2}	\cmidrule{6-6}
			Sample size	    & &  &  & & &  &  \\
		400 & [-2.445,\;2.445] & 0.640 & 0.966 &  & [0.376,\;2.495] & 0.761 & 0.944  \\
		800 & [-2.339,\;2.341] & 0.430 & 0.970 &  & [0.408,\;2.376] & 0.611 & 0.978  \\
		1,200 & [-2.297,\;2.300] & 0.347 & 0.962 &  & [0.460,\;2.329] & 0.512 & 0.994  \\
		2,400 & [-2.247,\;2.251] & 0.248 & 0.966 &  & [0.541,\;2.273] & 0.374 & 0.982 \\
		4,800 & [-2.206,\;2.213] & 0.170 & 0.954 &  & [0.603,\;2.229] & 0.268 & 0.976  \\
\midrule
		& 	\multicolumn{6}{c}{Parameter $\gamma_0$} \\	
			\cmidrule{2-8} 	
			Identified set	  & [-3.738,\;1.754]   & &  &  & [0,\;1.754]  & &    \\
			\cmidrule{2-2}	\cmidrule{6-6}
			Sample size	    & &  &  & & &  &  \\
	400 & [-4.578,\;2.590] & 1.676 & 0.98 &  & [0,\;2.729] & 0.975 & 0.996 \\
	800 & [-4.306,\;2.348] & 1.162 & 0.984 &  & [0,\;2.448] & 0.694 & 0.998 \\
	1,200 & [-4.197,\;2.214] & 0.919 & 0.990 & & [0,\;2.296] & 0.542 & 0.996 \\
	2,400 & [-4.062,\;2.076] & 0.646 & 0.976 &  & [0,\;2.135] & 0.381 & 0.988 \\
    4,800 & [-3.967,\;1.990] & 0.465 & 0.976 & & [0,\;2.032] & 0.278 & 0.992 \\
	\bottomrule
		\end{tabular}
	}
	\caption*{\footnotesize{Notes: results obtained with 500 simulations.  Column ``Bounds" reports either the identified set or the average of the bounds of the 95\% confidence intervals over simulations. ``Excess length" is the average length of the confidence region minus the length of the identified set.  Column ``Coverage'' displays the minimum of the estimated probability that $\gamma\in\CR(\gamma_0)$. We use 1,000 subsampling replications to compute the confidence intervals.}} 	\caption{Finite sample performances for $\beta_0$ and $\gamma_0$ with and without sign constraints}
	\label{tab:MCp1_Xnc}
\end{table}

\begin{table}[H]
	\centering
	\scalebox{0.7}{
		\begin{tabular}{rccccccccc}
			\toprule
			&	\multicolumn{3}{c}{Without sign constraint} &  & 	\multicolumn{3}{c}{With the constraint $\gamma_0 \geq 0$} & &  TSTSLS \\
			\cmidrule{2-4} 		\cmidrule{6-8} 	 	\cmidrule{10-10} 	
			Average & Bounds &  Excess length &  Coverage &  & Bounds &  Excess length &  Coverage& & Bounds   \\
			\cmidrule{2-4} 		\cmidrule{6-8} 	 	\cmidrule{10-10} 	
			Identified set	  &  [-2.125,\;2.125]   & &  &  & [1,\;2.125]  & &  &  & [1,\;1]  \\
			\cmidrule{2-2}	\cmidrule{6-6}
			Sample size	    & &  &  & & &  & &  &\\
			400 &   [-2.411,\;2.411] & 0.571 & 0.980 & & [0.642,\;2.457] & 0.689 & 0.922& & [0.634,\;1.426] \\
			800 & [-2.342,\;2.343] & 0.434 & 0.980  & & [0.67,\;2.377] & 0.582 & 0.964 & & [0.733,\;1.285]\\
			1,200 &  [-2.296,\;2.296] & 0.341 & 0.968 &  & [0.709,\;2.324] & 0.49 & 0.982 & &  [0.791,\;1.241] \\
			2,400 &  [-2.241,\;2.246] & 0.236 & 0.968 &  & [0.779,\;2.267] & 0.362 & 0.994& &  [0.844,\;1.159] \\
			4,800 & [-2.200,\;2.207] & 0.157 & 0.948 & & [0.836,\;2.223] & 0.261 & 0.970 &  &  [0.891,\;1.114] \\
			\bottomrule
		\end{tabular}
	}
	\caption*{\footnotesize{Notes: results obtained with 500 simulations.  Column ``Bounds" reports either the identified set or the average of the bounds of the 95\% confidence intervals over simulations. ``Excess length" is the average length of the confidence region minus the length of the identified set.  Column ``Coverage'' displays the minimum of the estimated probability that $\beta\in\CR(\beta_0)$. We use 1,000 subsampling replications to compute the confidence intervals.}} 	\caption{Finite sample performances for $\beta_0$ with one common regressor $\gamma_0=0$}
	\label{tab:MCp1_Xnc0}
\end{table}

\subsection{Computational time}
\label{subsubsec:CPU_time}

First, and following the discussion around Eq. \eqref{eq:direct_supp}, we compare our approach based on the radial function and the direct computation of the support function based on \eqref{eq:direct_supp}. We consider the DGP $Y = X_{nc}'\beta + \epsilon$, where $\beta=(1,\dots,1)'\in \R^p$, $X_{nc} \in \R^p$ with independent marginals $\mathcal{N}(0,2.25)$ and $\epsilon|X_{nc} \sim \mathcal{N}(0,1)$. We then compute $\sigma(\pm e_k,\widehat{F}_{Y_0},\widehat{F}_{X_0})$ for $k=1,...,p$ on 100 samples of size $n=2,000$. Our approach turns out to be 100 times faster when $p=1$, because it avoids the double optimization, and 11 (resp. 4) times faster when $p=2$ (resp. $p=3$).\footnote{All the computational times are obtained for a single simulation using our companion R package, on an Intel Xeon Gold 6130 CPU 2.10GHz with 382Gb of RAM and a single core. For the support function approach \eqref{eq:direct_supp}, we use the Constrained Optimization by Linear Approximations (COBYLA) algorithm for solving the linear optimization program under nonlinear constraints.}

\medskip
Next, we examine the computational time of our inference method and that of AS when $p$, the dimension of $X_{nc}$, is equal to either 1 or 2, for the DGPs considered in Sections~\ref{sec:simus1} and \ref{sec:simus2}, respectively, and for the five different sample sizes considered above. Table \ref{tab:time} below reports the computational time for $CR_{1-\alpha}(\beta_{0})$ when $p=1$, and for the two confidence intervals $CI_{1-\alpha}(\beta_{0,1})$ and $CI_{1-\alpha}(\beta_{0,2})$ when $p=2$.

\begin{table}[H]
	\centering
	\scalebox{0.9}{
	\begin{tabular}{rcccccc}
		\toprule
	Sample size	 & 400 & 800 & 1,200 & 2,400 & 4,800 \\
\cmidrule{2-6}
\cmidrule{1-1}
$p=1$ &  & &  &  &  \\
\cmidrule{1-1}
		AS (s)	 & 241.8  & 349.2 & 458.4 &  823.2 & 1137.0 \\
   DGM (s)	 & 	0.70 & 0.73 & 0.77 & 0.86 & 0.93 \\
	\cmidrule{1-1}
$p=2$  &  & &  &  &  \\
\cmidrule{1-1}
  	AS fast (min) 	 & 18.3 & 29.5 & 40.0 & 71.7 & 150.3 \\
  	AS recommended (min)  	 & 177.8& 296.5& 393.5& 702.8& 1500.2 \\
  DGM (s)	 &  18.9 & 19.8 &  20.8 &  23.9 & 30.6 \\
	\bottomrule
	\end{tabular}
}
\caption*{\footnotesize{Notes: The CPU time for the DGM method when $p=2$ corresponds to the computation of the 4 projections associated to $\pm e_k$, $k=1,2$.  For $p=2$, the ``AS fast'' approximation uses 25 directions to evaluate the computational time of the AS based method. The average over 50 replications of the excess length between the confidence intervals obtained with 250 directions and 25 directions over the length of the confidence intervals obtained with 250 directions (``AS recommended'') is 3.2\% for $n=1,200$. As in Sections \ref{sec:simus1}-\ref{sec:simus2}, we use 1,000 subsampling (resp. bootstrap) replications when $p=1$ and 200 replications when $p=2$ for the DGM (resp. AS) method.}}
	\caption{CPU time as function of sample size and dimension $p$ of $X_{nc}$.}
	\label{tab:time}
\end{table}

\medskip
In the univariate case ($p=1$), the computational gains of our method range from a factor of 342 to 1,217 compared to AS, for $n=400$ and $n=4,800$, respectively. While the computational time associated with our method increases with the sample size, it remains very modest (less than 1 second) for $n=4,800$.

\medskip
In the multivariate case ($p=2$), we compare our method with two alternative implementations of the AS method. ``AS fast'' corresponds to an approximation of the confidence intervals for both components of $\beta_0$ that uses 25 directions in $\mathcal{S}$ to implement the method, while ``AS recommended'' corresponds to the computational time associated with 250 directions. Since our method does not rely on any numerical approximation of this kind (as we exactly compute $1/\inf_{q\in\R^p:q_k=1}1/S_\eps(\widehat{F}_{Y_0}, \widehat{F}_{X_0'q})$), it is arguably more relevant to compare the computational times of our method and the ``AS recommended'' implementation. While the computational time of our method increases with $p$, it does remain tractable even with fairly large sample sizes, taking for instance $30.6$ seconds only to run for $n=4,800$. In the multivariate case also our method outperforms both implementations of the AS method. For instance, for $n=2,400$, our method runs 1,768 times faster than the recommended implementation of AS. In this case, computing $\varepsilon(q)$ for one direction with our method takes the same time as in the univariate case ($p=1$). The main difference and computational bottleneck with $p>1$ lies in the subsampling of the convex optimization in \eqref{eq:sig_hat}.

\medskip
To conclude, our approach can be implemented at a very limited computational cost, and achieves in our context considerable computational gains relative to the alternative method of AS.

\section{Additional results on the application}\label{app:appli}

\vspace{-0.4cm}
\begin{table}[H]
	\centering
	\scalebox{0.66}{
		\begin{tabular}{rlllllllll}
			\toprule
			Sample:		& \textbf{1850-1870} &  & \textbf{1860-1880} &  & \textbf{1880-1900} &  & \textbf{1900-1920} &  & \textbf{1910-1930} \\ 	
		\cmidrule{2-10}
		\multicolumn{1}{c}{\textbf{Baseline specification}} & & &   &  &  &  &  &  & \\
		DGM, set & 0.555 &  & 0.465 &  & 0.473 &  & 0.43 &  & 0.443 \\
		DGM, CI & 0.614 &  & 0.517 &  & 0.532 &  & 0.483 &  & 0.499 \\
		Number of names $X_c$ & 225 &  & 261 &  & 382 &  & 514 &  & 598 \\
		\midrule
	\multicolumn{7}{c}{ \hspace{-4.5cm} \textbf{Panel A: Robustness to the set of first names}}       &  &  & \\
		Threshold 0.005\% &  &  &  &  &  &  &  &  &  \\
		DGM, set  &0.555 &  & 0.465 &  & 0.473 &  & 0.43 &  & 0.443 \\
		DGM, CI  &0.611 &  & 0.521 &  & 0.529 &  & 0.484 &  & 0.499 \\
		Number of names $X_c$ & 225 &  & 261 &  & 382 &  & 515 &  & 626 \\
			\cmidrule{1-1}
		Threshold 0.02\% &  &  &  &  &  &  &  &  &  \\
		DGM, set  &0.555 &  & 0.465 &  & 0.493 &  & 0.511 &  & 0.477 \\
		DGM, CI &0.609 &  & 0.522 &  & 0.554 &  & 0.578 &  & 0.54 \\
			Number of names $X_c$  &225 &  & 261 &  & 332 &  & 378 &  & 415 \\
		\midrule
	\multicolumn{5}{c}{ \hspace{-2.5cm} \textbf{Panel B: Robustness to the choice of $\varepsilon$}}    &  &  &  &  & \\
	$\varepsilon/2$  &  &  &  &  &  &  &  &  &  \\
	DGM, set   & 0.555 &  & 0.442 &  & 0.473 &  & 0.419 &  & 0.415 \\
	DGM, CI & 0.612 &  & 0.49 &  & 0.534 &  & 0.471 &  & 0.467 \\
	Number of names $X_c$ & 224 &  & 259 &  & 380 &  & 512 &  & 596 \\
			\cmidrule{1-1}
		$2\varepsilon$ &  &  &  &  &  &  &  &  &  \\
		DGM, set  & 0.555 &  & 0.465 &  & 0.473 &  & 0.43 &  & 0.443 \\
		DGM, CI & 0.616 &  & 0.52 &  & 0.532 &  & 0.483 &  & 0.5 \\
	 	Number of names $X_c$ & 224 &  & 259 &  & 380 &  & 512 &  & 596 \\
	 		\midrule
	\multicolumn{7}{c}{\textbf{Panel C: Restricting the sample to the selected first names}}  &  &  & \\	 	
	 DGM, set  & 0.556 &  & 0.465 &  & 0.472 &  & 0.43 &  & 0.442 \\
	DGM, CI &  0.616 &  & 0.517 &  & 0.531 &  & 0.48 &  & 0.499 \\ [2mm]
Sample sizes $Y$ & 33,796 & & 46,296& & 73,961&  &99,874  & & 111,126\\ 		
Sample sizes $X_{nc}$ & 29,209 & &  40,431 & & 62,567&  & 85,202  & & 99,270\\ 		
			\bottomrule
			\multicolumn{10}{p{580pt}}{{\footnotesize Notes: $Y$=son's log income. The baseline specification restricts $X_c$ to be the dummies for the names appearing in the pooled dataset more than 0.01\%, and 10 times in both datasets. Panel A presents the results when we consider names appearing more than 0.005\% or 0.02\%  in the pooled dataset. In the baseline specification, the parameter $\varepsilon$ is chosen according to the data-driven rule \eqref{eq:tradeoff}. Panel B presents the results when using 0.5 or 2 times this choice of $\varepsilon$. Panel C presents results when we restrict the samples to the selected names based on our rule in the baseline case. We report the corresponding modified sample sizes.}}
		\end{tabular}
	}
	\caption{Robustness checks for the upper bound on intergenerational income correlation for sons.}
	\label{tab:robust_m}
\end{table}

\begin{table}[H]
	\centering
	\scalebox{0.66}{
		\begin{tabular}{rlllllllll}
			\toprule
			Sample:		& \textbf{1850-1870} &  & \textbf{1860-1880} &  & \textbf{1880-1900} &  & \textbf{1900-1920} &  & \textbf{1910-1930} \\ 	
	\cmidrule{2-10}
\multicolumn{1}{c}{\textbf{Baseline specification}}  & & &  &  &  &  &  &  & \\
DGM, set  & 0.531 &  & 0.442 &  & 0.481 &  & 0.454 &  & 0.452 \\
DGM, CI & 0.6 &  & 0.507 &  & 0.555 &  & 0.515 &  & 0.513 \\
	Number of names $X_c$  & 155 &  & 212 &  & 323 &  & 468 &  & 545 \\
	\midrule
\multicolumn{4}{c}{\hspace{-0.5cm}\textbf{Panel A: Robustness to the set of first names}}  &  &  &  &  &  & \\	
Threshold 0.02\%&  &  &  &  &  &  &  &  &  \\
DGM, set  & 0.531 &  & 0.442 &  & 0.48 &  & 0.573 &  & 0.452 \\
DGM, CI & 0.604 &  & 0.503 &  & 0.554 &  & 0.658 &  & 0.514 \\
	Number of names $X_c$   & 155 &  & 212 &  & 316 &  & 397 &  & 430 \\
\midrule
\multicolumn{5}{c}{\hspace{-2.5cm} \textbf{Panel B: Robustness to the choice of $\varepsilon$}}    &  &  &  &  & \\
$\varepsilon/2$ &  &  &  &  &  &  &  &  & \\
DGM, set  & 0.455 &  & 0.44 &  & 0.481 &  & 0.434 &  & 0.411 \\
DGM, CI & 0.51 &  & 0.505 &  & 0.553 &  & 0.495 &  & 0.466 \\
	Number of names $X_c$   & 155 &  & 212 &  & 323 &  & 468 &  & 545 \\
	\cmidrule{1-1}
$2\varepsilon$ &  &  &  &  &  &  &  &  & \\
DGM, set  &  0.531 &  & 0.442 &  & 0.481 &  & 0.454 &  & 0.452 \\
DGM, CI &   0.599 &  & 0.504 &  & 0.551 &  & 0.517 &  & 0.514 \\
	Number of names $X_c$   &155 &  & 212 &  & 323 &  & 468 &  & 545 \\
		\midrule
		\multicolumn{7}{c}{\textbf{Panel C: Restricting the sample to the selected first names}}  &  &  & \\	 	
		DGM, set  & 0.534 &  & 0.445 &  & 0.481 &  & 0.456 &  & 0.453 \\
		DGM, CI & 0.61 &  & 0.509 &  & 0.554 &  & 0.52 &  & 0.514 \\
		&  &  &  &  &  &  &  &  &  \\
		Sample sizes $Y$ & 20,375 & & 26,418 & & 41,212 &  & 61,742  & & 70,656\\ 		
		Sample sizes $X_{nc}$ & 27,096 & &  37,231 & & 57,474&  & 81,551  & & 94,706\\
		\bottomrule
			\multicolumn{10}{p{580pt}}{{\footnotesize Notes: $Y$=son-in-law's log income. The baseline specification restricts $X_c$ to be the dummies for the names appearing in the pooled dataset more than 0.01\%, and 10 times in both datasets. Panel A presents the results when we consider names appearing more than 0.02\%  in the pooled dataset. Results with considering names appearing more than 0.005\% in the pooled dataset are identical to the baseline, hence note reported. In the baseline specification, the parameter $\varepsilon$ is chosen according to the data-driven rule \eqref{eq:tradeoff}. Panel B presents the results when using 0.5 or 2 times this choice of $\varepsilon$. Panel C presents results when we restrict the samples to the selected names based on our rule in the baseline case. We report the corresponding modified sample sizes.}}
		\end{tabular}
	}
	\caption{Robustness checks for the upper bound on intergenerational income correlation for sons-in-law.}
		\label{tab:robust_w}
\end{table}

\begin{figure}[H]
	\begin{centering}
		\subfigure[For sons]{\includegraphics[width=15cm,height=6cm]{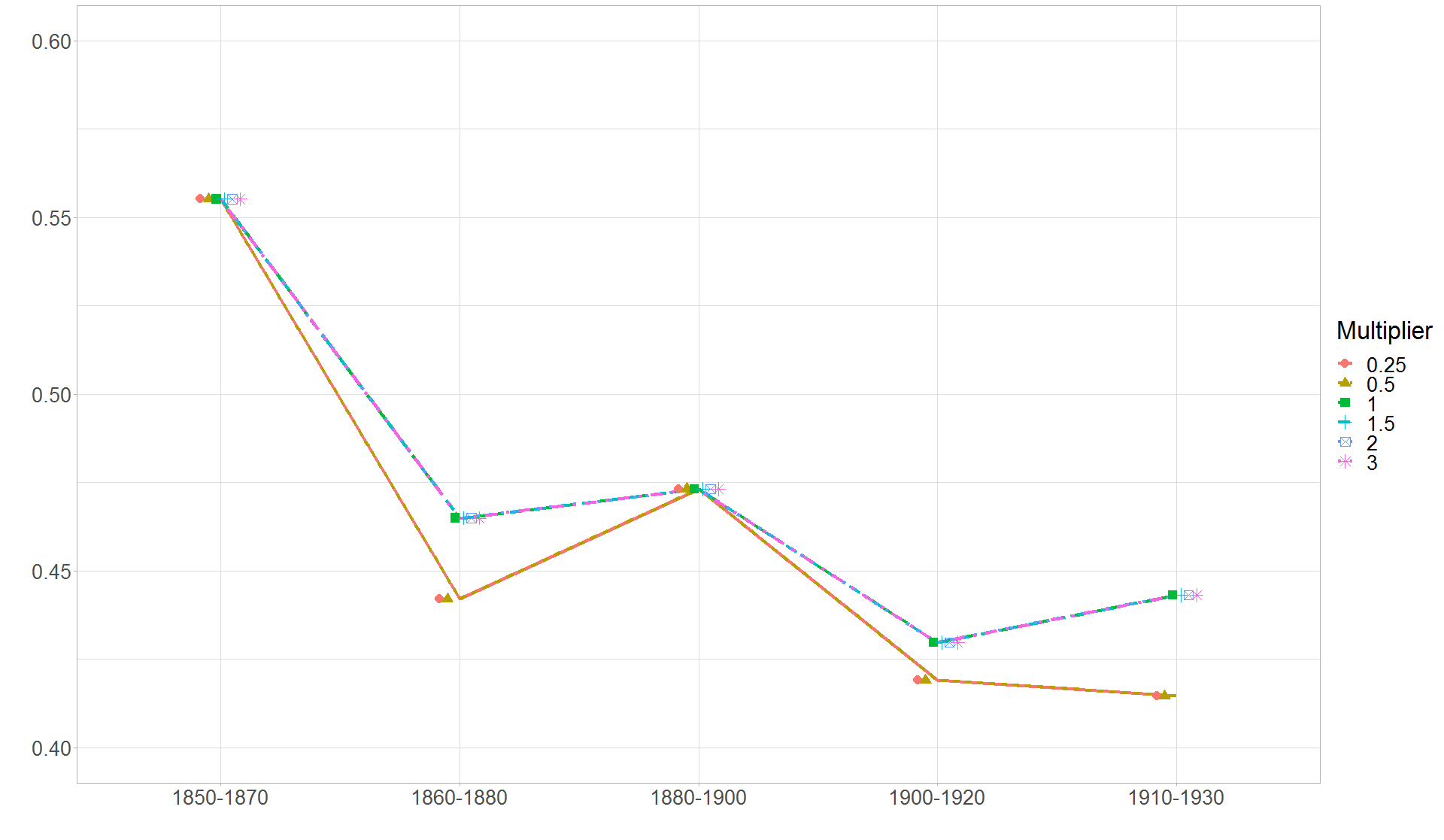}
			\label{fig:Seps}}
		\quad
		\subfigure[For sons-in-law]{\includegraphics[width=15cm,height=6cm]{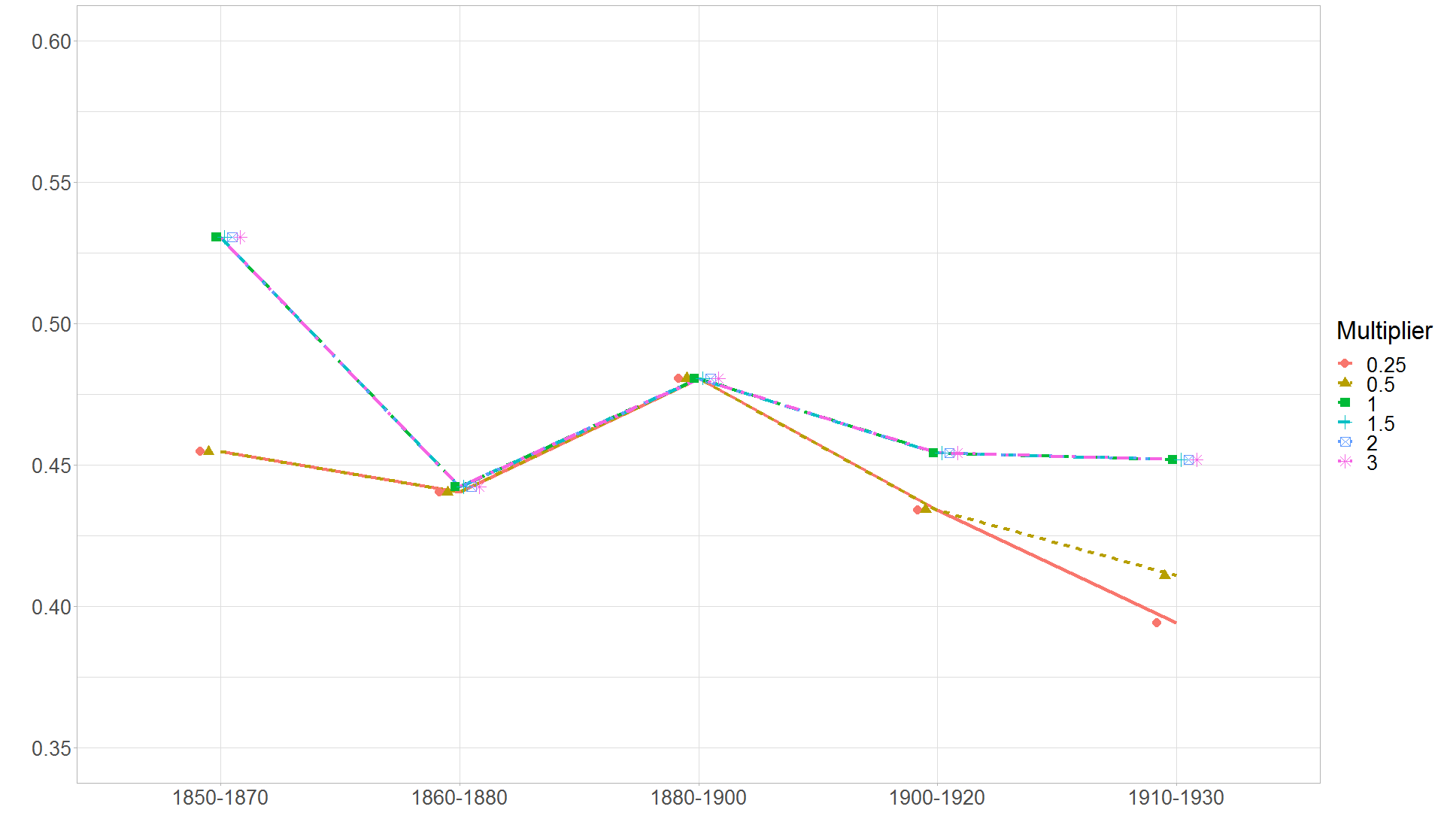}
			\label{fig:Seps0}}
	\end{centering}
	{\footnotesize \\ Note: the graphs display the value of $\overline{S}_{\eps'}(q,\widehat{F}_{Y,X_c}, \widehat{F}_{X_{nc},X_c})$ for $\eps'=c \eps$, where $\eps$ is selected via \eqref{eq:tradeoff} and $c\in\{0.25,0.5,1,1.5,2,3\}$.}
	\caption{$\overline{S}_\eps(q,\widehat{F}_{Y,X_c}, \widehat{F}_{X_{nc},X_c})$ for different $\eps$.}
	\label{fig:Seps1}
\end{figure}

\section{Proofs}\label{sec:proofs}

\subsection{Notation}

We denote by $\mathcal{P}_q(\R^p) $ the set of Borel probability measures on $\R^p$ with $q$ finite absolute moments. We assimilate herafter probability measures on $\R^p$ with their cdf, so we may write for instance $F\in \mathcal{P}_q(\R^p)$. We let $W_1$ denote the 1-Wasserstein distance and recall that for $(F,G)\in \mathcal{P}_1(\R)^2$,
\begin{equation}
W_1(F,G) = \inf_{U\sim F, V\sim G}  E\left[|U-V|\right]= \int_0^1 |F^{-1}(t)-G^{-1}(t)|dt = \int_{-\infty}^\infty |F(t)-G(t)|dt. 	
	\label{eq:Wasserstein}
\end{equation}
We denote by $\ell^{\infty}(\mathcal{X})$ the space of bounded functions on $\mathcal{X}$ for the uniform metric. Finally, $g(x) \lesssim h(x)$ means that $g(x) \le A h(x)$ for some universal constant $A>0$.

\subsection{Theorem \ref{thm:main}} 
\label{sub:theorem_ref_th_ellips}

Let $\B'$ denote the set on the right-hand side of \eqref{eq:caract_B}. We first show that $\B \subset \B'$. Then, we show the other inclusion. Finally, we show the other properties of $\B$.

\subsubsection*{1. $\B \subset \B'$} 

Let $F$ be such that $0<\int x^2dF(x)<\infty$ and $\int xdF(x)=0$ and define $g(\alpha)=\int_{\alpha}^1 F^{-1}(t)dt$, for any $\alpha \in [0,1]$. Since $F^{-1}$ is left-continuous, $g$ admits a left derivative equal to $-F^{-1}(\alpha)$. As it is decreasing, $g$ is concave. Moreover, $g(0)=g(1)=0$. For some $\alpha\in (0,1)$, $F^{-1}(\alpha) \ge \int xdF(x)=0$ so $g(\alpha) \ge (1-\alpha) F^{-1}(\alpha) \ge 0$. Assume that $g(\alpha)=0$. Then, by concavity, $g(x)=0$ for all $x\in[0,1]$. This implies that $F^{-1}(x)=0$ for all $x \in (0,1)$, which contradicts $\int x^2dF(x)>0$. Thus, for all $\alpha \in (0,1)$, $g(\alpha)>0$.

\medskip
Then, because $E(X_0'q)=0$ and $E[(X_0'q)^2]>0$ (as $E(X_0X_0')$ is nonsingular), $\int_{\alpha}^1 F^{-1}_{X_0'q}(t)dt>0$ for all $\alpha \in (0,1)$. This means that $0\leq \lambda \leq S(F_{Y_0},F_{X_0'q})$ is equivalent to
$$\int_\alpha^1 F^{-1}_{X_0'(\lambda q)}(t)dt \leq \int_{\alpha}^1 F_{Y_0}^{-1}(t)dt \quad \forall \alpha\in (0,1).$$
This, in turn, is equivalent to $F_{X_0'(\lambda q)}$ dominating $F_{Y_0}$ at the second order \citep[see, e.g.][]{de2006stochastic}. Then, by definition of second-order stochastic dominance,
$$\B'=\left\{\beta\in\R^p: E[\phi(Y_0)]\geq E[\phi(X_0'\beta)] \quad \forall \phi \; \text{ convex}\right\}.$$
Now, for any $\beta\in\B$, there exists $(\widetilde{X},\widetilde{Y})$ such that $E(\widetilde{Y}_0|\widetilde{X}_0)=\widetilde{X}_0'\beta$, $\widetilde{X} \eqd X$ and $\widetilde{Y} \eqd Y$. Then, for all convex function $\phi$, we have, by Jensen's inequality,
$$E[\phi(\widetilde{Y}_0)| \widetilde{X}_0] \geq \phi(E[\widetilde{Y}_0|\widetilde{X}_0]) = \phi(\widetilde{X}_0'\beta).$$
As a result, $\beta\in \B'$.

\subsubsection*{2. $\B'\subset \B$} 

For any $(F,G)\in \mathcal{P}_1(\R)\times \mathcal{P}_1(\R^{p+1})$, let $G_1$ denote the first marginal of $G$ and  define
\begin{align}
W_w(F, G_1) &:= \inf_{F_{U,V_1}: F_{U}=F, F_{V_1}=G_1} E\left[\left|V_1 - E[U | V_1]\right|\right], \notag \\	
W_c(F, G) & := \underset{F_{U,V_1,V_2}: \ F_{U} = F,\  F_{V_1,V_2} = G}{\inf} E\left[\left| V_1 - E(U|V_1,V_2) \right| \right].\label{eq:def_Wc}
\end{align}
We first  prove that $W_c(F, G)\leq W_w(F,G_1)$. To this end, let us define $c(x,H)=\left| x_1 - \int y dH(y)\right|$, for any $x=(x_1,x_2)\in \R\times \R^p$ and $H\in\mathcal{P}_1(\R)$. Because the function $c$ satisfies the assumptions of Theorem 1.3. in \cite{backhoff2019existence},\footnote{\label{foot:appli_BV} We use, with their notation, $t=1$ $Y=\R$ and $d_Y(y,y')=|y-y'|$, so that their $\mathcal{P}^t_{d_Y}$ is simply $\mathcal{P}_1(\R)$ here.} we have
$$W_c(F,G) =  \underset{f \in \Phi_{\text{bel}}}{\sup}\left\{ \int R_c(f)(x_1,x_2) dG(x_1,x_2)  - \int f(y)dF(y) \right\},$$
where we define
\begin{align*}
	\Phi_{\text{bel}}= & \left\{\psi:\R\to\R \text{ continuous s.t. }\exists (a,b,\ell,x_0)\in\R^4:\right. \\
	& \; \left. \forall x\in\R,\; \ell \le \psi(x)\le a + b|x-x_0| \right\},\\
	R_c(f)(x_1, x_2) = & \underset{H \in \mathcal{P}_1(\R) }{\inf} \int f(y)dH(y) + \left| x_1 - \int ydH(y)\right|.
\end{align*}
Let $U\sim F$ and $V=(V_1,V_2)\sim G$. By definition of  $R_c(f)$, we have, for almost all $x_1$,
$$R_c(f)(x_1,x_2) \leq E[f(U)| V_1=x_1] + \left|x_1 - E[U| V_1 = x_1]\right|.$$
As a result,
\begin{align*}
 \int R_c(f)(x_1,x_2) dG(x_1,x_2) - \int f(y)dF(y) \leq & E[f(U)]  + E\left[\left| V_1 - E(U|V_1)\right| \right]  - \int f(y)dF(y)  \\
 =  & E\left[\left| V_1 - E(U|V_1)\right| \right].
\end{align*}
Since this holds for all $(U,V_1)$ with $U\sim F$ and $V_1\sim G_1$,
$$\int R_c(f)(x_1,x_2) dG(x_1,x_2) - \int f(y)dF(y)\leq W_w(F,G_1).$$
Taking the supremum over $f\in \Phi_{\text{bel}}$, we obtain $W_c(F,G) \leq W_w(F,G_1)$.

\medskip
Now, let  $\beta\in\B'$. By Strassen's theorem (Theorem 8 in \citeauthor{strassen1965existence}, \citeyear{strassen1965existence}; see also Theorem 3.1 in \citeauthor{gozlan2015characterization}, \citeyear{gozlan2015characterization}), we have $W_w(F_{Y_0}, F_{X_0'\beta})=0$. As a result, $W_c(F_{Y_0}, F_{X_0'\beta,X_0})=0$. Because the function $c$ satisfies the assumptions of Theorem 1.2 in \cite{backhoff2019existence}, there exists a minimizer reaching the infimum in \eqref{eq:def_Wc}. This implies that there exist random variables $(\widetilde{Y},\widetilde{X}^\beta,\widetilde{X})$ with $F_{\widetilde{Y}}=F_{Y}$, $F_{\widetilde{X}^\beta,\widetilde{X}} = F_{X'\beta,X}$ and satisfying $\widetilde{X}^\beta_0 = E[\widetilde{Y}_0|\widetilde{X}_0^\beta, \widetilde{X}_0]$. The equality $F_{\widetilde{X}^\beta,\widetilde{X}} = F_{X'\beta,X}$ implies that $\widetilde{X}^\beta=\widetilde{X}'\beta$ almost surely. Then, $E[\widetilde{Y}_0|\widetilde{X}_0]=\widetilde{X}_0'\beta$ and in view of \eqref{eq:def_B}, $\beta\in\B$. The result follows.


\subsubsection*{3. Other properties of $\B$} 

Let $\widetilde{X}, \widetilde{Y}$ be independent variables such that $\widetilde{X}\eqd X$ and $\widetilde{Y}\eqd Y$. Then
$$E[\widetilde{Y}_0|\widetilde{X}_0]=E[\widetilde{Y}_0]=0=\widetilde{X}_0'0_p.$$
Hence, $0_p\in\B$. Now, let $(\beta_1,\beta_2)\in\B^2$ and $t\in[0,1]$. For any convex function $\phi$, we have
$$\phi(X_0'(t\beta_1+(1-t)\beta_2))\leq t\phi(X_0'\beta_1) + (1-t)\phi(X_0'\beta_2).$$
Hence, because $(\beta_1,\beta_2)\in\B'{}^2$,
$$E\left[\phi(X_0'(t\beta_1+(1-t)\beta_2))\right] \leq E\left[\phi(Y_0)\right],$$
which also implies that $t\beta_1+(1-t)\beta_2 \in \B'\subset \B$. Thus, $\B$ is convex. The inclusion $\B \subset \B^V$ follows from $\B \subset \B'$ and the convexity of $x\mapsto x^2$ which implies $\B' \subset \B^V$ .

\medskip
This last point also implies that $\B$ is bounded, as a subset of $\B^V$. Thus, to prove that $\B =\B'$ is compact, it suffices to show that it is closed. First, remark that in the definition of $\B'$, we can replace ``$\phi$ convex'' by ``$\phi$ continuous and convex'' (in fact, we can focus on the functions $x\mapsto \max(0,x-t)$ for $t\in\R$). Let $(\beta_n)_{n\in\N}$ be such that $\beta_n\in\B'$ and $\beta_n\to \beta$. By Fatou's lemma,
\begin{align*}
		E\left[\phi(X_0'\beta)\right] &= E\left[\liminf_n \phi(X_0'\beta_n)\right] \leq  \liminf_n E\left[\phi(X_0'\beta_n)\right] \leq E\left[\phi(Y_0))\right].	
\end{align*}
Thus, $\beta\in\B'=\B$, and $\B$ is closed.



\subsection{Corollary \ref{cor:sub}} 

By definition, $\B_k=\{b_k: \exists \beta \in \B: \, \beta_k=b_k\}$. Because $\B$ is convex and compact, $\B_k$ is a compact interval $[\underline{b}_k, \ \overline{b}_k]$, with $\underline{b}_k  = \inf_{\beta \in \B} e_k'\beta$ and $\overline{b}_k  = \sup_{\beta \in \B} e_k'\beta$. Thus, $\overline{b}_k= \sigma(e_k, F_{Y_0}, F_{X_0})$ and, similarly, $\underline{b}_k=-\sigma(-e_k, F_{Y_0}, F_{X_0})$.

\medskip
Next, remark that solutions $\beta$ of $ \sup_{\beta \in \B} e_k'\beta$ are at the boundary of $\B$ and are thus of the form $\beta=S(F_{Y_0}, F_{X_0'q}) q$ for some $q\in\S$ such that $q_k:=e'_k q> 0$. Thus,
\begin{align*}
 \sigma(e_k, F_{Y_0}, F_{X_0})  &= \sup_{q\in \S: q_k> 0} q_k  S(F_{Y_0}, F_{X_0'q})\\
 & =  \sup_{q\in \S: q_k> 0} S\left(F_{Y_0}, F_{X_0'q/q_k}\right) \\
 & =  \sup_{q\in \R^p: q_k> 0} S\left(F_{Y_0}, F_{X_0'q/q_k}\right) \\
 & =  \sup_{q\in \R^p: q_k=1} S\left(F_{Y_0}, F_{X_0'q}\right) \\
 & =  \frac{1}{\inf_{q\in \R^p: q_k=1} 1/S\left(F_{Y_0}, F_{X_0'q}\right)},
\end{align*}
where the second equality follows by definition of $S$. The same reasoning applies to $\sigma(-e_k, F_{Y_0}, F_{X_0})$.


\subsection{Proposition \ref{prop:point_ident}} 
\label{sub:proposition_ref_prop_ident}

\paragraph{Point 1.} 
\label{par:point_1}

Let $\psi(y)=\phi(y/2)$. By convexity of $\phi$, $\psi(Y_0)\le [\phi(Y)+\phi(-E(Y))]/2$. Thus, $E[\psi(Y_0)]<\infty$. Now, let $b\ne 0$. By convexity again, $\phi(X'b/4) \le \{\phi(X_0'b/2) +  \phi[E(X_0'b/2)]\}/2$. Since $E[\phi(X'b/4)]=\infty$, this implies $E[\psi(X_0'b)]=\infty$. Because $\B=\{\beta: Y_0\cvx X_0'\beta\}$, $b\not\in\B$. Thus $\B=\{0\}$. The result follows since $\beta_0\in\B$.


\paragraph{Point 2.}  
\label{par:point_2}

Let $\beta=(\beta_1,\beta_{-1})\in\B$. Since $\B=\{\beta: Y_0\cvx X_0'\beta\}$, we have, as above, $\infty >  E[\psi(Y_0)]\ge E[\psi(X_0'\beta)]$. Moreover, by convexity of $\psi$,
$$\psi(X_1 \beta_1/3) \le \frac{1}{3}\left\{\psi(X_0'\beta) + \psi(-X_{-1}' \beta_{-1}) + \psi[E(X'\beta)]\right\}.$$
Moreover, by assumption, $E[\psi(-X_{-1}' \beta_{-1})]<\infty$. Thus,
$$E[\phi(X_1 \beta_1/6)] = E[\psi(X_1 \beta_1/3)] < \infty,$$
which, by assumption, implies $\beta_1=0$. The result follows since $\beta_{0,1}\in\B_1$.



\subsection{Proposition \ref{prop:shape_f}} 
\label{sub:proposition_ref_prop_shape_f}

By Proposition \ref{prop:common} and linearity of $R$, we have
\begin{align*}
\Bcon=& \left\{\lambda q: q\in \S^+: -\overline{S}(F_{Y,X_c},F_{-X_{nc}'q,X_c}) \le \lambda \le \overline{S}(F_{Y,X_c},F_{X_{nc}'q,X_c}), \right. \\
& \left. \: \forall r\in\mathcal{R}: [Rm_Y - \underline{c}](r) \ge \lambda [Rm_{X_{nc}}'q](r) \right\}.	
\end{align*}
Remark that when $[Rm_{X_{nc}}'q](r)>0$, $[Rm_Y - \underline{c}](r) \ge \lambda [Rm_{X_{nc}}'q](r)$ is equivalent to $\lambda \le [Rm_Y - \underline{c}](r)/[Rm_{X_{nc}}'q](r)$. This implies that
$$\lambda \le \inf_{\substack{r\in \mathcal{R}:\\ [Rm'_{X_{nc}}q](r)> 0}} \frac{[Rm_Y - \underline{c}](r)}{[Rm'_{X_{nc}}q](r)}.$$
When $[Rm_{X_{nc}}'q](r)=0$, there are two cases: either $[Rm_Y - \underline{c}](r)\ge 0$, in which case we have no constraint on $\lambda$ (equivalently, $\lambda\le \infty$); or $[Rm_Y - \underline{c}](r)< 0$, in which case $\lambda q\not\in\Bcon$ for any $\lambda\in\R$ (equivalently, $\lambda\le -\infty$). This can be summarized by
$$\lambda \le \inf_{\substack{r\in \mathcal{R}:\\ [Rm'_{X_{nc}}q](r)\ge 0}} \lim_{u\downarrow 0} \frac{[Rm_Y - \underline{c}](r)+u}{[Rm'_{X_{nc}}q](r)+u^2}.$$
The reasoning is similar for the lower bound, yielding the final expression for $\Bcon$. The expression of $\mathcal{F}^{\con}$ follows as in Proposition \ref{prop:common}.

\medskip
Finally, $\Bcon$ is closed and convex, as the intersection of $\B^c$ and $\{\beta\in\R^p: \forall r\in\mathcal{R},\, [Rm_Y - \underline{c}](r)\ge [Rm_{X_{nc}}'\beta](r)\}$, which are both closed and convex. Since $\B^c$ is bounded, it is also bounded and thus compact. Finally, because $0_p\in\B^c$, $0_p\in \Bcon$ if and only if $[Rm_Y - \underline{c}](r)\ge 0$ for all $r\in\mathcal{R}$.


\subsection{Proposition \ref{prop:point_f}} 
\label{sub:proof_of_proposition_ref_prop_point_f}

First, $E(Y|X_c) = f(X_c) + m(X_c)'\beta_0$. Assume that $(\tilde{f},\tilde{\beta})$ also rationalizes the data and the model. Then
$$[f-\tilde{f}](X_c)= m(X_c)'[\tilde{\beta} - \beta_0].$$
Because $f-\tilde{f}\in\mathcal{G}$, we must have $\tilde{\beta} = \beta_0$ and in turn $\tilde{f}=f$.


\subsection{Proposition \ref{prop:point_ident2}} 
\label{sub:proposition_ref_prop_point_ident2}

\paragraph{Point 1.} 
Fix $c>0$. For any $M>0$, let
$$\phi_M(x) = \phi(x) \indic{|x|\le M} + \phi_+'(-M) (-M-x)^+ + \phi_-'(M) (x-M)^+,$$
where $\phi_+'$ (resp. $\phi_-'$) denotes the right (resp. left) derivative of $\phi$. Because $\phi_M(x)\le K_1 +K_2 |x|$ for some $K_1,K_2>0$, we have $E[\phi_M(X_0'\beta_0(1+c))]<\infty$. Also, $\phi_M(x)\uparrow \phi(x)$ as $M\uparrow\infty$. Then, by the monotone convergence theorem,
$$\lim_{M\to\infty} E[\phi_M(X_0'\beta_0(1+c))] = E[\phi(X_0'\beta_0(1+c))] = \infty.$$
On the other hand, $E[\phi_M((1/c+1) U)]\le E[\phi((1/c+1)U)]<\infty$. Thus, there exists $M_c$ such that
$$E[\phi_{M_c}((1/c+1) U)] < E[\phi_{M_c}(X_0'\beta_0(1+c))].$$
Moreover, using $Y_0=X_0'\beta_0+U$ and convexity of $\phi_{M_c}$, we obtain
$$\phi_{M_c}(Y_0) \le \frac{1}{1+c} \phi_{M_c}(X_0'\beta_0(1+c)) + \frac{c}{1+c} \phi_{M_c}((1/c+1)U).$$
Combining the  two inequalities, we obtain\footnote{Using $\phi$ instead of $\phi_{M_c}$ would not work: $E[\phi(Y_0)]\ge E[\phi(X_0'\beta_0)]$, so we may have $E[\phi(Y_0)]=\infty$.}
$$E[\phi_{M_c}(Y_0)] < E[\phi_{M_c}(X_0'\beta_0(1+c))].$$	
Because $\phi_{M_c}$ is convex, this implies that $\beta_0(1+c)\not\in\B$. Since $c>0$ was arbitrary, $\beta_0\in\partial\B$. The result follows.


\paragraph{Point 2.} 

By convexity, $\phi(X\lambda/2)\le [\phi(X_0\lambda)+\phi(E(X)\lambda)]/2$ for all $\lambda>0$. Therefore, for such $\lambda$, $E[\phi(X_0\lambda)]=\infty$. Since $X\in\R$ and $\beta_0>0$, this implies $E[\phi((X_0'\beta_0)\lambda)]=\infty$ for all $\lambda>1$. Thus, the condition of Point 1 holds and the identified set of $\beta_0$ is included in $\partial \B$, which is of the form $\{b,\beta_0\}$ for some $b\le 0$ (since $0\in\B$). Because it is known that $\beta_0>0$, the identified set is $\{\beta_0\}$.



\subsection{Proposition \ref{prop:epsilon}} 
\label{sub:proposition_epsilon}

\subsubsection*{1. $\B_\eps$ is compact and convex.} 
\label{ssub:_b__eps_is_compact_and_convex}

We showed in the proof of Theorem \ref{thm:main} that for all $\alpha\in(0,1)$, $ \int_{\alpha}^1 F_{X_0'q}^{-1}(t)dt>0$ and $\int_{\alpha}^1 F_{Y_0}^{-1}(t)dt>0$. Then, by continuity of $\alpha \mapsto \int_{\alpha}^1 F_{Y_0}^{-1}(t)dt/\int_{\alpha}^1F_{X_0'q}^{-1}(t)dt$,
$$S_\eps(F_{Y_0},F_{X_0'q}) = \min_{\alpha \in [\eps,1-\eps]} \frac{\int_{\alpha}^1 F_{Y_0}^{-1}(t)dt}{\int_{\alpha}^1F_{X_0'q}^{-1}(t)dt}>0.$$
Hence, $p_\eps(q):=1/S_\eps(F_{Y_0},F_{X_0'q})$ is well-defined  and
$$p_\eps(q) = \max_{\alpha \in [\eps,1-\eps]} \frac{\int_{\alpha}^1 F_{X_0'q}^{-1}(t)dt}{\int_{\alpha}^1 F_{Y_0}^{-1}(t)dt}.$$
Besides, for any random variables $U$ and $V$, and $\lambda \in [0,1]$,
\begin{align*}
	\int_\alpha^1 F_{\lambda U+ (1-\lambda) V}^{-1}(t)dt & \leq \int_\alpha^1 F_{\lambda U}^{-1}(t)dt + \int_\alpha^1 F_{(1-\lambda) V}^{-1}(t)dt \\
	& = \lambda \int_\alpha^1 F_{U}^{-1}(t)dt + (1-\lambda)  \int_\alpha^1 F_{V}^{-1}(t)dt,
\end{align*}
where the first inequality follows from Theorem 1.1 in \cite{embrechts2015seven}. As a result, for any $\alpha\in (0,1)$, the function $q \mapsto \int_\alpha^1 F_{X_0'q}^{-1}(t)dt$ is convex. Because the maximum of convex functions is also convex, the function $p_\eps$ is convex on $\R^p$. As such, it is also continuous. This implies that $\B_\eps=\{q\in\R^p: p_\eps(q)\leq 1\}$ is convex and closed. Finally, by continuity of $q\mapsto S_\eps(F_{Y_0},F_{X_0'q})$,
$$\sup_{q\in\S} S_\eps(F_{Y_0},F_{X_0'q}) = \max_{q\in\S} S_\eps(F_{Y_0},F_{X_0'q}) < \infty,$$
which implies that $\B_\eps$ is bounded, and thus compact.


\subsubsection*{2. For all $0<\eps < \eps'<1/2$, $\B \subset \B_\eps \subset \B_{\eps'}$ and $\cap_{\eps\in (0,1/2)} \B_\eps =\B$.} 

The first result follows since by definition, $S_\eps(F,G)\leq S_{\eps'}(F,G)$ for any $0<\eps<\eps'<1/2$. Now,
$$\cap_{\eps\in(0,1/2)} \B_\eps = \left\{\lambda q: \; q\in \S, \; 0 \leq \lambda \leq \inf_{\eps\in(0,1/2)} S_\eps(F_{Y_0},F_{X_0'q})\right\}.$$
Thus, to prove $\cap_{\eps\in(0,1/2)} \B_\eps= \B$, it suffices to show that $\inf_{\eps\in(0,1/2)} S_\eps(F, G)=S(F,G)$. First,  $\inf_{\eps\in(0,1/2)} S_\eps(F, G)\geq S(F,G)$ since
$S_\eps(F,G) \geq S(F,G) $ for all $\eps\in (0,1/2)$. Now, fix $\eta>0$. By definition, there exists $\alpha_0\in (0,1)$ such that
$$S(F,G) > R(\alpha_0,F,G)-\eta.$$
Hence, there exists $\eps\in (0,1/2)$ such that
$$S(F,G) > S_\eps(F,G) -\eta \geq \inf_{\eps\in(0,1/2)} S_\eps(F, G) -\eta.$$
Since $\eta$ is arbitrary, we have $S(F,G) \geq \inf_{\eps\in(0,1/2)} S_\eps(F, G)$. The result follows.


\subsubsection*{3. Under the stated conditions, there exists $0<\eps_0< 1/2$ such that $\B_{\eps_0}=\B$.} 
\label{ssub:point_2}

We first show that for all $q$, as $\alpha\to 1$, $R(\alpha, F_{Y_0}, F_{X_0'q}) \to \infty$. First,  $E(X_0'\beta_0)=0$ implies that $P(X_0'\beta_0\ge 0)>0$. Next, for all $\lambda$ and $M$, there exists $t_0$ such that for all $t\ge t_0$ and all $s$, $\overline{F}_{U|X_0'\beta_0}(t|s)> (M/P(X_0'\beta_0\ge 0)) \overline{F}_{\|X_0\|}(\lambda t)$. Then, for all $t\ge t_0$,
\begin{align}
\overline{F}_{Y_0}(t) = & E[\overline{F}_{U|X_0'\beta_0}(t-X_0'\beta_0|X_0'\beta_0) ] \notag \\
 	\ge & E[\overline{F}_{U|X_0'\beta_0}(t-X_0'\beta_0|X_0'\beta_0)\indic{X_0'\beta_0 \ge 0} ] \notag \\
	\ge & E[\overline{F}_{U|X_0'\beta_0}(t|X_0'\beta_0)\indic{X_0'\beta_0 \ge 0} ] \label{eq:supp_{Y_0}} \\
	\ge & M \overline{F}_{\|X_0\|}(\lambda t). \notag
\end{align}
In other words,
\begin{equation}
\forall \lambda>0,\; \lim_{t\to\infty} \frac{\overline{F}_{\|X_0\|}(\lambda t)}{\overline{F}_{Y_0}(t)}=0.	
	\label{eq:ratio_survies}
\end{equation}
If $\sup \Supp(X_0'q)<\infty$, \eqref{eq:supp_{Y_0}} together with $\Supp(U)=\R$ implies that $\sup \Supp(Y_0)=\infty$ and thus $F_{\|X_0\|}^{-1}(\alpha)= o(F_{Y_0}^{-1}(\alpha))$. Now, if $\sup \Supp(\|X_0\|)=\infty$, $F^{-1}_{\|X_0\|}(\alpha)\to \infty$ as $\alpha\to 1$. Thus,
$$\forall \lambda>0,\quad \lim_{\alpha\to 1} \frac{\overline{F}_{\|X_0\|}(F^{-1}_{\|X_0\|}(\alpha))}{\overline{F}_{Y_0}(\lambda F^{-1}_{\|X_0\|}(\alpha))}=0.$$
Now, remark that by continuity of $F_{Y_0}$, $\overline{F}_{\|X_0\|}(F^{-1}_{\|X_0\|}(\alpha)) \le 1-\alpha= \overline{F}_{Y_0}(F^{-1}_{Y_0}(\alpha))$. Therefore,
$$\forall \lambda>0,\quad \lim_{\alpha\to 1} \frac{\overline{F}_{Y_0}(F^{-1}_{Y_0}(\alpha))}{\overline{F}_{Y_0}(\lambda F^{-1}_{\|X_0\|}(\alpha))}=0.$$
Since $\overline{F}_{Y_0}$ is decreasing, this implies that there exists $\alpha_0(\lambda)$ such that, for all $\alpha \geq\alpha_0(\lambda)$,
$$F^{-1}_{Y_0}(\alpha) >  \lambda F^{-1}_{\|X_0\|}(\alpha).$$
Because $\lambda$ was arbitrary, this proves $F_{\|X_0\|}^{-1}(\alpha)= o(F_{Y_0}^{-1}(\alpha))$. Then, by integration, we obtain, as $\alpha\to 1$,
$$R(\alpha, F_{Y_0}, F_{\|X_0\|}) \to \infty.$$
The exact same reasoning shows that as $\alpha\to 0$, $R(\alpha, F_{Y_0}, F_{\|X_0\|}) \to \infty$. Now, let us define
$M:=\sup_{q\in \mathcal{S}} S_{1/4}(F_{Y_0}, F_{X_0'q})$. We proved in Point 1 above that $q\mapsto S_{1/4}(F_{Y_0}, F_{X_0'q})$ is continuous, implying that $M<\infty$. Then, by what precedes, there exists $\eps_0\in (0,1/4)$ such that
$$ \inf_{\alpha \in (0,\eps_0) \cup (1-\eps_0,1)} R(\alpha, F_{Y_0}, F_{\|X_0\|})>M.$$
Moreover, by the Cauchy-Schwarz inequality, $R(\alpha, F_{Y_0}, F_{X_0'q})\ge R(\alpha, F_{Y_0}, F_{\|X_0\|})$. As a result,
$$\inf_{q\in\mathcal{S}}\inf_{\alpha \in (0,\eps_0) \cup (1-\eps_0,1)} R(\alpha, F_{Y_0}, F_{X_0'q}) > M.$$
By definition, for all $q\in \mathcal{S}$, $S_{\eps_0}(F_{Y_0},F_{X_0'q})\le S_{1/4}(F_{Y_0}, F_{X_0'q}) \le M$. Then,
\begin{align*}
	S(F_{Y_0},F_{X_0'q}) & = \min\left(\inf_{\alpha \in (0,\eps_0) \cup (1-\eps_0,1)} R(\alpha, F_{Y_0}, F_{X_0'q}), S_{\eps_0}(F_{Y_0},F_{X_0'q})\right), \\
	& = S_{\eps_0}(F_{Y_0},F_{X_0'q}).
\end{align*}
This proves that $\B=\B_{\eps_0}$.



\subsection{Proposition \ref{prop:Hausdorff}} 
\label{sub:proposition_ref_prop_hausdorff}

In both cases, it suffices to prove the result for $\eps$ small enough.

\subsubsection*{Proof of Point 1} 
\label{par:proof_of_point_1}

The proof proceeds in two steps. First, we obtain an upper bound $S_\eps(F_{Y_0},F_{X_0'\beta_0}) -S(F_{Y_0},F_{X_0'\beta_0})$. Then, we obtain the bound on $d_H(\B,\B_\eps)$.

\medskip
\textbf{Step 1: upper bound on $S_\eps(F_{Y_0},F_{X_0'\beta_0}) -S(F_{Y_0},F_{X_0'\beta_0})$.}

\medskip
First, observe that $R(\cdot,F_{Y_0},F_{X_0'\beta_0})$ is differentiable and
\begin{align}
	& S_\eps(F_{Y_0},F_{X_0'\beta_0}) -S(F_{Y_0},F_{X_0'\beta_0}) \notag \\
	\le & \max\bigg(R(\eps,F_{Y_0},F_{X_0'\beta_0}) - \inf_{\alpha \in [0, \eps)} R(\alpha,F_{Y_0},F_{X_0'\beta_0}), \notag \\
	& \hspace{1.2cm} R(1-\eps,F_{Y_0},F_{X_0'\beta_0}) - \inf_{\alpha \in [1-\eps, 1)} R(\alpha,F_{Y_0},F_{X_0'\beta_0})\bigg) \notag \\
	\le & \int_{[0,\eps]\cup[1-\eps, 1]} \left|\Deriv{R}{\alpha}(\alpha,F_{Y_0},F_{X_0'\beta_0})\right|d\alpha. \label{eq:ss}
\end{align}
Now, $F^{-1}_{X_0'\beta_0}(\alpha_0)>0$ for some $\alpha_0<1$. Then, for $\alpha\ge\alpha_0$,
\begin{align}
\left|\Deriv{R}{\alpha}(\alpha,F_{Y_0},F_{X_0'\beta_0})\right| = & \frac{1}{\int_\alpha^1 F^{-1}_{X_0'\beta_0}(t)dt} \left|-F^{-1}_{Y_0}(\alpha) + R(\alpha, F_{Y_0}, F_{X'_0\beta_0}) F^{-1}_{X_0'\beta_0}(\alpha)\right| \notag \\
= & \frac{|F^{-1}_{X_0'\beta_0}(\alpha)|}{\int_\alpha^1 F^{-1}_{X_0'\beta_0}(t)dt} \left|R(\alpha, F_{Y_0}, F_{X'_0\beta_0}) -\frac{F^{-1}_{Y_0}(\alpha) }{F^{-1}_{X_0'\beta_0}(\alpha)}\right| \notag \\
\le & \frac{1}{1-\alpha}\left|R(\alpha, F_{Y_0}, F_{X'_0\beta_0}) -\frac{F^{-1}_{Y_0}(\alpha) }{F^{-1}_{X_0'\beta_0}(\alpha)}\right|. \label{eq:ineq_deriv_R}
\end{align}
Let $w_\alpha(t)=F^{-1}_{X_0'\beta_0}(t)/\int_{\alpha}^1 F^{-1}_{X_0'\beta_0}(u)du$. For $t\ge\alpha_0$, $w_\alpha(t)>0$. Then, for $\alpha\ge\alpha_0$,
\begin{align}
\left|R(\alpha, F_{Y_0}, F_{X'_0\beta_0}) - \frac{F^{-1}_{Y_0}(\alpha)}{F^{-1}_{X_0'\beta_0}(\alpha)}\right| = & \left|\int_\alpha^1 w_\alpha(t)\frac{F^{-1}_{Y_0}(t) }{F^{-1}_{X_0'\beta_0}(t)} dt - \frac{F^{-1}_{Y_0}(\alpha) }{F^{-1}_{X_0'\beta_0}(\alpha)}\right| \label{eq:diff_R_integ} \\
= & \left|\int_\alpha^1 w_\alpha(t)\left(\frac{F^{-1}_{Y_0}(t)}{F^{-1}_{X_0'\beta_0}(t)} - 1\right) dt - \left(\frac{F^{-1}_{Y_0}(\alpha) }{F^{-1}_{X_0'\beta_0}(\alpha)}-1\right)\right| \notag \\
\le &  2 \sup_{t\in[\alpha, 1]}  \left|\frac{F^{-1}_{Y_0}(\alpha) }{F^{-1}_{X_0'\beta_0}(\alpha)}-1\right| \notag \\
\lesssim & (1-\alpha)^{\frac{1/c - 1/d}{1+1/d}}, \label{eq:controle_diff_R}
\end{align}
where the last inequality follows from Lemma \ref{lem:upper_bound_ratio_Q} in the supplementary material. Combining \eqref{eq:ineq_deriv_R} and \eqref{eq:controle_diff_R}, we obtain, for $\eps\le 1-\alpha_0$,
\begin{equation}\label{eq:int1}
\int_{1-\eps}^1 \left|\Deriv{R}{\alpha}(\alpha,F_{Y_0},F_{X_0'\beta_0})\right|d\alpha \lesssim  \eps^{\frac{1/c - 1/d}{1+1/d}}.
\end{equation}
Similarly, note that $\int_\alpha^1 F^{-1}_{X_0'\beta_0}(t)dt=-\int_0^\alpha F^{-1}_{X_0'\beta_0}(t)dt\ge -\alpha F^{-1}_{X_0'\beta_0}(\alpha)$ and $F^{-1}_{X_0'\beta_0}(\alpha_1)<0$ for some $\alpha_1$. Then, for $\alpha\le \alpha_1$, we obtain, instead of \eqref{eq:ineq_deriv_R},
\begin{equation}\label{eq:alpha1}
	\left|\Deriv{R}{\alpha}(\alpha,F_{Y_0},F_{X_0'\beta_0})\right| \le \frac{1}{\alpha}\left|R(\alpha, F_{Y_0}, F_{X'_0\beta_0}) -\frac{F^{-1}_{Y_0}(\alpha) }{F^{-1}_{X_0'\beta_0}(\alpha)}\right|.
\end{equation}
The same reasoning as to get \eqref{eq:controle_diff_R} but using $R(\alpha, F_{Y_0}, F_{X'_0\beta_0})=\int_0^\alpha F^{-1}_{Y_0}(t)dt/\int_0^\alpha F^{-1}_{X'_0\beta_0}(t)dt$, $w_\alpha(t)=F^{-1}_{X_0'\beta_0}(t)/\int_{0}^\alpha F^{-1}_{X_0'\beta_0}(u)du$ and, again, Lemma \ref{lem:upper_bound_ratio_Q} yields, for $\alpha\le \alpha_1$,
\begin{equation}\label{eq:alpha1_1}
	\left|R(\alpha, F_{Y_0}, F_{X'_0\beta_0}) - \frac{F^{-1}_{Y_0}(\alpha)}{F^{-1}_{X_0'\beta_0}(\alpha)}\right| \lesssim \alpha^{\frac{1/c - 1/d}{1+1/d}}.
\end{equation}
Thus, for $\eps \le \alpha_1$,
\begin{equation}\label{eq:int2}
\int_0^\eps \left|\Deriv{R}{\alpha}(\alpha,F_{Y_0},F_{X_0'\beta_0})\right|d\alpha \lesssim  \eps^{\frac{1/c - 1/d}{1+1/d}}.
\end{equation}
Then, \eqref{eq:ss}, \eqref{eq:int1} and \eqref{eq:int2} imply that for $\eps\le \min(1-\alpha_0,\alpha_1)$,
\begin{equation}
S_\eps(F_{Y_0},F_{X_0'\beta_0}) -S(F_{Y_0},F_{X_0'\beta_0}) \lesssim  \eps^{\frac{1/c - 1/d}{1+1/d}}.	
	\label{eq:controle_diff_S_Seps}
\end{equation}

\textbf{Step 2: upper bound on $d_H(\B,\B_\eps)$.}

\medskip
$X$ has an elliptical distribution with nonsingular variance matrix $\Sigma$. As a result, for all $q\in\S$, there exists $\sigma(q)$ such that $X_0'q\eqd \sigma(q) X_0'\beta_0$, with
$$\sigma(q)^2 = \frac{q'\Sigma q}{\beta_0'\Sigma \beta_0} \ge \frac{\underline{\lambda}_\Sigma}{\beta_0'\Sigma \beta_0},$$
where $\underline{\lambda}_\Sigma>0$ denotes the smallest eigenvalue of $\Sigma$ and the inequality can be reached. Then,
\begin{align*}
d_H(\B,\B_\eps) \le & \, \sup_{q\in \S} S_\eps(F_{Y_0}, F_{X_0'q}) - S(F_{Y_0}, F_{X_0'q}) \\
= & \, \sup_{q\in \S} S_\eps(F_Y,F_{\sigma(q) X_0'\beta_0}) - S(F_Y,F_{\sigma(q) X_0'\beta_0}) \\
= & \, \left[S_\eps(F_{Y_0},F_{X_0'\beta_0}) - S(F_{Y_0},F_{X_0'\beta_0})\right] \sup_{q\in \S} [1/\sigma(q)] \\
= & \, \left[\frac{\beta_0'\Sigma \beta_0}{\underline{\lambda}_\Sigma}\right]^{1/2}
  \left[S_\eps(F_{Y_0},F_{X_0'\beta_0}) - S(F_{Y_0},F_{X_0'\beta_0})\right]\\
\lesssim & \, \eps^{\frac{1/c - 1/d}{1+1/d}},
\end{align*}
where the first inequality uses the definition of the Hausdorff distance and $\B\subset \B_\eps$, the  first equality follows since $X_0'q\eqd \sigma(q) X_0'\beta_0$, the second equality uses the definition of $R$, $S$ and $S_\eps$ and the last inequality is due to \eqref{eq:controle_diff_S_Seps}.

\subsubsection*{Proof of Point 2} 
\label{par:proof_of_point_2}

First, our assumptions imply that, for $\alpha \geq \alpha_0$ (resp. $\alpha \leq \alpha_1$) and all $q\in\S$, $F^{-1}_{X_0'q}(\alpha) \gtrsim (1-\alpha)^{-1/c}$  (resp. $F^{-1}_{X_0'q}(\alpha) \gtrsim \alpha^{-1/c}$)  and $F^{-1}_{U}(\alpha) \lesssim (1-\alpha)^{-1/d}$ (resp. $F^{-1}_{U}(\alpha) \lesssim \alpha^{-1/d}$), hence using $\beta_0=0_p$,
$$\forall \alpha \geq \alpha_0, \quad \sup_{q\in\mathcal{S}}\frac{F^{-1}_{Y_0}(\alpha) }{F^{-1}_{X_0'q}(\alpha)}\ \lesssim (1-\alpha)^{1/c - 1/d}, \quad \forall \alpha \leq \alpha_1, \quad \sup_{q\in\mathcal{S}}\frac{F^{-1}_{Y_0}(\alpha) }{F^{-1}_{X_0'q}(\alpha)}\ \lesssim \alpha^{1/c - 1/d}.$$
Now, remark that \eqref{eq:ss}, \eqref{eq:ineq_deriv_R}, and \eqref{eq:alpha1} still hold with $X_0'\beta_0$ replaced by $X_0'q$, $q\in\S$.
Then, using \eqref{eq:diff_R_integ}, we obtain, instead of \eqref{eq:controle_diff_R} and \eqref{eq:alpha1_1}, for $\alpha \in (0,\alpha_0]\cup[\alpha_1,1)$,
\begin{align*}
	\sup_{q\in\mathcal{S}} \left|R(\alpha, F_{Y_0}, F_{X_0'q}) - \frac{F^{-1}_{Y_0}(\alpha) }{F^{-1}_{X_0'q}(\alpha)}\right| & \le 2 \sup_{q\in\mathcal{S}} \sup_{t\in[\alpha,1]} \frac{F^{-1}_{Y_0}(t) }{F^{-1}_{X_0'q}(t)}\\
	& \lesssim   (1-\alpha)^{1/c - 1/d} \indic{\alpha \geq \alpha_0} +  \alpha^{1/c - 1/d} \indic{\alpha \leq \alpha_1}.
\end{align*}
As a result, for $\eps\le \min(1-\alpha_0,\alpha_1)$,
$$d_H(\B,\B_\eps) \le  \sup_{q\in\mathcal{S}} S_\eps(F_{Y_0}, F_{X_0'q})  \lesssim  \eps^{1/c - 1/d},$$
where in the first inequality we used $S_\eps(F_{Y_0}, F_{X_0'q})-S(F_{Y_0}, F_{X_0'q})\le S_\eps(F_{Y_0}, F_{X_0'q})$.



\subsection{Theorem \ref{thm:consistency}} 
\label{ssub:theorem_ref_thm_consistency}

Recall that $\widehat{F}_{Y_0}(t)=\frac{1}{n_Y}\sum_{i=1}^{n_Y} \indic{Y_i - \overline{Y} \leq t}$ and $\widehat{F}_{X'_0q}(t)$ is defined similarly. The proof proceeds in two steps. We first prove that for all $q\in\S$, $S_\eps(\widehat{F}_{Y_0},\widehat{F}_{X'_0q}) \convP S_\eps(F_{Y_0},F_{X_0'q})$. Then, we show that $d_H(\widehat{\B}_\eps,\B_\eps)\convP 0$.

\subsubsection*{Step 1: $S_\eps(\widehat{F}_{Y_0},\widehat{F}_{X'_0q}) \convP
S_\eps(F_{Y_0},F_{X_0'q})$, for all $q\in\S$.} 

The idea is to apply the continuous mapping theorem, with the metric
$$d((F, G), (F', G')) =  W_1(F,F') + W_1(G,G'),$$
where we recall that $W_1$ is the 1-Wasserstein distance. To this end, we first show that $(\widehat{F}_{Y_0},\widehat{F}_{X'_0q})$ converges to $(F_{Y_0}, F_{X'_0q})$ for this metric. It suffices to prove that $W_1(\widehat{F}_{Y_0}, F_{Y_0})\convP 0$, the proof being similar for $X_0'q$. Remark that $\widehat{F}_{Y_0}(t)=\widehat{F}_Y(t+\overline{Y})$ and $F_{Y_0}(y)=F_Y(y+E(Y))$. Then,
\begin{align*}
W_1(\widehat{F}_{Y_0}, F_{Y_0}) = & \int_{-\infty}^{\infty} |\widehat{F}_Y(t+\overline{Y}) - F_Y(t+\overline{Y}) + F_Y(t+\overline{Y}) - F_Y(t+E(Y))|dt \\
\leq & W_1(\widehat{F}_Y, F_Y) + \int_{-\infty}^{\infty} |F_Y(t+\overline{Y}) - F_Y(t+E(Y))|dt \\
= & W_1(\widehat{F}_Y, F_Y) + |\overline{Y}-E(Y)|,
\end{align*}
where the first equality follows by \eqref{eq:Wasserstein} and the last equality by Fubini's theorem. Because $E[|Y|]<\infty$, we have, by the law of large numbers $|\overline{Y}-E(Y)|\convP 0$ and also \citep[see (1.3) in][]{del1999central} $W_1(\widehat{F}_Y, F_Y)\convP 0$.

\medskip
Thus, the first step follows if we prove that $S_\eps$ is continuous for the metric $d$. First, by Lemma \ref{lem:continuous_R}, $R$ is continuous with respect to the metric $d'$ on $[\eps,1-\eps]\times \mathcal{D}^2$, where $\mathcal{D}$ denote the set of cdfs with mean 0 and $d'$ is defined by
\begin{equation}
	\label{eq:def_metric}
d'((\alpha, F, G), (\alpha', F', G')) = |\alpha'-\alpha| + W_1(F,F') + W_1(G,G').	
\end{equation}
Now, because the product topology is induced by $d'$, $R$ is continuous on the product $[\eps,1-\eps]\times \mathcal{D}^2$. Since $[\eps,1-\eps]$ is compact, it follows from Berge maximum theorem \citep[see, e.g., Theorem 9.14 in][]{sundaram1996first} that $S_\eps$ is also continuous with respect to the metric $d$. The result follows.


\subsubsection*{Step 2: Convergence of the set $\widehat{\B}_\eps$.}

We showed in the proof of Proposition \ref{prop:epsilon} that $S_\eps(F_{Y_0},F_{X_0'q})>0$ for all $q\in\S$. Then, let $p_\eps(q)=1/S_\eps(F_{Y_0},F_{X_0'q})$ and $\widehat{p}_\eps(q)=1/S_\eps(\widehat{F}_{Y_0}, \widehat{F}_{X_0'q})$. By the continuous mapping theorem, for all $q\in\S$, $\widehat{p}_\eps(q)\convP p_\eps(q)$. Moreover,
\begin{equation}
\widehat{p}_\eps(q)=\max_{\alpha \in [\eps,1-\eps]} 1/R\left(\alpha, \widehat{F}_{Y_0},\widehat{F}_{X'_0q}\right).	
	\label{eq:for_convexity_phat}
\end{equation}
Note that for any $(F_Y,F_X)$ and $\alpha \in  [\eps,1-\eps]$, $q\mapsto 1/R(\alpha,F_{Y_0},F_{X_0'q})$ is convex (see the proof of Point 1 in Proposition \ref{prop:epsilon}). Then, \eqref{eq:for_convexity_phat} implies that $\widehat{p}_\eps$ is also convex. As a result, by the convexity lemma of \cite{pollard1991asymptotics},
\begin{equation}\label{eq:gauge}
\sup_{q\in \S} \left|\widehat{p}_\eps(q) - p_\eps(q)\right| \convP 0.
\end{equation}
By construction, $\widehat{p}_\eps$ (resp. $p_\eps$) is the gauge function of the set $\widehat{\B}_\eps$ (resp.  $\B_\eps$). The gauge function of a nonempty, compact and convex set $H$ containing the origin is defined as  the support function of its polar set \citep[see, e.g., Corollary 3.2.5 p.149 in][]{hiriart2012fundamentals}. Thus, using Theorem 3.3.6 p.155 in \cite{hiriart2012fundamentals} and denoting respectively by $\widehat{\B}_\eps^{\circ}$ and $\B_\eps^{\circ}$ the polar sets of
$\widehat{\B}_\eps$ and $\B_\eps$, we obtain
$$ d_H\left(\widehat{\B}_\eps^{\circ} , \B_\eps^{\circ}\right) = \sup_{q\in \S}  \left|\widehat{p}_\eps(q) - p_\eps(q)\right|.$$
Thus, by \eqref{eq:gauge}, $d_H\left(\widehat{\B}_\eps^{\circ} , \B_\eps^{\circ}\right)\convP 0$. The result follows because convergence of polar sets for the Hausdorff distance implies convergence of the sets themselves for the same distance, see Theorem 7.2 in \cite{wijsman1966convergence}.


\subsection{Theorem \ref{thm:inference}}

\subsubsection*{1. Asymptotic validity of the confidence region}\label{sec:app_as} 

Let us define $\iota(G)=\inf_{\alpha\in [\eps,1-\eps]} G(\alpha)$. By definition,
$$S_\eps(\widehat{F}_{Y_0}, \widehat{F}_{X_0'q})=\iota\left[R(\cdot, \widehat{F}_{Y_0},\widehat{F}_{X_0'q})\right].$$
Moreover, by Theorem 2.1 of \cite{carcamo2019directional}, $\iota$ is Hadamard directionally differentiable.
Then, by Lemma \ref{lem:weak_cv_Fn} in the supplementary material and the functional delta method for Hadamard directionally differentiable functions \citep[see, e.g., Proposition 2.1 in][]{carcamo2019directional}, we have
\begin{equation}\label{eq:conv_SS}
	n^{1/2}\left(S_\eps(\widehat{F}_{Y_0}, \widehat{F}_{X_0'q})- S_\eps(F_{Y_0},F_{X_0'q})\right) \convD \iota'_{R(\cdot, F_{Y_0},F_{X_0'q})}(\mathbb{F}),
\end{equation}
where, in view of Corollary 2.3 in \cite{carcamo2019directional}, $\iota_f'(h) = \inf\{h(x): x\in \argmin_{\alpha \in [\eps,1-\eps]} f(\alpha)\}$ for any continuous functions $f$ and $h$.

\medskip
Now, let us show that $\widehat{c}_{\alpha,\eps}\convP c_{\alpha,\eps}$. Denote by  $H$ the cdf of $\iota'_{R(\cdot, F_{Y_0},F_{X_0'q})}(\mathbb{F})$. Note that $-\iota'_{R(\cdot, F_{Y_0},F_{X_0'q})}$ is convex.  Then, by Theorem 11.1 in \cite{davydov1998local}, its cdf $H$ is continuous and strictly increasing in a neighborhood of every point of its support except perhaps at $\underline{r} := \inf\{r \in \R: H(r) >0\}$. By Problem 11.3 in \cite{davydov1998local}, we also have that $H(r) >0$ for any $r \in \R$. Thus,  $H$ is continuous and strictly increasing on $\R$. Since $-c_{\alpha,\eps}$ is the quantile of order $1-\alpha$ of $-\iota'_{R(\cdot, F_{Y_0},F_{X_0'q})}(\mathbb{F})$ and using \eqref{eq:conv_SS}, it follows from Theorem 2.2.1 in \cite{politis1999subsampling}
 that $\widehat{c}_{\alpha,\eps}\convP c_{\alpha,\eps}$.

\medskip
Finally, fix $\beta\in \B_\eps$, so that $\beta=\lambda q$ with $\lambda\in [0, S_\eps(F_{Y_0},F_{X_0'q})]$. By definition, $\beta\in \CR(\beta_0)$ if and only if
\begin{equation}
n^{1/2}\left(S_\eps(\widehat{F}_{Y_0}, \widehat{F}_{X_0'q}) - \lambda \right)- \widehat{c}_{\alpha,\eps} \ge 0.
\label{eq:CNS_CR}
\end{equation}
Suppose first that $\lambda < S_\eps(F_{Y_0},F_{X_0'q})$. Since $S_\eps(\widehat{F}_{Y_0}, \widehat{F}_{X_0'q})$ is consistent for $S_\eps(F_{Y_0},F_{X_0'q})$ and $\widehat{c}_{\alpha,\eps}=O_P(1)$, \eqref{eq:CNS_CR} holds with probability approaching one and $\liminf_{n\to\infty} P(\beta\in \CR(\beta_0))=1$. Now, suppose that $\lambda = S_\eps(F_{Y_0},F_{X_0'q})$. Then, by what precedes,
$$n^{1/2}\left(S_\eps(\widehat{F}_{Y_0}, \widehat{F}_{X_0'q}) - \lambda \right)- \widehat{c}_{\alpha,\eps} \convD \iota'_{R(\cdot, F_{Y_0} ,F_{X_0'q})}(\mathbb{F}) - c_{\alpha,\eps}.$$
Moreover, by continuity of the cdf of $\iota'_{R(\cdot, F_{Y_0} ,F_{X_0'q})}(\mathbb{F})$ at $c_{\alpha,\eps}$,
$$P(\iota'_{R(\cdot, F_{Y_0}, F_{X_0'q})}(\mathbb{F}) - c_{\alpha,\eps}\ge 0) = 1-\alpha.$$
Thus, $\liminf_{n\to\infty} P(\beta\in \CR(\beta_0))=1-\alpha$. Equation \eqref{eq:valid_CR} follows since $\beta\in \B_\eps\subset \B$.

\medskip
Now, suppose that Assumption \ref{hyp:eps_q} holds and let us prove that \eqref{eq:valid_CR} is still true if $\eps$ is replaced by $\eps(q)$ (if $p=1$) or $\underline{\eps}$ (if $p>1$). We can focus on $\beta\in \partial\B_\eps$, $\beta= S_\eps(F_{Y_0},F_{X_0'q})q$. If $S_\eps(F_{Y_0},F_{X_0'q})>S(F_{Y_0},F_{X_0'q})$ for all $\eps \in \mathcal{E}$, we get, for $\eps'=\eps(q)$ or $\eps'=\underline{\eps}$,
$$S_{\eps'}(\widehat{F}_{Y_0}, \widehat{F}_{X_0'q}) \ge \min_{\eps\in\mathcal{E}} S_\eps(\widehat{F}_{Y_0}, \widehat{F}_{X_0'q}) \convP \min_{\eps\in\mathcal{E}}S_\eps(F_{Y_0},F_{X_0'q})>S(F_{Y_0},F_{X_0'q}),$$
where the convergence holds by the convergence in probability of $S_\eps(\widehat{F}_{Y_0}, \widehat{F}_{X_0'q})$ for any $\eps\in\mathcal{E}$ and the continuous mapping theorem. Equation \eqref{eq:valid_CR} follows. Suppose instead that $a\mapsto R(a,F_{Y_0},F_{X_0'q})$ admits a unique minimizer $a_0$ on $(0,1)$. Up to replacing $a_0$ by $1-a_0$, we can suppose without loss of generality that $a_0\le 0.5$. Let $\mathcal{E}=\{\eps_1,...,\eps_J\}$, with $\eps_1<...<\eps_J<1/2$. Reasoning as above, we have
$$\sqrt{n}\begin{pmatrix}
            S_{\eps_1}(\widehat{F}_{Y_0}, \widehat{F}_{X_0'q}) - S_{\eps_1}(F_{Y_0},F_{X_0'q}) \\
            \vdots \\
            S_{\eps_J}(\widehat{F}_{Y_0}, \widehat{F}_{X_0'q}) - S_{\eps_J}(F_{Y_0},F_{X_0'q})
          \end{pmatrix} \convD \begin{pmatrix}
            \iota'_{\eps_1,R(\cdot, F_{Y_0},F_{X_0'q})}(\mathbb{F})\\
            \vdots \\
            \iota'_{\eps_J,R(\cdot, F_{Y_0},F_{X_0'q})}(\mathbb{F})
          \end{pmatrix},
$$
where, compared to \eqref{eq:conv_SS}, we let the dependence of $\iota'$ on $\eps$ explicit. If $\eps_1>a_0$, then for any $\eps\in\mathcal{E}$, $S_\eps(F_{Y_0},F_{X_0'q})>S(F_{Y_0},F_{X_0'q})$, and the reasoning above applies. Otherwise, let $\eps_{j_0}=\max\{\eps\in\mathcal{E}: \eps\le a_0\}$  (where we simply let $\eps_{J+1}=1$ if $j_0=J$). Then, with probability approaching one, $\eps(q) \in \{\eps_1,...,\eps_{j_0}\}$. Moreover, the expression of $\iota'_{\eps,R(\cdot, F_{Y_0},F_{X_0'q})}$ and that $a\mapsto R(a,F_{Y_0},F_{X_0'q})$ admits a unique minimizer $a_0$ imply
that $\iota'_{\eps_1,R(\cdot, F_{Y_0},F_{X_0'q})}(\mathbb{F})= \dots = \iota'_{\eps_{j_0},R(\cdot, F_{Y_0},F_{X_0'q})}(\mathbb{F})=\mathbb{F}(a_0)$. As a result,
$$\left(\widehat{c}_{\alpha,\eps_1},...,\widehat{c}_{\alpha,\eps_{j_0}}\right)\convP \left(c_\alpha,...,c_\alpha\right),$$
where $c_\alpha$ is the quantile of order $\alpha$ of $\mathbb{F}(a_0)$. Combining these results yield, for all $j\in\{2,...,j_0\}$,
$$S_{\eps_j}(\widehat{F}_{Y_0}, \widehat{F}_{X_0'q}) - \widehat{c}_{\alpha,\eps_j} n^{-1/2} =  S_{\eps_1}(\widehat{F}_{Y_0}, \widehat{F}_{X_0'q}) - \widehat{c}_{\alpha,\eps_1} n^{-1/2} +o_P(n^{-1/2}).$$
In turn, this implies that
$$S_{\eps(q)}(\widehat{F}_{Y_0}, \widehat{F}_{X_0'q}) - \widehat{c}_{\alpha,\eps(q)} n^{-1/2} =  S_{\eps_1}(\widehat{F}_{Y_0}, \widehat{F}_{X_0'q}) - \widehat{c}_{\alpha,\eps_1} n^{-1/2} +o_P(n^{-1/2}),$$
which ensures that using $\eps(q)$ leads to asymptotically correct coverage in this case. Finally, remark that by definition of $\eps(q)$ (and letting the dependence of the confidence region on $\eps$ explicit),
$$P\left(S(F_{Y_0},F_{X_0'q})q\in \CR^{\underline{\eps}}(\beta_0)\right) \ge P\left(S(F_{Y_0},F_{X_0'q})q\in \CR^{\eps(q)}(\beta_0)\right),$$
which ensures the validity of using $\underline{\eps}$ instead of a fixed $\eps$.


\subsubsection*{2. Asymptotic validity of the confidence interval} 

Let $\beta_k\in \B_k$. First assume that $\beta_k\leq 0$. Because $0\in\CI(\beta_{0,k})$, $\beta_k \not\in\CI(\beta_{0,k})$ only if
$$\beta_k <  -\sigma_\eps(-e_k, \widehat{F}_{Y_0}, \widehat{F}_{X_0}) + n^{-1/2}\widetilde{c}_{\alpha,\eps}(-e_k).$$
In turn, this event implies that $\underline{E}_n$ holds, with
\begin{equation}
\underline{E}_n := \left\{n^{1/2}\left(-\sigma_\eps(-e_k, \widehat{F}_{Y_0}, \widehat{F}_{X_0}) +\sigma(-e_k, F_{Y_0}, F_{X_0})\right) >  -\widetilde{c}_{\alpha,\eps}(-e_k)\right\}.	
	\label{eq:def_Einf}
\end{equation}
Hence, $\sup_{\beta_k\in \B_k \cap \R^-}P(\beta_k\not \in \CI(\beta_{0,k}))
\leq  P\left(\underline{E}_n\right)$. Reasoning similarly for $\beta_k\geq 0$, we obtain
$$\sup_{\beta_k\in \B_k}P(\beta_k\not \in \CI(\beta_{0,k}))
\leq \max\left[P\left(\underline{E}_n\right),P\left(\overline{E}_n\right)\right],$$	
where we let $\overline{E}_n:=\left\{n^{1/2}\left(\sigma_\eps(e_k,\widehat{F}_{Y_0}, \widehat{F}_{X_0})  -\sigma(e_k, F_{Y_0}, F_{X_0})\right) <  \widetilde{c}_{\alpha,\eps}(e_k)\right\}$. As the reasoning is similar for $\underline{E}_n$ and $\overline{E}_n$, it suffices to prove that $\limsup_{n\to\infty} P(\underline{E}_n) \le \alpha$, with equality if $\sigma(-e_k, F_{Y_0}, F_{X_0})=\sigma_\eps(-e_k, F_{Y_0}, F_{X_0})$. To this end, first remark that
$$\sigma_\eps(-e_k, F_{Y_0},F_{X_0}) = \sup_{q\in \S} \inf_{\alpha\in [\eps,1-\eps]} \left[R(\alpha, F_{Y_0},F_{X_0'q})q\right]'(-e_k).$$
Let us define $\kappa(f):=\sup_{q\in \S} \inf_{\alpha\in [\eps,1-\eps]}  f(q,\alpha)$ and $G(q,\alpha) := [R(\alpha, F_{Y_0},F_{X_0'q})q]'(-e_k)$. By Lemma B.1 in \cite{firpo2021uniform}, $\kappa$ is Hadamard directionally differentiable. Moreover, by Lemma \ref{lem:weak_cv_Fn} in the supplementary material, the process $(q,\alpha) \mapsto [\mathbb{F}_n(q,\alpha)q]'$ $(-e_k)$ converges weakly to a Gaussian process ($\widetilde{\mathbb{F}}$, say). Then, as above,
\begin{equation}
n^{1/2}\left(\sigma_\eps(-e_k,\widehat{F}_{Y_0}, \widehat{F}_{X_0})- \sigma_\eps(-e_k, F_{Y_0},F_{X_0})\right) \convD \kappa'_{G}(\widetilde{\mathbb{F}}),	
	\label{eq:conv_as_support}
\end{equation}
where the expression of $\kappa'$ is given by (3.10) in \cite{firpo2021uniform}.

\medskip
Now, suppose that (i) in Assumption \ref{hyp:for_CI} holds: $\sigma_\eps(-e_k, F_{Y_0},F_{X_0})>\sigma(-e_k, F_{Y_0},F_{X_0})$. By, e.g., Theorem 2.2.1 in  \cite{politis1999subsampling}, the subsampling counterpart of \eqref{eq:conv_as_support} holds. This implies that $\widetilde{c}_{\alpha,\eps}(-e_k)=O_P(1)$. Combined with \eqref{eq:def_Einf} and \eqref{eq:conv_as_support}, this implies that $P(\underline{E}_n) \to 0$.

\medskip
Next, suppose that (ii) in Assumption \ref{hyp:for_CI} holds. Then, in view of (3.10) in \cite{firpo2021uniform} and since $G$ and $\widetilde{\mathbb{F}}$ are continuous,
$\kappa'_{G}(\widetilde{\mathbb{F}}) = \min_{\alpha\in[\eps, 1-\eps]} \widetilde{\mathbb{F}}(\widetilde{q}, \alpha)$, where $\widetilde{q}$ is the only $q\in\S$ such that $\inf_{\alpha\in[\eps,1-\eps]} G(\widetilde{q},\alpha)=\kappa(G)$. Because  $\widetilde{\mathbb{F}}(\widetilde{q}, \cdot)$ is Gaussian, the same reasoning as in Point 2 above applies, and  $\widetilde{c}_{\alpha,\eps}(-e_k)\convP c^s_{\alpha,\eps}(-e_k)$, the quantile of order $\alpha$ of $\kappa'_{G}(\widetilde{\mathbb{F}})$. Then, $P(E_n)\to \alpha$.

\medskip
Finally, suppose (iii) in Assumption \ref{hyp:for_CI} holds. We then obtain, still using (3.10) in \cite{firpo2021uniform},
\begin{equation}
\kappa'_{G}(\widetilde{\mathbb{F}}) = \max_{q_m\in\arg\max_{q\in\S} [qS_\eps(F_{Y_0},F_{X_0'q})]'(-e_k)} \widetilde{\mathbb{F}}(q_m, a_\eps(q_m)),
	\label{eq:expr_kappa_pr}
\end{equation}
where for each $q_m\in\arg\max_{q\in\S} [qS_\eps(F_{Y_0},F_{X_0'q})]'(-e_k)$, $a_\eps(q_m)$ is the only $a\in (\eps,1-\eps)$ such that $R(a_\eps(q_m), F_Y,F_{X'q_m}) = \inf_{\alpha\in[\eps,1-\eps]} R(\alpha, F_Y,F_{X'q_m})$. Because $\widetilde{\mathbb{F}}(\cdot, a_\eps(\cdot))$ is Gaussian, the same reasoning as above applies once more and again, $P(E_n)\to \alpha$.

\medskip
To conclude the proof, we show the validity of using $\eps(e)$ instead of a fixed $\eps$ under Assumption \ref{hyp:eps_e}. We do this by proving that we still have $\limsup_{n\to\infty} P(\underline{E}_n^{\eps(e)})\le \alpha$, now indexing $\underline{E}_n$ by $\eps$ to avoid any ambiguity. If (i) of Assumption \ref{hyp:eps_e} holds for all $\eps\in\mathcal{E}$, we have, by what precedes,
\begin{equation}
P(\underline{E}_n^{\eps(e)})\le  P\left(\cup_{\eps\in\mathcal{E}} \underline{E}_n^\eps\right)
\le  \sum_{\eps\in\mathcal{E}} P(\underline{E}_n^\eps)\to  0.	
	\label{eq:pr_En_0}
\end{equation}
Otherwise, (ii) in Assumption \ref{hyp:eps_e} holds. Let $\eps_{j_0}$ be as in Assumption \ref{hyp:eps_e} and let us first show that for all $j\in\{1,\cdots,j_0\}$, $\mathcal{Q}_j=\mathcal{Q}_{j_0}$, with $\mathcal{Q}_j:=\arg\max_{q\in\S} [S_{\eps_j}(F_{Y_0},F_{X_0'q}) q]'(-e_k)$. First, for all $q_m\in \mathcal{Q}_j$ and using that $ q_m'(-e_k) \geq 0$,
\begin{align*}
	[S_{\eps_{j_0}}(F_{Y_0},F_{X_0'q_m}) q_m]'(-e_k) & \ge [S_{\eps_{j}}(F_{Y_0},F_{X_0'q_m}) q_m]'(-e_k) \\
	& = \sigma_{\eps_j}(-e_k, F_{Y_0},F_{X_0}) \\
	& = \sigma_{\eps_{j_0}}(-e_k, F_{Y_0},F_{X_0}).
\end{align*}
Thus, $q_m\in \mathcal{Q}_{j_0}$ and $\mathcal{Q}_j\subset \mathcal{Q}_{j_0}$. Conversely, for any $q_m\in \mathcal{Q}_{j_0}$, by assumption,
\begin{align*}
	S_{\eps_j}(F_{Y_0},F_{X_0'q_m}) & = \min_{a\in [\eps_j, 1-\eps_j]} R(a, F_{Y_0}, F_{X_0'q_m}) \\
	& = R(a(q_m), F_{Y_0}, F_{X_0'q_m}) \\
	& = S_{\eps_{j_0}}(F_{Y_0},F_{X_0'q_m}).
\end{align*}
As a result,
\begin{align*}
	[S_{\eps_j}(F_{Y_0},F_{X_0'q_m}) q_m]'(-e_k) & = [S_{\eps_{j_0}}(F_{Y_0},F_{X_0'q_m}) q_m]'(-e_k) \\
	& = \sigma_{\eps_{j_0}}(-e_k, F_{Y_0},F_{X_0}) \\
	& = \sigma_{\eps_j}(-e_k, F_{Y_0},F_{X_0}),
\end{align*}
implying that $q_m\in \mathcal{Q}_{j}$. Thus, $\mathcal{Q}_{j_0}\subset \mathcal{Q}_j$ and then $\mathcal{Q}_{j_0}= \mathcal{Q}_j$. Now, by \eqref{eq:expr_kappa_pr} but making the dependence on $\eps$ explicit, we have, for all $j\le j_0$,
$$\kappa'_{\eps_j, G}(\widetilde{\mathbb{F}}) = \max_{q_m\in\mathcal{Q}_{j}} \widetilde{\mathbb{F}}(q_m, a_{\eps_{j}}(q_m)).$$
Hence, $\kappa'_{\eps_1, G}(\widetilde{\mathbb{F}})=\cdots=\kappa'_{\eps_{j_0}, G}(\widetilde{\mathbb{F}})$. Reasoning as in Point 2 above, we obtain $P(\underline{E}_n^{\eps_j})=P(\underline{E}_n^{\eps_1})+o(1)$. Then,
\begin{align*}
	P(\underline{E}_n^{\eps(e)}) & = \sum_{j=1}^J P\left(\underline{E}_n^{\eps_j}, \eps(e)=\eps_j\right) \\
	& = \sum_{j=1}^{j_0} P\left(\underline{E}_n^{\eps_j}, \eps(e)=\eps_j\right) + o(1) \\
	& = \sum_{j=1}^{j_0} P\left(\underline{E}_n^{\eps_1}, \eps(e)=\eps_j\right) + o(1) \\
	& \le P\left(\underline{E}_n^{\eps_1}\right) + o(1),
\end{align*}
where the second equality holds since when $\eps(e)>\eps_{j_0}$, $\sigma_{\eps(e)}(-e_k, F_{Y_0},F_{X_0})>\sigma(-e_k,$ $F_{Y_0},F_{X_0})$ and we can apply the same reasoning leading to \eqref{eq:pr_En_0}. The result follows since by what precedes, $\liminf_{n\to\infty} P\left(\underline{E}_n^{\eps_1}\right)\le \alpha$.


\newpage
\pagenumbering{arabic}
\begin{center}
{\huge Supplementary material } \\[2mm] {\large (not for publication)}	
\end{center}

\subsection*{1. Complements on the proof of Proposition \ref{prop:Hausdorff}} 

\begin{lem}\label{lem:ineq_q}
For any random variables $U_1$ and $U_2$, $\alpha\in (0,1)$ and $\beta\in(0,1-\alpha)$, we have:
\begin{equation}
F_{U_1+U_2}^{-1}(\alpha) \le F_{U_1}^{-1}(\alpha+\beta) + F_{U_2}^{-1}(1-\beta).	
	\label{eq:ineq_quantiles}
\end{equation}
\end{lem}

\textbf{Proof:} Fix $\alpha\in (0,1)$ and $\beta\in(0,1-\alpha)$. We have
\begin{align*}
	& P(U_1+U_2 \le F_{U_1}^{-1}(\alpha+\beta) + F_{U_2}^{-1}(1-\beta)) \\
 \ge & \: P(U_1\le F_{U_1}^{-1}(\alpha+\beta),\, U_2 \le F_{U_2}^{-1}(1-\beta)) \\
 \ge & \: P(U_1\le F_{U_1}^{-1}(\alpha+\beta)) + P(U_2 \le F_{U_2}^{-1}(1-\beta)) - 1 \\
 \ge & \: \alpha.
\end{align*}
Equation \eqref{eq:ineq_quantiles} follows by definition of quantiles.

\medskip
We now establish an upper bound on $|F^{-1}_{Y_0}(\alpha)/F^{-1}_{X_0'\beta_0}(\alpha) - 1|$ for $\alpha$ or $1-\alpha$ small:

\begin{lem}\label{lem:upper_bound_ratio_Q}
	Under the assumptions of Proposition \ref{prop:Hausdorff}.\ref{prop:H1} and with $\alpha_0$ and $\alpha_1$ as in the proof of Proposition \ref{prop:Hausdorff}, we have:
$$\forall \alpha\ge \alpha_0, \; \left|\frac{F^{-1}_{Y_0}(\alpha)}{F^{-1}_{X_0'\beta_0}(\alpha)} - 1\right| \lesssim (1-\alpha)^{\frac{1/c - 1/d}{1+1/d}}, \quad
\forall \alpha\le \alpha_1,\; \left|\frac{F^{-1}_{Y_0}(\alpha)}{F^{-1}_{X_0'\beta_0}(\alpha)} - 1\right| \lesssim \alpha^{\frac{1/c - 1/d}{1+1/d}}.$$
\end{lem}

\medskip
\textbf{Proof:} we focus hereafter on the case $\alpha\ge \alpha_0$; the other case can be treated similarly. Fix $\gamma>1$ and note that by Lemma \ref{lem:ineq_q}, we have
$$F^{-1}_{Y_0}(\alpha) \le F^{-1}_{X_0'\beta_0}(\alpha + (1-\alpha)^\gamma) + F^{-1}_{U}(1-(1-\alpha)^\gamma).$$
Moreover, since $X_0'\beta_0 = Y_0 - U$,
$$F^{-1}_{X_0'\beta_0}(\alpha - (1-\alpha)^\gamma) \le F^{-1}_{Y_0}(\alpha) + F^{-1}_{-U}(1-(1-\alpha)^\gamma). $$
Thus,
\begin{align}
   & \frac{F^{-1}_{X_0'\beta_0}(\alpha - (1-\alpha)^\gamma)- F^{-1}_{X_0'\beta_0}(\alpha)}{F^{-1}_{X_0'\beta_0}(\alpha)} - \frac{F^{-1}_{-U}(1-(1-\alpha)^\gamma)}{F^{-1}_{X_0'\beta_0}(\alpha)} \notag \\
   \le & \frac{F^{-1}_{Y_0}(\alpha)}{F^{-1}_{X_0'\beta_0}(\alpha)} - 1  \le \frac{F^{-1}_{X_0'\beta_0}(\alpha + (1-\alpha)^\gamma) - F^{-1}_{X_0'\beta_0}(\alpha)}{F^{-1}_{X_0'\beta_0}(\alpha)} + \frac{F^{-1}_{U}(1-(1-\alpha)^\gamma)}{F^{-1}_{X_0'\beta_0}(\alpha)}. \label{eq:lower_bounds_quantiles}
 \end{align}
Now, the tail conditions imply that $F^{-1}_{X_0'\beta_0}(\alpha) \gtrsim (1-\alpha)^{-1/c}$ and $F^{-1}_{U}(\alpha) \lesssim (1-\alpha)^{-1/d}$. As a result, we get, for some $h(\alpha) \in (0, (1-\alpha)^\gamma)$,
\begin{align*}
  \frac{F^{-1}_{Y_0}(\alpha)}{F^{-1}_{X_0'\beta_0}(\alpha)} - 1 & \le \frac{(1-\alpha)^\gamma}{f_{X'\beta_0}(F^{-1}_{X_0'\beta_0}(\alpha + h(\alpha))) \; F^{-1}_{X_0'\beta_0}(\alpha)} + \frac{F^{-1}_{U}(1-(1-\alpha)^\gamma)}{F^{-1}_{X_0'\beta_0}(\alpha)} \\
   & \lesssim  (1-\alpha)^\gamma(1-\alpha - h(\alpha))^{-1-1/c}(1-\alpha)^{1/c} + (1-\alpha)^{1/c} (1-\alpha)^{-\gamma/d} \\
   & \lesssim (1-\alpha)^{\gamma -1} +(1-\alpha)^{1/c - \gamma/d}.
\end{align*}
Choosing $\gamma=(1+1/c)/(1+1/d)$ yields
$$ \frac{F^{-1}_{Y_0}(\alpha)}{F^{-1}_{X_0'\beta_0}(\alpha)} - 1\lesssim (1-\alpha)^{\frac{1/c - 1/d}{1+1/d}}.$$
The exact same reasoning with the lower bound in \eqref{eq:lower_bounds_quantiles} finally yields the result.


\subsection*{2. Complements on the proof of Theorem \ref{thm:consistency}} 

\begin{lem}\label{lem:continuous_R}
$R$ is continuous for the metric $d'$ defined by \eqref{eq:def_metric}.
\end{lem}

\medskip
\textbf{Proof:} First, remark that for all $a, a', b, b'>0$, we have
\begin{equation}
\left|\frac{a'}{b'} - \frac{a}{b}\right| \leq \frac{1}{b}\left[\left|a' - a\right| + \left|\frac{a'}{b'} - \frac{a}{b}\right|\left|b' - b\right| + \frac{a}{b}\left|b' - b\right|\right].	
	\label{eq:ratio_ineq}
\end{equation}
Therefore, if $|b'-b|< b$,
$$\left|\frac{a'}{b'} - \frac{a}{b}\right| \leq \frac{\left|a' - a\right| + a/b\left|b' - b\right|}{b -|b'-b|}.$$
Fix $\alpha\in[\eps,1-\eps]$, $F$ and $G$ and let $G'$ be such that $W_1(G,G') < (1/4)\int_{\alpha}^1 G^{-1}(t)dt$. Let also $\alpha'\in[\eps,1-\eps]$ be such that
$$\left|\int_\alpha^{\alpha'} G^{-1}(t)dt\right|<\frac{1}{2} \int_\alpha^1 G^{-1}(t)dt.$$
Then,
$$\int_\alpha^1  G^{-1}(t)dt = \int_\alpha^{\alpha'}  G^{-1}(t)dt +\int_{\alpha'}^1 G^{-1}(t)dt < \frac{1}{2} \int_\alpha^1 G^{-1}(t)dt +\int_{\alpha'}^1 G^{-1}(t)dt.$$
Thus, $\int_\alpha^1 G^{-1}(t)dt < 2 \int_{\alpha'}^1 G^{-1}(t)dt$. Moreover, since $W_1(F,F')= \int_0^1 |F^{-1}(t)-F'{}^{-1}(t)|dt$, we have
$$\left|\int_{\alpha'}^1 G'{}^{-1}(t) - G^{-1}(t) dt \right| \leq W_1(G,G') < \frac{1}{2} \int_{\alpha'}^1 G^{-1}(t)dt.$$
Let $c_F=|F^{-1}(\eps)| \vee |F^{-1}(1-\eps)| $ and define $c_G$ similarly. Then, using \eqref{eq:ratio_ineq}, we get
\begin{align}
 \left|R(\alpha',F,G)-R(\alpha,F,G)\right| & \leq \frac{\left|\int_{\alpha}^{\alpha'} F^{-1}(t)dt\right| + R(\alpha,F,G)  \left|\int_{\alpha}^{\alpha'} G^{-1}(t)dt\right|}{\int_\alpha^1 G^{-1}(t)dt -  \left|\int_{\alpha}^{\alpha'} G^{-1}(t)dt\right|} \notag \\
	& \leq \frac{|\alpha'-\alpha| \left(|F^{-1}(\alpha)| \vee |F^{-1}(\alpha')| + R(\alpha,F,G)  |G^{-1}(\alpha)| \vee |G^{-1}(\alpha')|\right)}{1/2\int_\alpha^1 G^{-1}(t)dt} \notag \\
	&  \leq \frac{2|\alpha'-\alpha| \left(c_F + R(\alpha,F,G) c_G\right)}{\int_\alpha^1 G^{-1}(t)dt}.\label{eq:ineg1}
\end{align}
Next, for any $F'$, using again \eqref{eq:ratio_ineq},
\begin{align}
\left|R(\alpha',F',G')-R(\alpha',F,G)\right| & \leq \frac{\left|\int_{\alpha'}^1 F^{-1}(t) - F'{}^{-1}(t)dt\right| + R(\alpha',F,G) \left|\int_{\alpha'}^1 G^{-1}(t) - G'{}^{-1}(t)dt\right|}{\int_{\alpha'}^1 G^{-1}(t)dt - \left|\int_{\alpha'}^1 G'{}^{-1}(t) - G^{-1}(t) dt \right|} \notag \\
	& \leq \frac{W_1(F,F') + R(\alpha',F,G) W_1(G,G')}{1/4\int_{\alpha}^1 G^{-1}(t)dt} \notag \\
	& \leq \frac{4}{\int_{\alpha}^1 G^{-1}(t)dt}\left[W_1(F,F')+\left(\frac{2|\alpha'-\alpha| \left(c_F + R(\alpha,F,G) c_G\right)}{\int_\alpha^1 G^{-1}(t)dt} \right.\right. \notag \\
	& \hspace{2.5cm}+R(\alpha, F,G)\bigg) W_1(G,G') \bigg].\label{eq:ineg2}
\end{align}
The result follows by Inequalities \eqref{eq:ineg1} and \eqref{eq:ineg2} and  the triangle inequality.


\subsection*{3. Complements on the proof of Theorem \ref{thm:inference}}  

\begin{lem}\label{lem:weak_cv_Fn}
Fix $\eps\in (0,1/2)$ and suppose that $n_X/(n_X+n_Y)\to \mu\in (0,1)$ and Assumptions \ref{hyp:mom}-\ref{hyp:samples} and  \ref{hyp:for_CI} hold. Then, $\mathbb{F}_n$, as a process indexed by $(q,\alpha)\in\S\times [\eps,1-\eps]$, converges weakly to a Gaussian process $\mathbb{F}$. The same holds but for $\mathbb{F}_n$ indexed by $\alpha \in[\eps,1-\eps]$ only if Assumption \ref{hyp:for_CI} is replaced by Assumption \ref{hyp:for_CR}.
\end{lem}

\textbf{Proof:} First, $R(\alpha, F_{Y_0},F_{X_0'q})=\theta_1(q,\alpha)/\theta_2(q,\alpha)$, where $\theta_1(q,\alpha)=\int_\alpha^1 F_{Y_0}^{-1}(t)dt$, $\theta_2(q,\alpha)=\int_\alpha^1 F_{X_0'q}^{-1}(t)dt$ and we suppress the dependence of $\theta_1$ and $\theta_2$ in $F_{Y_0}$ and $F_{X_0'q}$ for simplicity. Moreover, $R(\alpha,\widehat{F}_{Y_0},\widehat{F}_{X_0'q})=\widehat{\theta}_1(q,\alpha)/\widehat{\theta}_2(q,\alpha)$ with $\widehat{\theta}_1(q,\alpha)=\int_\alpha^1 \widehat{F}_{Y_0}^{-1}(t)dt$ and $\widehat{\theta}_2(q,\alpha)=\int_\alpha^1 \widehat{F}_{X_0'q}^{-1}(t)dt$. The map $(U, V) \mapsto U/V$, from $\ell^\infty(\S\times[\eps,1-\eps])^2$ to $\ell^\infty(\S\times[\eps,1-\eps])$, is Hadamard differentiable at any $(U,V)$ such that $\inf_{(q,\alpha)\in \S\times[\eps,1-\eps]} V(q,\alpha)>0$. Now,  $\theta_2(\cdot, \alpha)$ is continuous (see the proof of Proposition \ref{prop:epsilon}). $\theta_2(q, \cdot)$ is also continuous. Thus,
$$\inf_{(q,\alpha)\in \S\times[\eps,1-\eps]} \theta_2(q,\alpha)=\min_{(q,\alpha)\in \S\times[\eps,1-\eps]} \theta_2(q,\alpha)>0.$$
Hence, by the functional delta method, $\mathbb{F}_n$ converges weakly as long as
$$n^{1/2}\left(\widehat{\theta}_1(q,\alpha) - \theta_1(q,\alpha), \widehat{\theta}_2(q,\alpha)- \theta_2(q,\alpha)\right)$$
converges weakly. By independence of the two samples, it suffices to show the weak convergence of each component. We focus on the second hereafter, as the proof is similar (and actually simpler) for the first. Also, it suffices to show the weak convergence of $n_X^{1/2}\left(\widehat{\theta}_2(q,\alpha)- \theta_2(q,\alpha)\right)$, as $n/n_X\to 1-\mu$ by assumption.

\medskip
Let us define
$$\widetilde{\theta}_2(q,\alpha) = \frac{1}{n_X}\sum_{i=1}^{n_X} \left(X_i'q - \overline{X'q}\right)\indic{\widehat{F}_{X'q}(X_i'q) > \alpha}. \label{eq:theta2}$$
Because $F_{X'q}$ is continuous, almost surely there are no ties and $\widehat{\theta}_2(q,\alpha)=\widetilde{\theta}_2(q,\alpha)$ for all $\alpha\in\{0/n_X,...,(n_X-1)/n_X\}$. Elsewhere, if $\alpha= [ti+(1-t)(i+1)]/n_X$, $t\in(0,1)$, we have $\widehat{\theta}_2(q,\alpha)=t\widetilde{\theta}_2(q,i/n_X)+(1-t) \widetilde{\theta}_2(q,(i+1)/n_X)$. As a result,
\begin{align*}
n_X^{1/2}\sup_{\alpha\in[\eps,1-\eps]} \left|\widehat{\theta}_2(q,\alpha)-\widetilde{\theta}_2(q,\alpha)\right| \le &
\frac{\sup_{i=\lfloor n\eps\rfloor,...,\lceil n(1-\eps)\rceil} \left|(X'q)_{(i)} - \overline{X'q}\right|}{n^{1/2}_X} \\
\le & \frac{\left|(X'q)_{(n)} - (X'q)_{(1)}\right|}{n_X^{1/2}} \\
\convP & 0,
\end{align*}
where the convergence follows by, e.g., Problem 2.3.4 in \cite{VdV_Wellner}. Hence, it suffices to show the weak convergence of $n_X^{1/2}(\widetilde{\theta}_2(q,\alpha) - \theta_2(q,\alpha))$. By, e.g. Lemma 21.1 in \cite{van2000asymptotic}, $$\theta_2(q,\alpha)=E\left[(X'q-E(X'q))\indic{F_{X'q}(X'q)\ge \alpha}\right].$$
As a result,
$$n_X^{1/2}\left(\widetilde{\theta}_2(q,\alpha) - \theta_2(q,\alpha)\right)= \mathbb{G}_{n_X} g_{q,\alpha}+ R_{n_X}(q,\alpha),$$
where $\mathbb{G}_{n_X}$ denotes the empirical process associated to $(X_1,...,X_{n_X})$ and
\begin{align*}
	g_{q,\alpha}(x) = & \left[F_{X'q}^{-1}(\alpha)-E(X'q) \right] \indic{F_{X'q}(x'q)\leq \alpha} -(1-\alpha) x'q \\
		& + (x'q - E(X'q))\indic{F_{X'q}(x'q) > \alpha}, \\
	R_{n_X}(q,\alpha) = & \frac{1}{n_X^{1/2}}\sum_{i=1}^{n_X} \left\{\left(X_i'q - \overline{X'q}\right)\left[\indic{F_{X'q}(X_i'q) \leq \alpha} - \indic{\widehat{F}_{X'q}(X_i'q) \leq \alpha} \right] \right. \\
	& \hspace{1cm}  - \left[F_{X'q}^{-1}(\alpha)-E(X'q) \right] \left(\indic{F_{X'q}(X_i'q)\leq \alpha} -\alpha\right)\bigg\} \\
	& \;  +\frac{n_X^{1/2}\left(\overline{X'q} - E(X'q)\right)}{n_X}\sum_{i=1}^{n_X} \left(\indic{F_{X'q}(X_i'q) \leq  \alpha} - \alpha \right).
\end{align*}
We first prove that the class $\mathcal{G}=\{g_{q,\alpha}: (q,\alpha) \in \S\times [\eps, 1-\eps]\}$ is Donsker. The class $\mathcal{I}_0=\{x\mapsto \indic{x'q\leq u}: (q,u)\in \mathcal{S}  \times \R\}$ is Donsker by Problem 2.6.14 and Theorem 2.6.8 in \cite{VdV_Wellner}. Then, $\mathcal{I}_1=\{x\mapsto \indic{F_{X'q}(x'q)\leq \alpha}: (q,\alpha) \in\S\times[\eps,1-\eps]\} \subset \mathcal{I}_0$ is also Donsker \citep[see, e.g., Theorem 2.10.1 in][]{VdV_Wellner}. Similarly, $\mathcal{I}_2=\{x\mapsto \indic{F_{X'q}(x'q)> \alpha}: (q,\alpha) \in\S\times[\eps,1-\eps]\}$ is Donsker. $\mathcal{I}_2$ also has a finite integral entropy and an envelope of 1. Since $\{x\mapsto x'q: q\in\S\}$ also has a finite integral entropy with envelope $x\mapsto \|x\|$, and $E[\|X\|^2]<\infty$, the class $\mathcal{I}_3=\{x\mapsto (x'q)\indic{F_{X'q}(x'q)> \alpha}: (q,\alpha) \in\S\times[\eps,1-\eps]\}$ is also Donsker \cite[see Example 19.19 in][]{van2000asymptotic}. Because $\{x\mapsto (1-\alpha)x'q: (q,\alpha)\in\S\times[\eps,1-\eps]\}$ is also Donsker and sums of Donsker classes are also Donsker, we finally get that $\mathcal{G}$ is Donsker.

\medskip
Next, we consider the remainder term $R_{n_X}(q,\alpha)$. Let $I_i(q,\alpha)= \indic{F_{X'q}(X_i'q) \leq \alpha}$ and $\widehat{I}_i(q,\alpha)=\indic{\widehat{F}_{X'q}(X_i'q) \leq \alpha}$. We have $R_{n_X}(q,\alpha) = R_{1n_X}+R_{2n_X}+R_{3n_X}$, with
\begin{align*}
	R_{1n_X}(q,\alpha)& = \frac{1}{n_X^{1/2}}\sum_{i=1}^{n_X} (I_i(q,\alpha) - \widehat{I}_i(q,\alpha)) \left[\left(X_i'q - \overline{X'q}\right) - \left(F_{X'q}^{-1}(\alpha)-E(X'q) \right) \right], \\
	R_{2n_X}(q,\alpha)& = \frac{\left(F_{X'q}^{-1}(\alpha)-E(X'q) \right) }{n_X^{1/2}}\sum_{i=1}^{n_X} \left[\alpha - \widehat{I}_i(q,\alpha)\right], \\
	R_{3n_X}(q,\alpha) & = \frac{n_X^{1/2}\left(\overline{X'q}-E(X'q)\right)}{n_X}\sum_{i=1}^{n_X} \left(I_i(q,\alpha)- \alpha \right).
\end{align*}
We now prove that for all $k\in\{1,2,3\}$,
\begin{equation}
\sup_{(q,\alpha)\in \S\times [\eps, 1-\eps]} R_{kn_X}(q,\alpha)=o_P(1).	
	\label{eq:reste_{n_X}eglig}
\end{equation}
Consider $R_{2n_X}$ first. By definition of the empirical cdf., we have, for all $(q,\alpha)$,
\begin{equation}
\left|\sum_{i=1}^{n_X} \left(\widehat{I}_i(q,\alpha)-\alpha\right)\right| = \lceil n_X\alpha\rceil - n_X\alpha <1.	
	\label{eq:Ihat}
\end{equation}
As a result,
\begin{align*}
\sup_{(q,\alpha)\in \S\times [\eps, 1-\eps]} |R_{2n_X}(q,\alpha)| \leq & \frac{F_{\|X\|}^{-1}(1-\eps)+E(\|X\|)}{n_X^{1/2}}  \times \sup_{(q,\alpha)\in \S\times [\eps, 1-\eps]} \left|\sum_{i=1}^{n_X} \left( \widehat{I}_i(q,\alpha)-\alpha \right) \right| \\
\leq & \frac{F_{\|X\|}^{-1}(1-\eps)+E(\|X\|)}{n_X^{1/2}},
\end{align*}
where the first inequality follows from the triangle and Cauchy-Schwarz inequalities and $|F_{X'q}^{-1}(\eps)| \vee |F_{X'q}^{-1}(1-\eps)| \leq F_{\|X\|}^{-1}(1-\eps)$. Hence, \eqref{eq:reste_{n_X}eglig} holds for $k=2$.

\medskip
Next, consider $R_{3n_X}$. We have
$$\sup_{(q,\alpha)\in \S\times [\eps, 1-\eps]} |R_{3n_X}(q,\alpha)| \leq n_X^{1/2}\|\overline{X}-E(X)\| \times \sup_{(q,\alpha)\in \S\times [\eps, 1-\eps]} \left|\frac{1}{n_X}\sum_{i=1}^{n_X} \left( I_i(q,\alpha)- \alpha \right) \right|.$$
The first term is an $O_P(1)$. Recall that the class $\mathcal{I}_1$ is Donsker; hence it is also Glivenko-Cantelli. Therefore, the second term is an $o_P(1)$. Therefore, \eqref{eq:reste_{n_X}eglig} holds for $k=3$.

\medskip
Finally, consider $R_{1n_X}$. We first decompose it further into $R_{11n_X}+R_{12n_X}$, with
\begin{align*}
	R_{11n_X}(q,\alpha)& = \frac{-n_X^{1/2}(\overline{X'q} - E(X'q))}{n_X}\sum_{i=1}^{n_X} [I_i(q,\alpha) - \widehat{I}_i(q,\alpha)], \\
	R_{12n_X}(q,\alpha)& = \frac{1}{n_X^{1/2}}\sum_{i=1}^{n_X} (I_i(q,\alpha) - \widehat{I}_i(q,\alpha))\left(X_i'q - F_{X'q}^{-1}(\alpha)\right).	
\end{align*}
That $R_{11n_X}$ is uniformly negligible follows by writing $I_i(q,\alpha) - \widehat{I}_i(q,\alpha)=I_i(q,\alpha) - \alpha + \alpha -  \widehat{I}_i(q,\alpha)$, reasoning as for $R_{3n_X}$ and using \eqref{eq:Ihat}. For $R_{12n_X}$, remark that by definition of $I_i(q,\alpha)$ and continuity of $X_i'q$, $I_i(q,\alpha)=\indic{X_i'q \le F^{-1}_{X'q}(\alpha)}$. Similarly, but accounting for the discontinuity of $\widehat{F}_{X'q}$, we have
$$\widehat{I}_i(q,\alpha)= \left|\begin{array}{ll} \indic{X_i'q < \widehat{F}^{-1}_{X'q}(\alpha)} & \text{ if } n_X\alpha\not\in\N, \\[2mm]
\indic{X_i'q \le \widehat{F}^{-1}_{X'q}(\alpha)} & \text{ otherwise.}	
\end{array} \right.$$
As a result,
\begin{align*}
\sum_{i=1}^{n_X} \left|I_i(q,\alpha)-\widehat{I}_i(q,\alpha)\right| = & \left(2\indic{F^{-1}_{X'q}(\alpha) \ge \widehat{F}^{-1}_{X'q}(\alpha)} - 1\right) \left(\sum_{i=1}^{n_X} I_i(q,\alpha)-\widehat{I}_i(q,\alpha)\right) \\
	= & \left|\sum_{i=1}^{n_X} I_i(q,\alpha)-\widehat{I}_i(q,\alpha)\right|.
\end{align*}
Moreover, $|I_i(q,\alpha)-\widehat{I}_i(q,\alpha)|=1$ only if $X'_iq\in J$, the interval $[\widehat{F}^{-1}_{X'q}(\alpha), F^{-1}_{X'q}(\alpha)]$ if $\widehat{F}^{-1}_{X'q}(\alpha) < F^{-1}_{X'q}(\alpha)$ and $[F^{-1}_{X'q}(\alpha),\widehat{F}^{-1}_{X'q}(\alpha)]$ otherwise. As a result,
\begin{align}
	|R_{12n_X}| \le & \frac{1}{n_X^{1/2}}\sum_{i=1}^{n_X} \left|I_i(q,\alpha)-\widehat{I}_i(q,\alpha)\right| \left|X_i'q - F^{-1}_{X'q}(\alpha)\right| \notag \\
	\le & \left|\widehat{F}_{X'q}^{-1}(\alpha)- F_{X'q}^{-1}(\alpha)\right| \times \left|\frac{1}{n_X^{1/2}}\sum_{i=1}^{n_X} (I_i(q,\alpha) - \widehat{I}_i(q,\alpha))\right|. \label{eq:ineq_R12n}
\end{align}
By \eqref{eq:Ihat} and the fact that $\mathcal{I}_1$ is a Donsker class,
$$\sup_{(q,\alpha)\in \S\times [\eps, 1-\eps]} \left|\frac{1}{n_X^{1/2}}\sum_{i=1}^{n_X} (I_i(q,\alpha) - \widehat{I}_i(q,\alpha))\right|=O_P(1).$$
Thus, the result holds as long as
\begin{equation}
\sup_{(q,\alpha)\in\S\times[\eps,1-\eps]} \left|\widehat{F}_{X'q}^{-1}(\alpha)- F_{X'q}^{-1}(\alpha)\right| = o_P(1).	
	\label{eq:controle_unif_q}
\end{equation}
To prove this, note first that the class $\{x\mapsto \indic{x'q\leq \alpha}: (q,\alpha)\in\S\times[\eps,1-\eps]\}$ is Glivenko-Cantelli (as it is Donsker). Hence,
\begin{equation}
\sup_{(q,\alpha)\in\S\times [\eps,1-\eps]}\left|F_{X'q}(\alpha)-\widehat{F}_{X'q}(\alpha)\right|=o_P(1).
\label{eq:controle_unif_F}
\end{equation}
Now, let $U_q=F_{X'q}(X'q)$ and $U_{q,1}<...<U_{q,n_X}$ denote the corresponding order statistic. Remark that $\widehat{F}_{X'q}^{-1}(\alpha) = F_{X'q}^{-1}(U_{q, \lceil n_X\alpha\rceil})$. Also, note that  $\inf_{(q,\alpha)\in\S\times[\eps,1-\eps]} U_{q, \lceil n_X\alpha\rceil} < \eps'$ implies that for some $q_0\in\S$, $\widehat{F}_{X'q_0}(F_{X'q_0}^{-1}(\eps'))\geq \lceil n_X\alpha\rceil/n_X$ and thus
$$\sup_{(q,\alpha)\in\S\times [\eps,1-\eps]}\left|F_{X'q}(\alpha)-\widehat{F}_{X'q}(\alpha)\right| > \eps -\eps'.$$
In view of \eqref{eq:controle_unif_F}, this occurs with probability approaching zero. The same is true for the event $\sup_{(q,\alpha)\in\S\times[\eps,1-\eps]} U_{q, \lceil n_X\alpha\rceil} >1-\eps'$. Hence, with probability approaching one,
\begin{equation}
\eps' \leq \inf_{(q,\alpha)\in\S\times[\eps,1-\eps]} U_{q, \lceil n_X\alpha\rceil} \leq \sup_{(q,\alpha)\in\S\times[\eps,1-\eps]} U_{q, \lceil n_X\alpha\rceil} \leq 1-\eps'.	
	\label{eq:inv_in_compact}
\end{equation}
Moreover, under this event,
\begin{align*}
 \left|\widehat{F}_{X'q}^{-1}(\alpha)- F_{X'q}^{-1}(\alpha)\right|
	= & \left|F_{X'q}^{-1}(U_{q, \lceil n_X\alpha\rceil})- F_{X'q}^{-1}(\alpha)\right| \\
 < & m\left(|U_{q, \lceil n_X\alpha\rceil}-\alpha|\right) \\
	\leq & m\left(|F_{X'q}((X'q)_{\lceil n_X\alpha\rceil})-\widehat{F}_{X'q}((X'q)_{ \lceil n_X\alpha\rceil})|\right. \\
	& \quad \left. +\left|\widehat{F}_{X'q}((X'q)_{ \lceil n_X\alpha\rceil}) - \alpha\right|\right) \\
	 < & m\left(\sup_{q\in\S}\sup_{t\in\R}|F_{X'q}(t)-\widehat{F}_{X'q}(t)| +\frac{1}{n_X}\right).
\end{align*}
Using \eqref{eq:controle_unif_F} and the continuity of $m$ finally yields \eqref{eq:controle_unif_q}.

\medskip
Finally, let us prove the weak convergence of $\mathbb{F}_n$ as a process indexed by $\alpha \in[\eps,1-\eps]$ only, but under the weaker Assumption \ref{hyp:for_CR}. It suffices to remark that all steps above still hold, except \eqref{eq:controle_unif_q}. Now, given that $q$ is fixed, we only need to establish the weaker
\begin{equation}
\sup_{\alpha\in[\eps,1-\eps]} \left|\widehat{F}_{X'q}^{-1}(\alpha)- F_{X'q}^{-1}(\alpha)\right| = o_P(1).	
	\label{eq:controle_unif_q2}
\end{equation}
Because $F_{X'q}^{-1}$ is continuous on $[\eps,1-\eps]$ (as the inverse of $F_{X'q}$ is strictly increasing on its support by Assumption \ref{hyp:for_CR}), it is uniformly continuous on $[\eps,1-\eps]$. Now, note that  $$\left|\widehat{F}_{X'q}^{-1}(\alpha)- F_{X'q}^{-1}(\alpha)\right| = \left|F_{X'q}^{-1}(U_{q, \lceil n_X\alpha\rceil})- F_{X'q}^{-1}(\alpha)\right|.$$
Moreover, $\sup_{\alpha \in [\eps,1-\eps]} \left|U_{q, \lceil n_X\alpha\rceil} - \alpha\right|=o_P(1)$. This implies that \eqref{eq:controle_unif_q2} holds.


\subsection*{4. Proof of Proposition \ref{prop:test_bord}} 
\label{sub:proposition_ref_prop_test_bord}

Our proof heavily draws on and use the same notation as in Theorem \ref{thm:inference}. It proceeds in four steps. First, we show that $h(\beta,\alpha):=E\left[X_{v0}\indic{F_{X_{v0}'\beta}(X_{v0}'\beta)\ge \alpha}\right]$ is continuous. Second, we prove that $\theta_3(t,\alpha):=\theta_2(t\widehat{\beta}_v+(1-t)\beta_v)$ ($t\in[0,1]$) is differentiable as a function of $t\in(0,1)$. Third, we show that $\sqrt{n}\left(\widehat{\theta}(\widehat{\beta}_v,\alpha)- \theta(\beta_v,\alpha)\right)$ converges to a Gaussian process. Finally, we prove the two points of the proposition.

\subsubsection*{Step 1: Continuity of $h$.} 
\label{ssub:1_continuity_of_h}

More precisely, we prove below that $h$ is continuous at any $(\beta_1,\alpha_1)\in K\times[\eps,1-\eps]$, with $K$ convex compact including $\beta_v$ in its interior and such that $\{\beta_1/\|\beta_1\|: \beta_1\in K\}\subset \mathcal{V}$. By the triangle inequality, for any $(\beta_1,\alpha_1)\in K\times[\eps,1-\eps]$ and  $(\beta_2,\alpha_2)\in K\times[\eps,1-\eps]$,
\begin{equation}
\|h(\beta_1,\alpha_1)-h(\beta_2,\alpha_2)\| \le \|h(\beta_1,\alpha_1)-h(\beta_1,\alpha_2)\| +
\|h(\beta_1,\alpha_2)-h(\beta_2,\alpha_2)\|.	
	\label{eq:decomp_h}
\end{equation}
Regarding the first term, and assuming without loss of generality that $\alpha_1\le \alpha_2$, we have
\begin{align}
\|h(\beta_1,\alpha_1)-h(\beta_1,\alpha_2)\| = & \left\|E\left[X_{v0}\indic{\alpha_2\ge F_{X_{v0}'\beta_1}(X_{v0}'\beta_1)\ge \alpha_1}\right]\right\| \notag \\
\le & E\left[\|X_{v0}\|^2\right]^{1/2}(\alpha_2-\alpha_1)^{1/2}. \label{eq:h_first}
\end{align}
Turning to the second term, we have
\begin{align*}
\|h(\beta_1,\alpha_2)-h(\beta_2,\alpha_2)\| \le & E\left[\|X_{v0}\|^2\right]^{1/2}\left[P\left(F_{X_{v0}'\beta_1}(X_{v0}'\beta_1)\ge \alpha_2 > F_{X_{v0}'\beta_2}(X_{v0}'\beta_2)\right) \right.\\
& \left. \hspace{2.2cm}+ P\left(F_{X_{v0}'\beta_2}(X_{v0}'\beta_2)\ge \alpha_2 > F_{X_{v0}'\beta_1}(X_{v0}'\beta_1)\right)\right]^{1/2}.	
\end{align*}
As both probabilities are similar, we only consider the first one, $P_1$ say. To simplify notation, let $\delta=\beta_2-\beta_1$, $U_k:=X_{v0}'{\beta_k}$, $F_k:=F_{X_{v0}'\beta_k} (k=1,2)$ and $F_\delta:=F_{X_{v0}'\delta}$. Fix $\eta\in(0,1-\alpha_2)$ and let $\delta$ be such that
\begin{equation}
\|\delta\| \le \frac{c\|\beta_1\|\eta^2}{2(\eta F^{-1}_{\|X_{v0}\|}(1-\eta/2)+E\left[\|X_{v0}\|\right])},
	\label{eq:choice_delta}
\end{equation}
where $c$ is defined in Assumption \ref{hyp:test_bord}. Then, we have
\begin{align}
P_1 & \le P\left(U_1\in [\alpha_2, \alpha_2+ \eta)\right) + P\left(U_1 \ge F_1^{-1}(\alpha_2+\eta), U_2 < F_2^{-1}(\alpha_2) \right) \notag \\
& = \eta +  P\left(U_1 \ge F_1^{-1}(\alpha_2+\eta), X_{v0}'\delta < F_2^{-1}(\alpha_2) - F_1^{-1}(\alpha_2+\eta)\right) \notag \\
& \le \eta +  P\left(X_{v0}'\delta < F_1^{-1}(\alpha_2 +\eta/2) + F^{-1}_\delta(1-\eta/2) - F_1^{-1}(\alpha_2+\eta)\right) \notag \\
& \le \eta +  P\left(X_{v0}'\delta < -c\|\beta_1\| \eta/2 + F^{-1}_{\|X_{v0}\|}(1-\eta/2) \|\delta\|\right) \notag \\
& \le \eta + P\left(\|X_{v0}\| > c\|\beta_1\|\eta/(2\|\delta\|) - F^{-1}_{\|X_{v0}\|}(1-\eta/2)\right) \notag \\
& \le \eta + \frac{E\left[\|X_{v0}\|\right]}{c\|\beta_1\|\eta/(2\|\delta\|) - F^{-1}_{\|X_{v0}\|}(1-\eta/2)} \notag \\
& \le 2\eta. \label{eq:h_second}
\end{align}
The second inequality follows from Lemma \ref{lem:ineq_q}. The third uses $F_\delta(x)\le F_{\|X_{v0}\|}(x/\|\delta\|)$, which implies $F_{\delta}^{-1}(1-\eta/2) \le F^{-1}_{\|X_{v0}\|}(1-\eta/2) \|\delta\|$, and $F_1^{-1}(y)-F_1^{-1}(x) > c\|\beta_1\|(y-x)$ for $y>x$, which follows from Assumption \ref{hyp:test_bord} and $\beta_1/\|\beta_1\|\in\mathcal{V}$. The fourth inequality follows from the Cauchy-Schwarz inequality, and the fifth uses Markov's inequality and the fact that by \eqref{eq:choice_delta},
$ c\|\beta_1\|\eta/(2\|\delta\|) - F^{-1}_{\|X_{v0}\|}(1-\eta/2)>0$. The last inequality follows from \eqref{eq:choice_delta}. By combining \eqref{eq:decomp_h}, \eqref{eq:h_first} and \eqref{eq:h_second}, we obtain that  $h$ is continuous.


\subsubsection*{Step 2: Differentiability of $t\mapsto \theta_3(t,\alpha)$ on $(0,1)$.} 
\label{ssub:differentiability_of_t_mapsto_theta_3_t_alpha}

Specifically, we prove the result  with probability approaching one. We show it by applying  the envelope theorem in \cite{milgrom2002envelope}. To this end, first remark that by Proposition 3 in \cite{horowitz1995identification},
$$\theta_2(\beta,\alpha)=\max_{F_{X_{v0},W}:W\sim \text{Be}(1-\alpha)} E[(X_{v0}'\beta)W],$$
where Be denotes Bernoulli distributions. As a result,
$$\theta_3(t,\alpha)=\max_{F_{X_{v0},W}:W\sim \text{Be}(1-\alpha)} \int x'(t\widehat{\beta}_v+(1-t)\beta_v) w dF_{X_{v0}, W}(x,w).$$
By the dominated convergence theorem, the function $t\mapsto f_\alpha(t,F_{X_{v0},W}):= \int x'(t\widehat{\beta}_v+(1-t)\beta_v) w dF_{X_{v0}, W}(x,w)$ is differentiable and
$$\Deriv{f_\alpha}{t}(t,F_{X_{v0},W})=\left[\int xw dF_{X_{v0}, W}(x,w)\right]'(\widehat{\beta}_v-\beta_v).$$
Since $t\mapsto \deriv{f_\alpha}{t}(t,F_{X_{v0},W})$ is constant, the family $\{\deriv{f_\alpha}{t}(\cdot,F_{X_{v0},W}): W\sim \text{Be}(1-\alpha)\}$ is equicontinuous and thus the  family of functions  $\{f_\alpha(\cdot,F_{X_{v0},W}): W\sim \text{Be}(1-\alpha)\}$ is equidifferentiable at any $t\in(0,1)$ \citep[see][p.587]{milgrom2002envelope}. Moreover, by the Cauchy-Schwarz inequality,
$$\sup_{F_{X_{v0},W}:W\sim \text{Be}(1-\alpha)}\left|\Deriv{f_\alpha}{t}(t,F_{X_{v0},W})\right|\le \left(E[\|X_{v0}\|^2](1-\alpha)\right)^{1/2}\|\widehat{\beta}_v-\beta_v\|.$$
Because $\widehat{\beta}_v$ is consistent, with probability approaching one, $\widehat{\beta}_v\in K$ and since $K$ is convex, $\{t\widehat{\beta}_v+(1-t)\beta_v\}\subset K$. Then,
the first step above implies that
$$t\mapsto \left[\int x \indic{F_{X_{v0}'(t\widehat{\beta}_v+(1-t)\beta_v)}[x'(t\widehat{\beta}_v+(1-t)\beta_v)]\ge \alpha}
dF_{X_{v0}}(x)\right]'\left(\widehat{\beta}_v - \beta_v\right).$$
is continuous on $[0,1]$. Hence, the conditions in Theorem 3 of \cite{milgrom2002envelope} hold. Combined with Theorem 1 therein, this implies that $t\mapsto \theta_3(t,\alpha)$ is differentiable and
$$\Deriv{\theta_3}{t}(t,\alpha)=\left[\int x \indic{F_{X_{v0}'(t\widehat{\beta}_v+(1-t)\beta_v)}[x'(t\widehat{\beta}_v+(1-t)\beta_v)]\ge \alpha}
dF_{X_{v0}}(x)\right]'\left(\widehat{\beta}_v - \beta_v\right).$$


\subsubsection*{Step 3: Convergence to a Gaussian process of $\sqrt{n}\left(\widehat{\theta}(\widehat{\beta}_v,\alpha)- \theta(\beta_v,\alpha)\right)$.} 
\label{ssub:convergence_to_a_gaussian_process_of_sqrt_n_left_widehat_theta_widehat_beta__v_alpha_theta_beta_v_alpha_right}

First, note that
\begin{equation}
n^{1/2}\left(\widehat{\beta}_v - \beta_v\right)=V(X_v)^{-1} \left(\frac{1}{n^{1/2}} \sum_{i=1}^n X_{vi} \eps_{vi}\right) + o_P(1),	
	\label{eq:lin_beta}
\end{equation}
where $\eps_{vi} := Y_{v0i} - X_{v0i}'\beta_v$. Let $\theta(q,\alpha)=(\theta_1(q,\alpha),\theta_2(q,\alpha))$ and define $\widehat{\theta}(q,\alpha)$ accordingly. By \eqref{eq:lin_beta}, the Cram\'er-Wold device, stability of Donsker classes by addition and the first part of the proof of Theorem \ref{thm:inference}, the process $\mathbb{G}_n:=\sqrt{n}\left(\widehat{\theta}(.,.) - \theta(.,.),\widehat{\beta}_v - \beta_v\right)$ converges weakly to a Gaussian process on $\mathcal{V}\times[\eps,1-\eps]$. Then, when $\norm{\widehat{\beta}_v}\ne 0$, which occurs with probability approaching one, we have
  \begin{align}
	\sqrt{n}\left(\widehat{\theta}(\widehat{\beta}_v,\alpha)-\theta(\beta_v,\alpha)\right) = &  \norm{\widehat{\beta}_v} \sqrt{n}\left(\widehat{\theta}(\widehat{q},\alpha)-\theta(\widehat{q},\alpha)\right) \notag \\
	& + \sqrt{n}\left(\theta(\widehat{\beta}_v,\alpha) - \theta(\beta_v,\alpha)\right), \label{eq:decomp_test}
\end{align}
where we let $\widehat{q}=\widehat{\beta}_v/\norm{\widehat{\beta}_v}$. First, consider the second term. By the second step and the mean value theorem,
\begin{align*}
\theta_2(\widehat{\beta}_v,\alpha) - \theta_2(\beta_v,\alpha) = & \theta_3(1,\alpha) - \theta_3(0,\alpha) \\
=& h(\widetilde{\beta},\alpha)'\left(\widehat{\beta}_v - \beta_v\right),
\end{align*}
with $\widetilde{\beta}=t\widehat{\beta}_v + (1-t)\beta_v$ for some $t\in[0,1]$. Now, by the first step, $h$ is continuous on the compact set $K\times[\eps,1-\eps]$, which includes $\widetilde{\beta}$ with probability approaching one. Thus, by the maximum theorem and the continuous mapping theorem, $\sup_{\alpha\in[\eps,1-\eps]}|h(\widetilde{\beta},\alpha) - h(\beta_v,\alpha)|\convP 0$.  As a result,
\begin{align}
 \sqrt{n}\left(\theta_2(\widehat{\beta}_v,\alpha) - \theta_2(\beta_v,\alpha)\right)
 = h(\beta_v,\alpha)'  \sqrt{n}\left(\widehat{\beta}_v - \beta_v\right) + \eps'_n(\alpha),
\label{eq:for_cv2}	
\end{align}
where $\sup_{\alpha\in[\eps,1-\eps]} |\eps'_n(\alpha)|\convP 0$.

\medskip
Now let us turn to the first term in \eqref{eq:decomp_test}. We show below that
\begin{equation}
\sup_{\alpha\in[\eps,1-\eps]} \int [g_{\widehat{\beta}_v,\alpha}(x)-g_{\beta_v,\alpha}(x)]^2 dF_X(x)\convP 0.	
	\label{eq:condit_vdV}
\end{equation}
Then, $\norm{\widehat{\beta}_v}\convP \norm{\beta_v} $ and the proof of Theorem 19.26 in \cite{van2000asymptotic} imply that
 \begin{equation}\label{eq:for_cv1}
 \norm{\widehat{\beta}_v} \sqrt{n}\left(\widehat{\theta}(\widehat{q},\alpha)-\theta(\widehat{q},\alpha)\right) = \norm{\beta_v} \sqrt{n}\left(\widehat{\theta}(q_0,\alpha) - \theta(q_0,\alpha)\right)+ \eps_n(\alpha),
 \end{equation}
where $\sup_{\alpha\in[\eps,1-\eps]}  |\eps_n(\alpha)|\convP 0$. Convergence of $\mathbb{G}_n$ combined with equations \eqref{eq:decomp_test}, \eqref{eq:for_cv2} and \eqref{eq:for_cv1} imply that $\sqrt{n}\left(\widehat{\theta}(\widehat{\beta}_v,\alpha)- \theta(\beta_v,\alpha)\right)$ converges in distribution to a Gaussian process $\mathbb{G}$.

\medskip
To prove \eqref{eq:condit_vdV}, given the definition of $g_{\beta,\alpha}$, it suffices to prove
{\small \begin{align}
\sup_{\alpha\in[\eps,1-\eps]} & \left[F^{-1}_{X_v'\widehat{\beta}_v}(\alpha) - F^{-1}_{X_v'\beta_v}(\alpha)\right]^2  \convP 0,
\label{eq:conv_quant_unif} \\
\sup_{\alpha\in[\eps,1-\eps]}& \int \left|\indic{F_{X_v'\widehat{\beta}_v}(x_v'\widehat{\beta}_v)\le \alpha}-
\indic{F_{X_v'\beta_v}(x_v'\beta_v)\le \alpha}\right| dF_X(x) \convP 0, \label{eq:conv_indic} \\
\sup_{\alpha\in[\eps,1-\eps]} & \int \left(x_v'\widehat{\beta}_v\indic{F_{X_v'\widehat{\beta}_v}(x_v'\widehat{\beta}_v)> \alpha}-
x_v'\beta_v\indic{F_{X_v'\beta_v}(x_v'\beta_v)> \alpha}\right)^2 dF_X(x) \convP 0. \label{eq:conv_product}
\end{align}}
We prove that the three terms inside the three suprema are continuous as functions of $(\widehat{\beta}_v,\alpha)$. The results then follow by the maximum and continuous mapping theorems. First remark that since $F_{X_v'\beta}$ is strictly increasing that for all $(\beta,\alpha)\in K\times[\eps,1-\eps]$,
$$F^{-1}_{X_v'\beta}(\alpha)=\argmin_{a\in[-M,M]} E[\rho_\alpha(X_v'\beta - a)],$$
for some $M>0$ large enough and $\rho_\alpha(x)=(\alpha-\indic{x\le 0})x$. By the dominated convergence theorem, the function $(\beta,\alpha,a)\mapsto E[\rho_\alpha(X_v'\beta - a)]$ is continuous. Hence, by the maximum theorem, $(\beta,\alpha)\mapsto F^{-1}_{X_v'\beta}(\alpha)$ is continuous. Then, let $\lambda(\beta):=\max_{\alpha\in[\eps,1-\eps]} (F^{-1}_{X_v'\beta}(\alpha) - F^{-1}_{X_v'\beta_v}(\alpha))^2$. By what precedes, $(\beta,\alpha) \mapsto (F^{-1}_{X_v'\beta}(\alpha) - F^{-1}_{X_v'\beta_v}(\alpha))^2$ is continuous, which implies \eqref{eq:conv_quant_unif}.

\medskip
The continuity of $(\beta,\alpha) \mapsto E\left[\left|\indic{F_{X_v'\beta}(X_v'\beta)\le \alpha}-
\indic{F_{X_v'\beta_v}(X_v'\beta_v)\le \alpha}\right|\right]$ follows from the exact same reasoning as the continuity of $h$. Finally, we prove the continuity of
$$j: (\beta,\alpha)\mapsto E\left[\left(X_v'\beta\indic{F_{X_v'\beta}(X_v'\beta)> \alpha}- X_v'\beta_v\indic{F_{X_v'\beta_v}(X_v'\beta_v)>\alpha}\right)^2\right]$$
on $K\times[\eps,1-\eps]$. Using $a^2 - b^2=(a-b)(a+b)$, the Cauchy-Schwarz inequality and $(\sum_{i=1}^k a_i)^2\le k\sum_{i=1}^k a^2_i$, we obtain
\begin{align*}
& |j(\beta_1,\alpha_1) - j(\beta_2,\alpha_2)| \\
\le & 6^{1/2}\left\{E\left[\left(X_v'\beta_1\indic{F_{X_v'\beta_1}(X_v'\beta_1)> \alpha_1}- X_v'\beta_2\indic{F_{X_v'\beta_2}(X_v'\beta_2)> \alpha_2}\right)^2\right] \right. \\
& \left. +E\left[(X_v'\beta_v)^2\left|\indic{F_{X_v'\beta_v}(X_v'\beta_v)>\alpha_1}- \indic{F_{X_v'\beta_v}(X_v'\beta_v)> \alpha_2}\right|\right] \right\}^{1/2} \\
& \times \left\{E\left[(X_v'\beta_1)^2+(X_v'\beta_2)^2 + 2(X_v'\beta_v)^2 \right]\right\}^{1/2}.	
\end{align*}
Thus, it suffices to bound the first and second terms, corresponding to the first and second lines. Regarding the second, by applying H\"older's inequality and using $E[\|X\|^{2+\delta}]<\infty$, we just need to bound
$$E\left[\left|\indic{F_{X_v'\beta_v}(X_v'\beta_v)>\alpha_1}- \indic{F_{X_v'\beta_v}(X_v'\beta_v)> \alpha_2}\right|\right],$$
which can be done as in Step 1 above. Regarding the first term, we also reason as in Step 1, with the sole difference that because of the square, we use again H\"older's inequality and $E[\|X\|^{2+\delta}]<\infty$.


\subsubsection*{Step 4: Conclusion.} 
\label{ssub:conclusion}

 Because $(F_1,F_2)\mapsto F_1/F_2$ is Hadamard differentiable for all $(F_1,F_2)$ such that $F_2$ does not vanish, the functional delta method implies that the process
$$\mathbb{H}_n(\alpha):=n^{1/2}\left(R(\alpha, \widehat{F}_{Y_{v0}}, \widehat{F}_{X_{v0}'\widehat{\beta}_v}) - R(\alpha, F_{Y_{v0}}, F_{X_{v0}'\beta_v})\right)$$
defined on $[\eps,1-\eps]$, also converges to a Gaussian process $\mathbb{H}$. By the directional Hadamard differentiability of $\iota$, we obtain
$$n^{1/2}\left(S_\eps(\widehat{F}_{Y_{v0}}, \widehat{F}_{X_{v0}'\widehat{\beta}_v})-S_\eps(F_{Y_{v0}}, F_{X_{v0}'\beta_v})\right)  \convD L:=\iota'_{R(\cdot, F_{Y_{v0}}, F_{X_{v0}'\beta_v})}(\mathbb{H}).$$
Moreover, by the same argument as in the proof of Theorem \ref{thm:inference}, the distribution of $L$ is continuous. Combined with Theorem 2.2.1 in \cite{politis1999subsampling}, this implies that $q_{1-\alpha}(T^*)\convP c_{1-\alpha}$, the quantile of order $1-\alpha$ of $L$. Finally, under the null hypothesis, because $S_\eps(F_{Y_{v0}}, F_{X_{v0}'\beta_v})=S(F_{Y_{v0}}, F_{X_{v0}'\beta_v})=1$, we have
$$T = n^{1/2}\left(S_\eps(\widehat{F}_{Y_{v0}}, \widehat{F}_{X_{v0}'\widehat{\beta}_v})-S_\eps(F_{Y_{v0}}, F_{X_{v0}'\beta_v})\right).$$
As a result, $P(T>q_{1-\alpha}(T^*)) \to P\left(L>c_{1-\alpha}\right)=\alpha$. The second result also follows since $T\to\infty$ under the alternative.



\subsection*{5. Proof of Proposition \ref{prop:ME}} 
\label{sub:proposition_ref_prop_me}

Remark that for any random variables $A, B$ and $C$ such that  $A\cvx B$, $A\indep C$ and $B\indep C$, we have $A+C\cvx B+C$. Fix $\beta\in \B^*$. By assumption, $\xi_{Y_0}\cvx \xi_{X_0}'\beta$. Thus,
\begin{equation}
X_0^*{}'\beta + \xi_{Y_0}\cvx X_0^*{}'\beta + \xi_{X_0}'\beta=X_0'\beta.
	\label{eq:forME2}
\end{equation}
Now, because $\beta\in\B^*$, we also have, by Theorem \ref{thm:main}, $Y_0^*\cvx X_0^*{}'\beta$. Hence, by independence, $Y_0^*+\xi_{Y_0}\cvx X_0^*{}'\beta + \xi_{Y_0}$. Combined with \eqref{eq:forME2}, this yields $Y_0\cvx X_0'\beta$. Hence, $\beta\in\B$ and $\B^*\subset \B$.

\end{document}